\documentclass{article}

\usepackage{authblk}
\usepackage[square]{natbib}
\usepackage{mathtools, amsmath, amsthm, amssymb, amsbsy, amsfonts, amscd}
\usepackage{euscript}
\usepackage{bm,relsize}
\usepackage{breqn}
\usepackage{xcolor, soul}
\usepackage{empheq}
\usepackage{rotating}
\usepackage{enumerate}

\usepackage{ftnxtra}
\usepackage{fnpos,bigints}
\usepackage{appendix}

\def\Fh{\widehat{F}}
\usepackage{graphicx}
\graphicspath{{Figures/}}
\usepackage{pstool}
\def\W{\mathcal{W}}
\def\vr{\varrho}
\def\A{\mathcal{A}}
\def\D{\mathcal{D}}
\def\Wh{\widehat{\W}}
\def\wh{\widehat{w}}
\def\ph{\widehat{\psi}}
\def\Qc{\mathcal{Q}}
\def\Gc{\widehat{\mathcal{Q}}}
\usepackage{psfrag}
\def\tauh{\widehat{\tau}}
\def\kh{\widehat{\kappa}}

\numberwithin{equation}{section}

\theoremstyle{plain}

\theoremstyle{definition}	
 
 \newtheorem{remark}{Remark}[section]
 \newtheorem{example}{Example}[section]

\usepackage{subcaption}

\usepackage{hyperref}
\hypersetup{colorlinks=true, linkcolor=blue}
\hypersetup{colorlinks=true,citecolor=blue}

\DeclareMathAlphabet{\mathpzc}{OT1}{pzc}{m}{it}

\makeatletter
\newsavebox{\@brx}
\newcommand{\llangle}[1][]{\savebox{\@brx}{$\m@th{#1\langle}$}%
 \mathopen{\copy\@brx\mkern2mu\kern-0.9\wd\@brx\usebox{\@brx}}}
\newcommand{\rrangle}[1][]{\savebox{\@brx}{$\m@th{#1\rangle}$}%
 \mathclose{\copy\@brx\mkern2mu\kern-0.9\wd\@brx\usebox{\@brx}}}%
\let\oldabs\abs
\def\abs{\@ifstar{\oldabs}{\oldabs*}}
\makeatother
\def\d{\mathrm{d}}
\newcommand{\g}{\mathbf g}
\newcommand{\G}{\mathbf G}
\newcommand{\Go}{\mathring{\G}}
\newcommand{\F}{\mathbf{F}}
\newcommand{\C}{\mathbf{C}}
\newcommand{\fb}{\mathbf{b}}
\newcommand{\ic}{\mathbf{c}}
\def\veps{\varepsilon}
\newcommand{\cg}{\mathrm g}
\newcommand{\cG}{\mathrm G}
\newcommand{\cGo}{\mathring{\cG}}
\newcommand{\cF}{\mathrm F}
\newcommand{\cC}{\mathrm{C}}
\def\Wc{\overline{\mathcal{W}}}

\usepackage{accents}
\newcommand{\Fe}{\accentset{e}{\F}}
\newcommand{\cFe}{\accentset{e}{\mathrm{F}}}
\newcommand{\Fa}{\accentset{a}{\F}}
\newcommand{\cFa}{\accentset{a}{\mathrm{F}}}
\newcommand{\Ce}{\accentset{e}{\C}}
\newcommand{\cCe}{\accentset{e}{\mathrm{C}}}
\newcommand{\Be}{\accentset{e}{\mathbf{B}}}
\newcommand{\cBe}{\accentset{e}{\mathrm{B}}}
\newcommand{\ce}{\accentset{e}{\mathbf{c}}}
\newcommand{\cce}{\accentset{e}{\mathrm{c}}}
\newcommand{\be}{\accentset{e}{\mathbf{b}}}
\newcommand{\cbe}{\accentset{e}{\mathrm{b}}}

\def\Xv{\mathbf{X}}

\newcommand{\sharpo}{\mathring\sharp}
\newcommand{\flato}{\mathring\flat}

\begin{document}
\bibliographystyle{abbrvnat}

\title{On the Mathematical Structure of Helical Shear in Anelastic Continua}

\author[1]{Souhayl Sadik\thanks{Corresponding author, e-mail: sosa@mpe.au.dk}}
\author[2]{Giuseppe Saccomandi}
\author[2]{Luigi Vergori}

\affil[1]{\small \textit{Department of Mechanical and Production Engineering, Aarhus University, 8000~Aarhus~C, Denmark}}
\affil[2]{\small \textit{Dipartimento di Ingegneria, Università degli Studi di Perugia, via Goffredo Duranti 93, Perugia, 06125, Italy}}

\maketitle


\begin{abstract}
We study helical shear of a thick-walled tube made of an incompressible, isotropic hyper-anelastic solid carrying a radially symmetric distribution of axial screw dislocations. Through the Bilby–Kröner–Lee decomposition, the dislocations are encoded in a Weitzenböck material manifold, whose underlying Riemannian metric alone enters the elastic response. Equilibrium reduces at each radius to an algebraic system for the azimuthal and axial shear strains, whose cross terms are proportional to the dislocation strength and to the dependence of the strain energy on the second invariant; in every material, the two shear modes therefore couple only in the presence of dislocations. Exact solutions are obtained for neo-Hookean, Fung–Demiray, Gent, Knowles, and Mooney–Rivlin materials, and numerical ones for Gent–Gent materials. The universal relations of Horgan and Saccomandi are extended to dislocated generalized neo-Hookean tubes. Dislocations stiffen the tube when the strain energy stiffens with the first invariant, couple the two shear modes chirally when it depends on the second, and, in some cases, restrict admissibility: in Gent and Gent–Gent materials they cap the dislocation strength, and in Knowles materials they contract the admissible traction region when the power-law index is below the critical exponent $n=1/2$. A weakly nonlinear analysis, to third order in the applied tractions and dislocation strength, shows that, at first order, i.e. within the linear theory of elasticity, the dislocations leave the shear response unaffected while carrying a residual stress. The chiral coupling enters only at second order, through the dependence of the strain energy on the second invariant, so that, in the presence of axial screw dislocations, every material couples the modes as the Mooney–Rivlin solid with the same second- and third-order elastic moduli. At third order, no further coupling arises, but the modes interact through stiffness corrections involving the fourth-order elastic modulus, with the dislocations further correcting each mode independently of the other.
\end{abstract}

\begin{description}
\item[Keywords:] Controllable Deformations, Anelasticity, Helical Shear, Screw Dislocations, Nonlinear Elasticity.
\end{description}

\tableofcontents

\section{Introduction}

The explosion of modern nonlinear continuum mechanics took place in the post–World War II period, mainly thanks to the celebrated works of Ronald Rivlin in the mathematical theory of hyperelasticity \citep{destrade2019rivlin}. The remarkable results obtained by Rivlin and his collaborators clearly and convincingly demonstrated that the time was ripe for bold developments in nonlinear theories.
Among the elastic results of that particularly prolific period, two stand out for their generality and mechanical significance: universal solutions and universal relations. Specifically we define:
\begin{itemize}
\item Universal solution. A deformation or motion is said to be universal if it satisfies the equilibrium equations with zero body force for all materials in a given class, and if it is sustained in equilibrium solely by appropriate surface tractions.
\item Universal relation. For a given deformation or motion, a local universal relation is an equation relating the stress components to the position vector, valid at any point of the body and identical for any material belonging to the considered class.
\end{itemize}
The importance of these universal results was clearly emphasized by \cite{beatty1987topics}:
\begin{quote}
“Universal deformations and universal results of various kinds are road signs posted to direct and to warn the experimenter in his exploration of the constitutive properties of real materials. If these signs are ignored, he can only wander aimlessly along an otherwise uncharted labyrinth in the vast realm of materials science. If they are thoughtfully evaluated, he will discover the rich rewards locked in the smaller domain of the science of highly elastic materials.”
\end{quote}
The first universal solutions were discovered, by trial, by Rivlin himself; however, the first systematic characterization of these results can be found in \citep{Ericksen1954} for incompressible, uniform, isotropic elastic materials and in \citep{Ericksen1955} for unconstrained, uniform, isotropic elastic materials. The latter case turned out to be much less complicated than its incompressible counterpart: the only universal deformations admitted by this class of materials are homogeneous deformations. On the other hand, for incompressible materials Ericksen’s results were not conclusive, and two possibilities were left open. One of these possibilities turned out to be void, as shown by \cite{marris1970universal}; the other case, namely when the principal stretches are constant but the local rotation does not reduce to the identity, has had a more intricate and intriguing history. A counter-example effectively reopened the problem and quickly led to the discovery of a new family of universal deformations independently by \cite{singh1965note} and \cite{klingbeil1966class}. 

A comprehensive review of universal solutions, updated up to the year 2000 and also addressing universal relations, can be found in \citep{saccomandi2001universal, saccomandi2001universalb}.

Universal deformations have since been extended to anelasticity. \citet{yavari2016anelastic} showed that, in homogeneous compressible isotropic anelastic solids, the universal deformations are precisely the covariantly homogeneous ones, and \citet{goodbrake2020anelastic} extended the analysis to the incompressible case, recovering the anelastic counterparts of Ericksen's families and of the Klingbeil--Pipkin--Shield--Singh solutions. These results have reopened the universal program of nonlinear elasticity \citep{yavari2023universal}.

In 1980 at the IUTAM symposium on Finite Elasticity held at the Lehigh University, Currie and Hayes presented a short note. The incipit in \citep{currie1981non} is quite interesting 
\begin{quote}
"Ericksen's theorem [1] has had a profound influence on the development of the theory of finite deformations of compressible elastic materials. \ldots Being proved early in the modern period of
resurgence of finite elasticity, Ericksen's theorem shaped the subsequent
development of the theory. The result has been that until recently little
emphasis has been placed on non-homogeneous deformations. The corresponding
theorem proved by Ericksen [2] for incompressible materials has had
a similar effect on the theory for those materials; most deformations
considered belong to the small class of universal deformations."
\end{quote}

In fact, Ericksen’s important results “froze” research for years on other types of solutions which, while not universal, still had good reasons to be investigated. It should be noted, however, that some work had already been carried out in the immediate postwar period—for instance \citep{adkins1954some}—but this line of research was almost entirely abandoned until it was revitalized by Currie and Hayes. 

To clarify this point, it is useful to revisit the terminology introduced for universal solutions, with the aim of providing a rigorous classification of the different types of admissible solutions. Towards this end we using the following definitions:
\begin{itemize}
\item A controllable solution is a deformation which satisfies the equilibrium equations of a field theory with zero body force, that is the deformation is controlled by means of the application of surfaces traction alone.
\item A general solution is a controllable solution which is admissible for any constitutive equation in a given class of materials.
\item A controllable solution that is the same for all materials in a given class is an universal solution. Therefore, a universal solution is a general solution with the same kinematics for all materials in the given class.
\item A partial solution is a controllable solution with a constitutive dependent geometry (or kinematics) that is possible only for a subset of the given class of materials.
\end{itemize}

Beatty very effectively explained the centrality of universal solutions, especially in a rational experimental campaign designed to determine the approximate form of the strain-energy function that describes the constitutive behavior of a given material. Indeed, knowing the deformation kinematics a priori allows one to focus solely on determining the stresses or, more precisely, on determining the derivatives of the elastic potential.

It would be complicated in this context not to know the geometry of the deformation, as one would have to deform the material according to modes that depend on the chosen strain-energy function and thus on the mechanical quantities that need to be measured.

By applying this known universal deformation in a controlled experiment, the researcher can focus entirely on measuring the forces (or pressures) required to sustain it. These forces are directly related to the Cauchy stresses.

On the other hand, for appreciating the difference in behavior between constitutive behaviours, generalized solutions are optimal. That is, identifying specific boundary-value problems that can find an ideal solution for a certain class of deformations which, in this specific case, also differ in their geometry from material to material.

Using generalized solutions (known boundary conditions) for complex problems to test how well a calibrated constitutive model can predict real-world behavior and differentiate itself from other materials. This is a comparative and validating process which can teach us a lot, both on a mechanical level and also regarding the mathematics of the problem.

The search for non-universal solutions flourished following \citep{currie1981non}, also because by the 1990s the computing power at our disposal had become such that even numerical approaches became easily feasible. Unfortunately, even the class of general solutions is not particularly large. In this restricted field, a class of general solutions is possible in the case of incompressible materials for circular cylindrical sectors. These solutions are deformations of axial and azimuthal shear or their combination, which is referred to as helical shear. 

The main purpose of this work is to address the problem of helical shear and its particular cases in an anelastic context. To the best of our knowledge, this is the first time a problem of this kind has been addressed in the literature. 

As has already been mentioned, in \citep{adkins1954some} within the context of incompressible isotropic nonlinear elastic theory, the problem of axial and azimuthal shear was addressed for the first time. The problem of helical shear, on the other hand, is studied in \citep{ogden1973combined}. Interesting additions to the results of such paper in the case of generalized neo-Hookean material has been obtained in \citep{horgan2003helical} and the case of non convex strain elastic energy density has been considered in \citep{fosdick1986helical}.

This work constitutes, to the best of our knowledge, the first systematic investigation of the solutions associated with a simple but mechanically significant class of deformations in nonlinear anelasticity. By extending the helical shear problem to solids with distributed defects, it opens a new chapter in nonlinear elasticity, in which the kinematics of the deformation remains simple enough to be analysed in closed form, yet rich enough to reveal the interplay between eigenstrains, constitutive nonlinearity, and applied tractions.

The remainder of this paper is organized as follows.
In~\S\ref{S:Anelasticity} we recall the geometric and kinematic foundations of nonlinear anelasticity, introducing the Bilby--Kröner--Lee multiplicative decomposition of the deformation gradient, the material metric induced by the eigenstrains, the governing equations, and the constitutive assumptions for incompressible isotropic hyper-anelastic solids; we also review the strain-energy models adopted in the sequel.
In~\S\ref{S:Problem} we formulate the helical shear problem for a thick-walled circular tube containing a distributed field of axial screw dislocations: we describe the geometry of the dislocated tube, derive the kinematics of the deformation, compute the stress components, and obtain the reduced equilibrium equations in both dimensional and dimensionless form.
In~\S\ref{S:Examples} we solve the equilibrium problem in closed form for neo-Hookean, generalized neo-Hookean (Fung--Demiray, Gent, and Knowles), and Mooney--Rivlin materials, and numerically for Gent--Gent materials, discussing existence, uniqueness, admissibility conditions, and the mechanical role of the dislocation field in each case.
In~\S\ref{S:WNL} we develop a weakly nonlinear theory of the fourth order via a perturbation expansion in the dislocation strength and the applied tractions, identifying the order at which, and the elastic moduli through which, the dislocation field couples the two shear modes, and we verify the expansion against the numerical Gent--Gent solutions.
Finally, in~\S\ref{S:Conclusion} we summarize our findings and outline possible directions for future research.

\section{Nonlinear Anelasticity}\label{S:Anelasticity}

\subsection{Kinematics}

\paragraph{Body, Motion, and Euclidean geometry.}
Consider a deformable elastic body, and identify it with a $3$-dimensional submanifold $\mathcal{B}$ of the Euclidean ambient space $\mathcal{S}=\mathbb R^3$. A motion of the body in the time interval $[0,T)$ is defined as a mapping $\varphi: \mathcal{B}\times [0,T) \to \mathcal{S}$, such that, for any $t\in(0,T)$, the mapping $\varphi_t(.)\equiv\varphi(\cdot,t)$ is one-to-one and assigns to each material point $ X \in \mathcal{B}$ the spatial position $x = \varphi_t(X) \in \mathcal{S}$ that it occupies at time~$t$.
We refer to $ \mathcal{B}_t\equiv \varphi_t(\mathcal{B}) \subset \mathcal{S}$ as the deformed body at time $t$.
We adopt the conventional notation such that quantities relating to the reference state are denoted with uppercase letters (e.g., $ X \in \mathcal{B} $), and those relating to the spatial (deformed) state with lowercase letters (e.g., $ x \in \mathcal{B}_t $). Local coordinate charts are denoted $ \{X^A\} $ and $ \{x^a\} $ on $ \mathcal{B} $ and $ \mathcal{S} $, respectively. Coordinate bases are denoted by $ \partial_A \equiv \frac{\partial}{\partial X^A} $ and $ \partial_a \equiv \frac{\partial}{\partial x^a} $, with dual bases $ \{dX^A\} $ and $ \{dx^a\} $. We employ Einstein's index summation notation throughout, e.g., $ u^i v_i \equiv \sum_i u^i v_i $.

The ambient space $ \mathcal{S} $ is endowed with the Euclidean metric $ \g = g_{ab}\, dx^a \otimes dx^b $. In a standard Cartesian coordinate, the metric $\g$ is represented in components by the standard Kronecker delta symbol ${g_{ab}=\delta_{ab}}$. For spatial vector fields $ \mathbf{u}, \mathbf{w} \in \chi(\mathcal{S}) $,\footnote{$\chi(\mathcal{S})$ is the set of spatial vector fields, i.e. vector fields over the spatial manifold $\mathcal{S}$.} the $\g$-inner product is denoted $ \llangle \mathbf{u}, \mathbf{w} \rrangle_{\g} = u^a w^b g_{ab} $. For a spatial covector field $ \boldsymbol{\omega} \in \chi^\star(\mathcal{S}) $,\footnote{$\chi^\star(\mathcal{S})$ is the set of spatial covector fields, i.e. 1-forms over the spatial manifold $\mathcal{S}$.} we write the natural pairing as $ \langle \boldsymbol{\omega}, \mathbf{u} \rangle = \omega_a u^a $. The corresponding volume form on $ \mathcal{S} $ is $ dv = \sqrt{\det \g}\, dx^1 \wedge dx^2 \wedge dx^3 $. We denote by $ \bar\nabla $ the Levi-Civita connection of $ (\mathcal{S}, \g) $, with local Christoffel symbols $ {\gamma^a}_{bc} $.

Recalling that $\mathcal{B}$ is a submanifold of $\mathcal{S}$, the spatial Euclidean metric $ \g $ induces a Euclidean reference metric, denoted $\Go = \cGo_{AB}\, dX^A \otimes dX^B$, on $\mathcal{B}$. It may be thought of as the restriction of $\g$ to $\mathcal B$ or, equivalently, as the pullback by the inclusion map $ \iota: \mathcal{B} \hookrightarrow \mathcal{S}$, i.e. $ \Go = \iota^* \g $.
We define the $\Go$-inner product and dual pairing on $ \mathcal{B} $ analogously to $ \mathcal{S} $, and denote the corresponding volume form by $ d\mathring{V} = \sqrt{\det \Go} \, dX^1 \wedge dX^2 \wedge dX^3 $. The Levi-Civita connection of $ (\mathcal{B}, \Go) $ is written $ \mathring{\nabla} $, with Christoffel symbols $ {\mathring{\Gamma}^A}_{BC} $.
Note that, while $ \Go $ reflects the inherited Euclidean geometry, it does not generally represent the intrinsic geometry of the material, which may differ due to (anelastic) eigenstrains.

\paragraph{Deformation Gradient.}
To describe the local strain in a deforming solid, we define the deformation gradient given by the tangent map of the deformation, i.e.~$ \F(X,t) \equiv T\varphi_t(X): T_X\mathcal{B} \to T_{\varphi_t(X)}\varphi_t(\mathcal{B}) $. It is a two-point tensor that maps tangent vectors from the reference configuration $ \mathcal{B} $ to the current configuration $ \varphi_t(\mathcal{B}) $. In local coordinates, it has the following representation
\begin{equation}
	\F(X) = \frac{\partial \varphi^a}{\partial X^A} \, \partial_a \otimes dX^A\,.
\end{equation}

\paragraph{BKL Decomposition and Material Configuration.}
To model anelastic effects, particularly the presence of defects, we adopt the Bilby--Kr{\"o}ner--Lee (BKL) multiplicative decomposition of the deformation gradient \citep{Bilby1957,Kroner1959,leeliu1967,Lee1969,Sadik2017}:
\begin{equation}
\F = \Fe \Fa\,.
\end{equation}
In this decomposition, $ \Fa: T_X\mathcal{B} \to T_X\mathcal{B} $ captures the local anelastic distortions due to defect-induced eigenstrains, while $ \Fe: T_X\mathcal{B} \to T_x\mathcal{S} $ accounts for the reversible elastic distortions---both contributions are generally incompatible, i.e. unlike the deformation gradient, they may not be written as a tangent of a smooth map. The resulting locally stress-free state achieved by $ \Fa $ is also, in general, incompatible and distinct from the initial undeformed state. This abstract manifold, known as the material manifold, is where the natural, stress-free distances are realized, and its geometry is encoded by the material metric $ \G $, given as the pull-back of the reference metric $ \Go $ by $ \Fa $, i.e. $\G = \Fa^* \Go$ \citep{Yavari2021Eshelby}, which in components reads $\cG_{AB}=\cFa^I{}_A \cGo_{IJ} \cFa^J{}_B$.
The material metric provides a unified geometric description of eigenstrains of various origins~\citep{SaYa2025GenColemanNoll}, including thermal strains~\citep{SadYav2015Thermo}, growth~\citep{Sad_etal2016geometric} and viscous strains~\citep{SadYav2023ViscoElast, SadYav2026ViscoAnElast}; here it encodes a distribution of screw dislocations~\citep{YavariGoriely2012Dislocations, OzakinYavari2014}.
We define the $\G$-inner product in $ T_X\mathcal{B} $ analogously to the $\Go$-inner product, and denote the corresponding material volume form by $ d{V} = \sqrt{\det {\G}} \, dX^1 \wedge dX^2 \wedge dX^3 \, $. The Levi-Civita connection of $ (\mathcal{B}, {\G}) $ is written $ {\nabla} $, with Christoffel symbols $ {{\Gamma}^A}_{BC} $.
We define the Jacobian $J$ of the motion as the quantity relating the material (stress-free) and spatial volume elements by $\varphi^* dv = J dV$. It can be shown that \citep{SaYa2026PhysComp}
\begin{equation}
J = \sqrt{\frac{\det\g}{\det\G}} \det\F = \sqrt{\frac{\det\g}{\det\Go}}\, \det\Fe\,.
\end{equation}

\paragraph{Musical Isomorphisms and Index Operations.}
On a Riemannian manifold $(\mathcal{M}, \mathbf{m})$, the metric induces a canonical bundle isomorphism $\flat: \chi(\mathcal{M}) \to \chi^\star(\mathcal{M})$, called the \emph{flat operator}, defined by $\mathbf{v}^\flat \equiv \mathbf{m}(\mathbf{v}, \cdot)$. In components, this lowers an index as $\mathrm{v}_j = \mathrm{v}^i \mathrm{m}_{ij}$. Its inverse $\sharp: \chi^\star(\mathcal{M}) \to \chi(\mathcal{M})$, called the \emph{sharp operator}, defined by $\omega^\sharp \equiv \mathbf{m}^{-1}(\omega,.)$, raises an index via $\omega^j = \omega_i \mathrm{m}^{-ij}$, where $\mathrm{m}^{-ij}$ denotes the components of $\mathbf{m}^{-1}$. These operations extend naturally to tensors of arbitrary type $\binom{n}{m}$: the flat operator $\flat$ contracts the contravariant indices with $\mathbf{m}$, while $\sharp$ contracts the covariant indices with $\mathbf{m}^{-1}$. In particular, applying $\sharp$ to the metric itself yields its inverse: $\mathbf{m}^\sharp=\mathbf{m}^{-1}\mathbf{m}\mathbf{m}^{-1}=\mathbf{m}^{-1}$.

In the reference configuration $\mathcal{B}$, we have defined two metrics, $\Go$ and $\G$, giving rise to two pairs of musical isomorphisms, which we denote $\flato$, $\sharpo$ and $\flat$, $\sharp$, respectively. In the spatial configuration $\mathcal{S}$, only the Euclidean metric $\g$ is defined, so there is a unique pair of musical isomorphisms; we denote these simply by $\flat$ and $\sharp$, as no ambiguity arises.

\paragraph{Adjoint.}
The adjoint of the deformation gradient is the unique linear map $ \F^\star $ satisfying $ {\langle \mathbf{u}, \F\mathbf{U} \rangle = \langle \F^\star \mathbf{u}, \mathbf{U} \rangle,}$ ${\forall\,\mathbf{U} \in T_X\mathcal B}$, ${\forall\mathbf{u} \in T^\star_{\varphi(X)}\mathcal S}$. It reads in components as $ (\mathrm{F}^\star)^A{}_a = \mathrm{F}^a{}_A $, and hence
${\F^\star(X) = \cF^a{}_A(X)\, dX^A \otimes \partial_a}$.\footnote{Note that while the adjoint is denoted by~``$\star$'', the pull-back and push-forward operations are denoted by~``$*$'' and must not be conflated with it, though they are related: for example, pulling back the spatial metric reads $\F^*\g = \F^\star \g \F$, and pushing forward the material metric reads $\F_*\G^\sharp = \F \G^\sharp \F^\star$.}

\paragraph{Transpose.}
Defining the transpose of $ \F $ requires a metric, and thus depends on whether $ \Go $ or $ \G $ is used in the reference configuration. The transpose with respect to $ \Go $ and $ \g $ is defined by the relation:
\begin{equation}
\llangle \F \mathbf{U}, \mathbf{u} \rrangle_{\g} = \llangle \mathbf{U}, \F^{\mathring{\mathsf{T}}} \mathbf{u} \rrangle_{\Go}\,,\quad
\forall\,\mathbf{U} \in T_X\mathcal B\,,\;\forall\mathbf{u} \in T^\star_{\varphi(X)}\mathcal S\,,
\end{equation}
leading to $\F^{\mathring{\mathsf{T}}}=\Go^\sharpo \F^\star \g$, or in components $(\mathrm{F}^{\mathring{\mathsf{T}}})^A{}_a = \cGo^{AB} \cF^b{}_B \cg_{ba}$.
Similarly, using $ \G $ instead of $ \Go $, we define $ \F^{\mathsf{T}} $ by:
\begin{equation}
\llangle \F \mathbf{U}, \mathbf{u} \rrangle_{\g} = \llangle \mathbf{U}, \F^{\mathsf{T}} \mathbf{u} \rrangle_{\G}\,,\quad
\forall\,\mathbf{U} \in T_X\mathcal B\,,\;\forall\mathbf{u} \in T^\star_{\varphi(X)}\mathcal S\,,
\end{equation}
and find $\F^{\mathsf{T}}=\G^\sharp \F^\star \g$, or is components $(\mathrm{F}^{\mathsf{T}})^A{}_a = \cG^{AB} \cF^b{}_B \cg_{ba}$.

\paragraph{Derived Strain Measures.}
Using the geometric tools defined above, we introduce the following measures of strain with respect to the total deformation
\begin{itemize}
  \item {Piola deformation tensor}: $ \mathbf{B} = \varphi^* \g^\sharp = \F^{-1} \g^\sharp \F^{-\star} $, in components, ${\mathrm B^{AB} = (\cF^{-1})^A{}_a \cg^{ab} (\cF^{-1})^B{}_b}$.
  \item {Right Cauchy--Green tensor}: $ \C = \varphi^* \g = \F^\star \g \F $, in components, ${\mathrm C_{AB} = \cF^a{}_A \cg_{ab} \cF^b{}_B}$.
  \item {Left Cauchy--Green (Finger) tensor}: $ \fb = \varphi_* \G^\sharp = \F \G^\sharp \F^\star $, in components, ${\mathrm b^{ab} = \cF^a{}_A \cG^{AB} \cF^b{}_B}$.
  \item {Inverse Finger tensor}: $ \ic = \varphi_* \G = \F^{-\star} \G \F^{-1} $, in components, ${\mathrm c_{ab} = (\cF^{-1})^A{}_a \cG_{AB} (\cF^{-1})^B{}_b}$.
\end{itemize}
Similarly, we define their elastic distortion counterparts:
\begin{itemize}
  \item {Elastic Piola deformation tensor}: $ \Be = \Fe^* \g^{-1} = \Fe^{-1} \g^{-1} \Fe^{-\star} $, in components, $\cBe^{AB} = {(\cFe^{-1})^A{}_a \cg^{ab} (\cFe^{-1})^B{}_b}$.
  \item {Elastic right Cauchy--Green tensor}: $ \Ce = \Fe^* \g = \Fe^\star \g \Fe $, in components, $\cCe_{AB} = {\cFe^a{}_A \cg_{ab} \cFe^b{}_B}$.
  \item {Elastic left Cauchy--Green (Finger) tensor}: $ \be = \Fe_* \Go^{-1} = \Fe \Go^{-1} \Fe^\star $, in components, $\cbe^{ab} = {\cFe^a{}_A \cGo^{AB} \cFe^b{}_B}$.
  \item {Elastic inverse Finger tensor}: $ \ce = \Fe_* \Go = \Fe^{-\star} \Go \Fe^{-1} $, in components, $\cce_{ab} = {(\cFe^{-1})^A{}_a \cGo_{AB} (\cFe^{-1})^B{}_b}$.
\end{itemize}

\begin{remark}
Note that since $\mathbf{G}=\Fa^*\Go$, it follows that $\mathbf{B}=\Fa^*\Be$ and $\mathbf{C}=\Fa_*\Ce$, while $\fb=\be$ and $\ic=\ce$.
\end{remark}

\subsection{Governing Equations}

\paragraph{Balance of mass.}
Let $\rho_m=\rho_m(X,t)$ be the material mass density, i.e. the mass density measured with respect to the material volume element $ d{V} = \sqrt{\det {\G}} \, dX^1 \wedge dX^2 \wedge dX^3$. Material balance of mass is given by
\begin{equation}
\label{eq:mass_bal}
\dot{\rho}_m + \rho_m\,\dot{\Fa}\!:\!\Fa^{-1} = 0
\,.
\end{equation}
Let $\rho_s=\rho_m(\varphi_t^{-1}(x),t)J^{-1}$ be the spatial mass density, i.e. the mass density per unit spatial volume. Then, the spatial form of the equation of balance of mass reads
\begin{equation}
\dot{\rho}_s + \rho_s \operatorname{div}\mathbf{v}=0
\,.
\end{equation}

\paragraph{Constitutive assumptions.}
We consider a hyper-anelastic constitutive model, i.e. we assume that there exists a strain energy density (per unit stress-free material volume) functional $\mathcal W=\mathring{\mathcal W}(\Xv,\Fe, \mathring{\G}, \g\circ\varphi)$.
Objectivity implies that the elastic energy may be written as the functional $\mathcal W=\hat{\mathring{\mathcal W}}(\Xv,\Ce, \mathring{\G})$.
The Cauchy stress tensor may be expressed in the following equivalent representation forms
\begin{equation}
\boldsymbol\sigma
= \frac{1}{J}\,\g^\sharp \frac{\partial \mathring{\mathcal W}}{\partial \Fe}\Fe^\star
= \frac{2}{J}\, \Fe \frac{\partial \hat{\mathring{\mathcal W}}}{\partial \Ce} \Fe^\star
\,.
\end{equation}

\paragraph{Conservation of momenta.}
In the absence of inertial and body forces, the balance of linear momentum, in terms of the Cauchy stress tensor, reads
\begin{equation}
\operatorname{div}\boldsymbol\sigma = \mathbf 0\,,
\end{equation}
and the balance of angular momentum
\begin{equation}
\boldsymbol\sigma^{\mathsf T} = \boldsymbol\sigma\,.
\end{equation}

\paragraph{Material symmetry.}
The material symmetry group $\mathring{\mathcal G}_X$ of a hyper-anelastic body—whose constitutive behavior is described by a free energy functional—at a point $X \in \mathcal{B}$, relative to the Euclidean reference configuration $(\mathcal{B}, \Go)$, is defined as the collection of $\mathring{\mathbf{K}} \in \mathrm{Orth}(\Go) = \{ \mathbf{Q}: T_X\mathcal{B} \to T_X\mathcal{B} ~|~ \mathbf{Q}^\star \Go \mathbf{Q} = \Go \}$ that leave $\mathring{\mathcal W}$ invariant under the following transformation:
\begin{equation} \label{Elasticity-Sym-Group-Reform}
\mathring{\mathcal W}(X, \mathring{\mathbf{K}}^* \Fe, \mathring{\mathbf{K}}^* \Go, \g)
= \mathring{\mathcal W}(X, \Fe, \Go, \g)\,,
\end{equation}
for every choice of elastic distortion tensor $\Fe$.
To account for the material symmetries encoded by $\mathring{\mathcal{G}}_X$, one may introduce a set of structural tensors $\mathring{\boldsymbol{\Lambda}}$ that generate the space of tensors invariant under the group action.\footnote{A finite set of structural tensors $\mathring{\boldsymbol{\Lambda}}_i$, $i = 1, \dots, N$, can characterize the subgroup $\mathring{\mathcal{G}}_X \subseteq \mathrm{Orth}(\Go)$; see, e.g., \citep{liu1982, boehler1987, zheng1993, zheng1994theory, lu2000covariant, MazzucatoRachele2006}.} When these tensors are incorporated into the free energy as independent arguments, the constitutive law becomes materially covariant \citep{Lu2012MatCov}. Specifically, for any linear isomorphism $\boldsymbol{\mathsf{T}}: T_X\mathcal{B} \to T_X\mathcal{B}$, material covariance requires:
\begin{equation}
\mathring{\mathcal W}(\Xv,\Fe, \mathring{\boldsymbol{\Lambda}}, \Go, \g)
= \mathring{\mathcal W}(\Xv,\boldsymbol{\mathsf{T}}^* \Fe, \boldsymbol{\mathsf{T}}^* \mathring{\boldsymbol{\Lambda}}, \boldsymbol{\mathsf{T}}^* \Go, \g)\,.
\end{equation}
Choosing $\boldsymbol{\mathsf{T}} = \Fa$ and using the identities $\Fa^* \Fe = \Fe \Fa = \F$ and $\G = \Fa^* \Go = \Fa^\star \Go \Fa$, we obtain the following equivalent representation of the free energy \citep{yavari2016anelastic}:
\begin{equation} \label{eq:FreeAnelast-Reform}
\mathcal{W}
= \mathcal{W}(X, \F, \boldsymbol{\Lambda}, \G, \g)\,,
\end{equation}
where $\boldsymbol{\Lambda} := \Fa^* \mathring{\boldsymbol{\Lambda}}$ represents the structural tensors expressed in the configuration $(\mathcal{B}, \G)$. Equation~\eqref{eq:FreeAnelast-Reform} thus illustrates how the material metric $\G = \Fa^* \Go$ arises naturally from the underlying hyper-anelastic framework.
Objectivity further implies that the energy functional may be again alternatively written as $\mathcal W=\widetilde{\mathcal W}(X, \C, \boldsymbol{\Lambda}, \G)$. In terms of the representations $\mathcal W$ and $\widetilde{\mathcal W}$, the Cauchy stress may be recast as
\begin{equation}
\boldsymbol\sigma
= \frac{1}{J}\,\g^\sharp \frac{\partial {\mathcal W}}{\partial \F}\F^\star
= \frac{2}{J}\, \F \frac{\partial \widetilde{\mathcal W}}{\partial \mathbf C} \F^\star
\,.
\end{equation}

\paragraph{Isotropic representation formula.}
For an isotropic material, the elastic energy depends only on the principal invariants of $\mathbf C$, namely $\mathcal{W}=\overline{\mathcal{W}}(X,I_1,I_2,I_3)$, where
\begin{equation}
\label{eq:iso_inv}
I_1	=\operatorname{tr}_{\G}\C
	=\cC_{AB}\,\cG^{BA}\,,
\quad
I_2	=\frac{1}{2}\left(I_1^2-\operatorname{tr}_{\G}\C^2\right)
	=\frac{1}{2}\left(I_1^2-\cG^{AB}\,\cC_{BC}\,\cG^{CD}\,\cC_{DA}\right)\,,
\quad
I_3	=\frac{\det\mathbf C}{\det \textrm G}=J^2\,.
\end{equation}
Denoting $\Wc_i \equiv \partial \overline{\mathcal{W}}/ \partial I_i$, for $i \in \{1,2,3\}$, the Cauchy stress in an isotropic anelastic solid may be written as
\begin{equation}\label{eq:sig_iso_represent}
\begin{aligned}
	\boldsymbol{\sigma} &= \frac{2}{J}\left[
	\Wc_1\fb
	-\Wc_2J^2\ic^{\sharp}
	+ \left(
	\Wc_2I_2
	+\Wc_3J^2
	\right)\mathbf{g}^{\sharp}
	\right]
	\,.
\end{aligned}
\end{equation}
If we further assume an incompressible material, i.e. $I_3=J^2=1$, it follows that the Cauchy stress may be written as
\begin{equation}
\label{eq:sig_iso_represent_inc}
\begin{aligned}
	\boldsymbol{\sigma} &= 
	- p \mathbf{g}^{\sharp}
	+ 2\left[
	\Wc_1 \fb
	-\Wc_2 \ic^{\sharp}
	\right]
	\,,
\end{aligned}
\end{equation}
where $p$ is the Lagrange multiplier associated with the incompressibility constraint $J=1$.

\subsection{Mathematical models for the strain-energy function}
\label{S:Const_Mod}

In what follows, we assume an incompressible isotropic homogeneous hyper-anelastic solid, i.e. the strain energy density may be written as
\begin{equation}\label{Wchom}
\W = \Wc(I_1, I_2)\,.
\end{equation}
Since the strain energy is defined up to an additive constant, it is customary to normalise it so that it vanishes in the stress-free material state, i.e. when $\mathbf{C}=\mathbf{G}$, for which $I_1=I_2=3$, whence
\begin{equation}\label{refw}
\Wc(3,3)=0\,.
\end{equation}
Note that in the presence of eigenstrains this state is not the Euclidean reference configuration $(\mathcal{B},\Go)$ but the material reference configuration $(\mathcal{B},\G)$: in the Euclidean configuration $\C=\Go$, the invariants exceed $3$, and the body stores residual eigenstrain energy in the absence of any applied load.
Requiring $\Wc$ to be compatible with the linearized theory of elasticity (which is valid at infinitesimal deformations), it is necessary that \citep{saveza}
\begin{equation}\label{eq:mu}
  \Wc_1(3,3)+\Wc_2(3,3)=\dfrac{\mu}{2},
\end{equation}
where $\mu$ is the infinitesimal shear modulus.
Requiring the strain energy function $\Wc$ to be compatible with the weakly nonlinear theory of the fourth order (which is valid for small but not necessarily infinitesimal deformations), it is necessary for ~\eqref{eq:mu} and the following two identities to hold \citep{hamilton, saveza}
\begin{subequations}\label{ad}
\begin{gather}
  \label{alan}
  \Wc_1(3,3)+2\Wc_2(3,3)=-\dfrac{\A}{8},
  \\
  \label{alan2}
  \Wc_1(3,3)+3\Wc_2(3,3)+\Wc_{11}(3,3)+2\Wc_{12}(3,3)+\Wc_{22}(3,3)=\dfrac\D2\,,
\end{gather}
\end{subequations}
with $\A$ and $\D$ being the Landau elastic moduli of the third and fourth order, respectively. 

For large deformations, the search for the most appropriate functional form of the strain energy which predicts reasonable material behaviour opened a debate that is still open. For this reason, there are a plethora of mathematical models for the strain energy function of isotropic elastic solids. A choice of a suitable functional form of $\Wc$ may be achieved by testing theoretical predictions against experimental data. 

Subject to the analyticity of the strain-energy in a neighbourhood of the
reference configuration, one can write $\Wc$ as a power series of $I_1 - 3$ and $I_2 - 3$, that is,
\begin{equation}\label{expansion}
\Wc=\dfrac12\sum_{p,q=0}^{+\infty}C_{pq}(I_1-3)^p(I_2-3)^q,
\end{equation}
where the coefficients $C_{pq}$ do not depend on the deformation and, in view of the requirement~\eqref{refw}, $C_{00} = 0$. While~\eqref{expansion} is an infinite series, truncating it at a sufficiently high order approximates the strain energy of a given material as closely as desired.
The two simplest such truncations are those most widely adopted in the development of the non-linear theory of elasticity and its early applications. Retaining only the first order term in~$I_1$ yields the neo-Hookean strain-energy function
\begin{equation}\label{nHmod}
  \Wc^{\textrm{nH}}=\dfrac{\mu}{2}(I_1-3)\,,\quad \mu>0,
\end{equation}
with $C_{10}=\mu$ for consistency with the linear theory, while retaining both first-order terms yields the Mooney--Rivlin model
\begin{equation}\label{mrmod}
\Wc^{\textrm{MR}} =\dfrac{\mu}{2}\Big[\alpha (I_1-3)+(1-\alpha)(I_2-3)\Big]\,,\quad \mu>0 \quad \textrm{and} \quad \alpha\in]0,1]\,,
\end{equation}
where, in view of~\eqref{eq:mu} and~\eqref{alan}, one finds $C_{10}=\mu\alpha=2\mu+\A/4$ and $C_{01}=\mu(1-\alpha)=-\mu-\A/4$, with the dimensionless parameter $\alpha=2+\A/(4\mu)$ set by the infinitesimal shear modulus $\mu$ and the Landau elastic modulus of the third order $\A$. Further, the empirical inequalities $\Wc_1>0$ and $\Wc_2\geq0$ \citep{TrueNoll2004Field} require $C_{10}>0$ and $C_{01}\geq0$, whence $\alpha\in\,]0,1]$.

Applications of non-linear elasticity to soft tissues make extensive use of the so-called generalized neo-Hookean models, for which the strain energy depends on the first invariant alone, i.e. $\Wc=\Wc_{gnH}(I_1)$, and is increasing on a right neighbourhood of $I_1=3$ (isochoric deformations necessarily satisfy $I_1\geq3$ and $I_2\geq3$, so a right neighbourhood suffices). Their appeal is twofold: they are mathematically tractable, and they reproduce the J-shaped curves typical of uniaxial tension tests on biological tissues \citep{Horgan2006}. Since $\Wc_2\equiv0$ for these models, \eqref{eq:mu} reduces to
\begin{equation}
  \dfrac{\d \Wc^{\textrm{gnH}}}{\d I_1}(3)=\dfrac\mu2.
\end{equation}
We consider three notable examples of generalized neo-Hookean models:
\begin{itemize}
  \item the Fung--Demiray model \citep{fung,demiray}:
  \begin{equation}\label{FDmod}
    \Wc^{\textrm{FD}}=\dfrac{\mu}{2c}\left\{\exp\left[c(I_1-3)\right]-1\right\},
  \end{equation}
  where the positive (dimensionless) constant $c$ accounts for the effects due to the stiffening of the material;
  \item the Gent model \citep{gent1996new}:
  \begin{equation}\label{gent}
    \Wc^{\textrm{G}}=-\dfrac\mu2J_m\ln\left(1-\dfrac{I_1-3}{J_m}\right),
  \end{equation}
  where the positive (dimensionless) modulus $J_m$ provides a measure of the limited extensibility of the chain elastomers as only deformations for which
  \begin{equation}\label{domainG}
    3\leq I_1 < 3+J_m
  \end{equation}
  can be sustained by these materials;
  \item the Knowles model \citep{knowles1977finite}:
  \begin{equation}\label{knowlesn}
    \Wc^{\textrm{K}}=\dfrac{\mu}{2 b}\left\{\left[1+\dfrac bn(I_1-3)\right]^n-1\right\},
  \end{equation}
  where the stiffening rate $b$ and the power-law index $n$ are positive dimensionless constants.
\end{itemize}
The Fung--Demiray, Gent, and Knowles models all converge to the neo-Hookean strain energy~\eqref{nHmod} as $c\to0$, $J_m\to\infty$ and $b\to0$, respectively. Moreover, the convergence is uniform for $I_1$ in any compact subset of $[3,+\infty[$. The Knowles model~\eqref{knowlesn} further reduces to the neo-Hookean model exactly at $n=1$, and tends to the Fung--Demiray model~\eqref{FDmod} with $c=b$ as $n\to\infty$, again uniformly for $I_1$ in any compact subset of $[3,+\infty[$.

We conclude this short list of mathematical models for the strain energy density with the extension proposed by \cite{Pucci2002Note} to resolve the limitations of the Gent model~\eqref{gent} at small-to-moderate deformations. Colloquially known in the continuum mechanics community as the Gent--Gent model, it adds a logarithmic term dependent on the second principal invariant $I_2$ and reads
\begin{equation}\label{GGmodel}
\Wc^{\textrm{GG}}=-\dfrac{C_1}{2}J_m\ln\left(1-\dfrac{I_1-3}{J_m}\right)+\dfrac32C_2\ln\left(\dfrac{I_2}{3}\right),
\end{equation}
where $\Wc^{\textrm{GG}}_1(3,3)=C_1/2$, $\Wc^{\textrm{GG}}_2(3,3)=C_2/2$, $\Wc^{\textrm{GG}}_{11}(3,3)=C_1/(2J_m)$, $\Wc^{\textrm{GG}}_{12}(3,3)=0$ and $\Wc^{\textrm{GG}}_{22}(3,3)=-C_2/6$, so that~\eqref{eq:mu} and~\eqref{ad} place the three constitutive parameters in one-to-one correspondence with the Landau moduli,
\begin{equation}\label{ggcal}
  C_1=2\mu+\dfrac{\A}{4}\,,\qquad C_2=-\mu-\dfrac{\A}{4}\,,\qquad \D=C_1\left(1+\dfrac{1}{J_m}\right)+\dfrac83 C_2\,.
\end{equation}
In particular, $C_1$ and $C_2$ coincide with the Mooney--Rivlin coefficients $C_{10}$ and $C_{01}$, so that $C_1=\mu\alpha$ and $C_2=\mu(1-\alpha)$ with $\alpha=2+\A/(4\mu)$ as in~\eqref{mrmod}, and the empirical inequalities again require $\alpha\in\,]0,1]$.

\section{Shearing of a thick tube with distributed axial screw dislocations}\label{S:Problem}

Henceforth, we will be interested in the shearing deformation of a hollow cylinder $\mathcal{B}$, with inner radius $A$ and outer radius $B$, made of an incompressible hyper-anelastic material with a continuous distribution of axial screw dislocations. We assume that, in the absence of eigenstrains, the material is homogeneous and isotropic. The inner surface of the tube is attached to a rigid solid cylinder and the outer surface is subjected to uniformly distributed helical shear traction $\boldsymbol{\tau}$.

\subsection{Geometry of the dislocated tube}

We equip the reference and spatial manifolds $\mathcal B$ and $\mathcal S$ with cylindrical coordinates $\{R,\Theta,Z\}$ and $\{r,\theta,z\}$, respectively, where the Euclidean reference and spatial metrics have the following representations
\begin{equation}\label{ghel}
\left[\mathring{\G}\right]=\left[\iota^*\g\right]\begin{bmatrix}
	1&0&0\\
	0&R^2&0\\
	0&0&1
	\end{bmatrix}\,, \quad \textrm{and} \quad
	[\g]=\begin{bmatrix}
	1&0&0\\
	0&r^2(R)&0\\
	0&0&1
	\end{bmatrix}\,.
\end{equation}

For a nonlinear solid with distributed dislocations, the material manifold is known to be a Weitzenb\"{o}ck manifold, i.e. a manifold with torsion, a flat connection and vanishing non-metricity \citep{YavariGoriely2012Dislocations, OzakinYavari2014}. It is shown in \citep{YavariGoriely2012Dislocations}, by virtue of Cartan's moving frames, that the plastic distortion $\Fa$ associated with the dislocation distribution given by a radially-symmetric Burgers vector density $b(R)$ may be written as
\begin{equation}
	\left[\Fa\right]=\begin{bmatrix}
	1&0&0\\
	0&1&0\\
	0&f(R)&1
	\end{bmatrix}\,,
\end{equation}
where $f(R)$ is such that $f'(R)=Rb(R)/(2\pi)$, the prime denoting differentiation with respect to $R$. Since the bonded rigid core carries no net dislocation content within, $f(A)=0$, so that
\begin{equation}\label{fint}
  f(R)=\dfrac{1}{2\pi}\int_A^R \xi\,b(\xi)\,\d\xi\,.
\end{equation}
Recalling that $\G=\Fa^*\mathring{\G}=\Fa^\star \mathring{\G} \Fa$, it follows that the material metric has the following representation
\begin{equation}\label{Ghel}
	\left[\G\right]=\begin{bmatrix}
	1&0&0\\
	0&R^2+f^2(R)&f(R)\\
	0&f(R)&1
	\end{bmatrix}\,.
\end{equation}

\subsection{Kinematics}
We seek equilibrium configurations of $\mathcal{B}$ within the class of helical shear deformations, combining azimuthal and axial shear, i.e. a deformation ansatz of the form
\begin{equation}
\label{eq:eg4_ansatz}
\left\{\begin{array}{ll}
r = r(R)\,,\\
[3mm]
\theta=\Theta+\psi(R)\,,\\
[3mm]
z=Z+w(R)\,,
\end{array}\right.
\end{equation}
where, since the inner boundary is fixed, the fields $r$, $\psi$ and $w$ must satisfy the boundary conditions
\begin{equation}\label{bcin}
  r(A)=0\,, \quad \psi(A)=0\,, \quad \textrm{and} \quad w(A)=0.
\end{equation}
From~\eqref{eq:eg4_ansatz} the deformation gradient is found to have the following representation \footnote{Note that, following \citep{truesdell1953physical, SaYa2026PhysComp}, the physical components of the deformation gradient $\F$ are given in local coordinates by $\hat{\mathrm{F}}^a{}_A = \cF^a{}_A \sqrt{\cg_{aa}/\cG_{AA}}$ (with no summation on any of the indices). In physical components, we then have
\begin{equation}
\left[\hat\F\right] =
\begin{bmatrix}
r'(R) & 0 & 0 \\
r(R)\psi'(R) & r(R)/(R\sqrt{1+\tilde{f}^2(R)}) & 0 \\
w'(R) & 0 & 1
\end{bmatrix}
\,.
\end{equation}
}
\begin{equation}\label{Fhel}
[\F] =
\begin{bmatrix}
r'(R) & 0 & 0 \\
\psi'(R) & 1 & 0 \\
w'(R) & 0 & 1
\end{bmatrix}
\,.
\end{equation}
Due to the incompressibility of the material, from~\eqref{ghel}, \eqref{Ghel} and~\eqref{Fhel}, we have the following restriction  
\begin{equation}
  J =\dfrac{r(R)}{R} r'(R)= 1\,,
\end{equation} 
which by integration, and taking into account the first boundary condition in~\eqref{bcin}, yields 
\begin{equation}\label{rR}
  r(R) = R\,.
\end{equation} 

The right Cauchy--Green deformation tensor reads 
\begin{equation}
[\C] =
\begin{bmatrix}
1 + {w'}^2(R) + R^2 {\psi'}^2(R)& R^2 \psi'(R) & w'(R) \\
R^2 \psi'(R) & R^2 & 0 \\
w'(R) & 0 & 1
\end{bmatrix}
\,,
\end{equation}
and its first two principal invariants calculated with respect to the material metric~\eqref{Ghel} read
\begin{equation}
\label{eq:invariants}
I_1 = 3 + \tilde{f}^2 + \kappa_z^2 + \kappa_\theta^2
	\,,\quad
I_2 = 3 + \tilde{f}^2 + \left(\kappa_z + \tilde{f} \kappa_\theta \right) ^2 + \kappa_\theta^2
	\,,
\end{equation}
where, for brevity, we have introduced the azimuthal and axial strains 
\begin{equation}
\label{eq:shear_nota}
 \kappa_\theta(R) \equiv R\, \psi'(R), \quad 	\kappa_z(R) \equiv w'(R),
	\end{equation}
and set	$\tilde{f}(R) \equiv f(R)/R$.


\subsection{Stress components and reduced equilibrium equations}
For a homogeneous incompressible isotropic material with the stored elastic energy density of the form~\eqref{Wchom}, we denote
\begin{equation}\label{eq:Wi->Qi}
\Qc_i(\kappa_\theta,\kappa_z;\tilde{f})
	\equiv \Wc_i\left(I_1, I_2\right)\,,\quad i=1,2\,,
\end{equation}
recalling $I_1$ and $I_2$ as given by~\eqref{eq:invariants}.
Following~\eqref{eq:sig_iso_represent_inc}, the Cauchy stress tensor has components
\begin{subequations}\label{eq:eg3_stress1}
\begin{align}
\label{eq:eg3_stress1_rr}
\sigma^{rr} &= -p + 2\Qc_1 - 2\Qc_2\left[1 + \left(\kappa_z + \tilde{f} \kappa_\theta \right) ^2 + \kappa_\theta^2\right] \,, \\
\label{eq:eg3_stress1_tt}
\sigma^{\theta\theta} &= \dfrac{1}{r^2} \left[ -p + 2\Qc_1\left(1 + \kappa_\theta^2\right) - 2\Qc_2(1 + \tilde{f}^2) \right]\,, \\
\label{eq:eg3_stress1_zz}
\sigma^{zz} &= -p + 2\Qc_1\left(1 + \tilde{f}^2 + \kappa_z^2\right) - 2\Qc_2 \,, \\
\label{eq:eg3_stress1_rt}
\sigma^{r\theta} &= \frac{1}{r}\left\{2\Qc_1 \kappa_\theta + 2\Qc_2 \left[\tilde{f}\kappa_z + (1 + \tilde{f}^2) \kappa_\theta\right]\right\} \,, \\
\label{eq:eg3_stress1_rz}
\sigma^{rz} &= 2 \Qc_1 \kappa_z + 2 \Qc_2 \left( \kappa_z + \tilde{f} \kappa_\theta \right) \,, \\
\label{eq:eg3_stress1_tz}
\sigma^{\theta z} &= \dfrac{1}{r} \left[ 2\Qc_1\kappa_z \kappa_\theta -2\left(\Qc_1 + \Qc_2\right) \tilde{f} \right].
\end{align}
\end{subequations}
Note that in terms of curvilinear coordinates, the components of the Cauchy stress tensor do not necessarily have the corresponding physical dimensions due to the fact that the radial and axial cylindrical coordinates have the same physical dimension, whereas the azimuthal coordinate is dimensionless. Following \citep{truesdell1953physical, SaYa2026PhysComp}, the physical components of the Cauchy stress tensor read in local coordinates $\hat{\sigma}^{ab} = \sigma^{ab} \sqrt{\cg_{aa}\cg_{bb}}$ (without summation over any of the indices), from which the physical components of $\boldsymbol{\sigma}$ are related to those in~\eqref{eq:eg3_stress1} by the connections
\begin{equation}\label{connections}
\begin{aligned}
\hat{\sigma}^{rr} &= \sigma^{rr}\,,\quad
&\hat{\sigma}^{\theta\theta} &= r^2 \sigma^{\theta\theta} \,,\quad
&\hat{\sigma}^{zz} &= \sigma^{zz} \,,\\
\hat{\sigma}^{r\theta} &= r \sigma^{r\theta} \,,\quad
&\hat{\sigma}^{\theta z} &= r \sigma^{\theta z} \,,\quad
&\hat{\sigma}^{rz} &= \sigma^{rz}.
\end{aligned}
\end{equation}
Since a purely tangential stress acts on the outer boundary of $\mathcal{B}$, we impose the traction boundary conditions on $r(B)=B$
\begin{equation}\label{bcout}
  \hat{\sigma}^{rr}(B)=0\,, \quad \hat{\sigma}^{r\theta}(B)=\tau_\theta\,, \quad \hat{\sigma}^{r z}(B)=\tau_z\,,
\end{equation}
where $(\tau_\theta,\tau_z)$ are the components of the helical shear boundary traction $\boldsymbol{\tau}$ with respect to the spatial orthonormal frame~${\{\mathbf{e}_\theta\equiv\partial_\theta/\sqrt{\cg_{\theta\theta}}},{\mathbf{e}_z\equiv\partial_z\}}$, i.e. $\boldsymbol{\tau}=\tau_\theta\boldsymbol{e}_\theta+\tau_z\boldsymbol{e}_z$.
Using the comma notation for differentiation, the equilibrium equations read
\begin{subequations}
\label{eq:Cyl_Balance_Physic}
\begin{align}
\hat\sigma^{rr}{}_{,r}+\dfrac{1}{r}\left(\hat\sigma^{rr}-\hat\sigma^{\theta\theta}\right) = 0\,,\\
-p_{,\theta}+\hat\sigma^{\theta r}{}_{,r}+\dfrac{2}{r}\hat\sigma^{\theta r} = 0\,,\\
-p_{,z}+\hat\sigma^{zr}{}_{,r}+\dfrac{1}{r}\hat\sigma^{zr} = 0\,.
\end{align}
\end{subequations}
Recalling~\eqref{rR}, it can be seen from~\eqref{bcout} and~\eqref{eq:Cyl_Balance_Physic} that the Lagrange multiplier depends only on the radial coordinate $R$, \emph{viz.} $p=p(R)$. Next, using~\eqref{rR}, \eqref{eq:eg3_stress1} and~\eqref{connections}, and introducing the auxiliary function
\begin{equation}
\begin{split}
\mathcal{I}_\sigma(R) \equiv &-\frac{2\Qc_1(\kappa_\theta(R),\kappa_z(R);\tilde{f}(R))}{R} \kappa_\theta^2(R)\\
&+ \frac{2 \Qc_2(\kappa_\theta(R),\kappa_z(R);\tilde{f}(R))}{R} \left[
\tilde{f}^2(R)
- \left(\kappa_z(R) + \tilde{f}(R) \kappa_\theta(R) \right) ^2
- \kappa_\theta^2(R)
\right]\,,
\end{split}
\end{equation}
the equilibrium equations~\eqref{eq:Cyl_Balance_Physic} simplifies to
\begin{subequations}
\begin{align}
\hat\sigma^{rr}{}_{,R} + \mathcal{I}_\sigma &= 0
\,,\\
\left(R^2 \hat\sigma^{\theta r} \right)_{,R} &= 0
\,,\\
\left(R \hat\sigma^{zr} \right)_{,R} &= 0\,,
\end{align}
\end{subequations}
whose integration combined with the boundary conditions~\eqref{bcout} yields
\begin{subequations}
\label{eq:eg3_stress2}
\begin{align}
\label{eq:eg3_stress2_rr}
\hat\sigma^{rr}(R) &= \int_R^B \mathcal{I}_\sigma(\xi) \, d\xi
\,,\\
\label{eq:eg3_stress2_tt}
\hat\sigma^{r\theta}(R) &=\dfrac{B^2\tau_\theta}{R^2}
\,,\\
\label{eq:eg3_stress2_zz}
\hat\sigma^{rz}(R) &=\dfrac{B\tau_z}{R}
\,.
\end{align}
\end{subequations}

Finally, from~\eqref{eq:eg3_stress1_rr}, \eqref{eq:eg3_stress1_rt}, \eqref{eq:eg3_stress1_rz}, and~\eqref{eq:eg3_stress2}, we deduce a system of two nonlinear algebraic equations in the shear strains
\begin{subequations}
\label{eq:Cramer}
\begin{alignat}{5}
\label{eq:sol2}
\left[\Qc_1+\Qc_2\left(1+\tilde{f}^2\right)\right]
\kappa_\theta
+
\Qc_2\tilde{f}
\,\kappa_z
&=&
\dfrac{B^2\tau_\theta}{2R^2}\,,\\[0.8ex]
\label{eq:sol3}
\Qc_2\tilde{f}
\;\kappa_\theta
+
\left(\Qc_1+\Qc_2\right)
\kappa_z
&=&
\dfrac{B\tau_z}{2R}\,,
\end{alignat}
\end{subequations}
which, once solved, allows for the determination of the Lagrange multiplier associated with the constraint of incompressibility through the formula
\begin{equation}
\label{eq:sol1}
p = 2\Qc_1 
- 2\Qc_2\left[1 + \left(\kappa_z + \tilde{f} \kappa_\theta \right) ^2 + \kappa_\theta^2 \right] 
- \int_R^B \mathcal{I}_\sigma(\xi) \, d\xi\,.
\end{equation}

For the subsequent analysis, it is convenient to non-dimensionalise the problem by introducing the following dimensionless quantities
\begin{equation}\label{nondimz}
\left.\begin{array}{cc}
\varrho_0 \equiv \dfrac AB\,,
\quad
\varrho \equiv \dfrac RB \in[\varrho_0,1]\,,
\quad
\Wh \equiv \dfrac{\W}{\mu}\,,
\quad
\widehat{\sigma}^{kl} \equiv \dfrac{\hat{\sigma}^{kl}}{\mu}\,,\; \textrm{for}\;k,l\in\{r,\theta,z\}\,,\\
[3mm]
\widehat{p} \equiv \dfrac{p}{\mu}\,,
\quad
\wh \equiv \dfrac{w}{B}\,,
\quad
\widehat{b} \equiv \dfrac{b}{b_\max}\,,
\quad
\tauh_\alpha \equiv \dfrac{\tau_\alpha}{\mu}\,,\; \textrm{for}\; \alpha\in\{\theta,z\} \,,
\end{array}\right.
\end{equation}
where $b_\max \equiv \displaystyle\max_{R\in[A,B]}b(R)$.
It follows that the azimuthal and axial strains recast, respectively, as
\begin{equation}
  \kappa_\theta(R) = \kappa_\theta(B\vr) =\vr\ph'(\vr)\equiv \kh_\theta(\vr) \quad \textrm{and} \quad\kappa_z(R)=\kappa_z(B\vr) = \wh'(\vr)\equiv \kh_z(\vr)\,,
\end{equation}
with the prime now denoting differentiation with respect to the dimensionless spatial variable $\vr$. On the other hand, setting $\beta=Bb_\max/(4\pi)$, the dislocations' density function $\tilde{f}$ rescales as
\begin{equation}
  \tilde{f}(R)=\tilde{f}(B \varrho) = \dfrac{2\beta}{\vr}\int_{\vr_0}^\vr \bar\vr\, \widehat{b}(\bar\vr) \mathrm{d} \bar\vr \equiv \beta \Fh(\vr),
\end{equation}
where the auxiliary function $\Fh$ satisfies the inequalities
\begin{equation}\label{fhin}
  0 \leq \Fh(\vr) \leq \vr-\vr_0^2/\vr \leq 1-\vr_0^2 <1 \quad \forall \vr\in[\vr_0,1].
\end{equation}
In dimensionless form, the invariants~\eqref{eq:invariants} recast as
\begin{equation}\label{invdim}
I_1=3+\beta^2\Fh^2+\kh_\theta^2+\kh_z^2 \equiv \widehat{I}_1(\kh_\theta,\kh_z;\Fh), \quad I_2=3+\beta^2\Fh^2+\left(\beta\Fh\kh_\theta+\kh_z\right)^2+\kh_\theta^2 \equiv \widehat{I}_2(\kh_\theta,\kh_z;\Fh),
\end{equation}
and the dimensionless version of the reduced equilibrium equations~\eqref{eq:Cramer} is given by the following system
\begin{equation}\label{eqdim}
\left\{ \begin{array}{ll}
\left[\Gc_1+\Gc_2\left(1+\beta^2\Fh^2\right)\right]\kh_\theta+\beta\Gc_2 \Fh \kh_z=\dfrac{\tauh_\theta}{2\vr^2},\\
 [5mm]
   \beta\Gc_2 \Fh \kh_\theta +\left(\Gc_1+\Gc_2\right)\kh_z=\dfrac{\tauh_z}{2\vr},
\end{array}\right.
\end{equation}
recalling~\eqref{eq:Wi->Qi}
\begin{equation}
\Gc_i=\Gc_i(\kh_\theta,\kh_z;\Fh)\equiv\Wh_i\left(\widehat{I}_1(\kh_\theta,\kh_z;\Fh),\widehat{I}_2(\kh_\theta,\kh_z;\Fh)\right) \,,\quad\textrm{for} \; i\in\{1,2\}\,.
\end{equation}

The search for solutions of the equilibrium equations~\eqref{eqdim} is equivalent to determining the zeroes of the vector equation
\begin{subequations}\label{eq:eqdim_sys}
\begin{equation}
 \boldsymbol{f}(\vr,\widehat{\boldsymbol{\kappa}};\Fh)=  \boldsymbol{A}\left(\widehat{\boldsymbol{\kappa}};\Fh\right)\widehat{\boldsymbol{\kappa}}-\widehat{\boldsymbol{\tau}}(\vr)=\boldsymbol{0}\,,
\end{equation}
with
\begin{equation}
\widehat{\boldsymbol{\kappa}} =
	\begin{bmatrix}
	\kh_\theta\\[3mm]
	\kh_z
	\end{bmatrix}\,,
	\qquad
\boldsymbol{A}\left(\widehat{\boldsymbol{\kappa}};\Fh\right) =
	\begin{bmatrix}
	\Gc_1+\Gc_2(1+\beta^2\Fh^2) & \beta\Gc_2\Fh\\[3mm]
	\beta\Gc_2\Fh & \Gc_1+\Gc_2
	\end{bmatrix}\,,
	\quad \textrm{and} \quad 
\widehat{\boldsymbol{\tau}}(\vr)=\dfrac12
	\begin{bmatrix}
	{\tauh_\theta}/{\vr^2}\\[3mm]
	{\tauh_z}/{\vr}
	\end{bmatrix}\,.
\end{equation}
\end{subequations}
By the implicit function theorem, there exists a unique smooth solution for $\widehat{\boldsymbol{\kappa}}$ in a neighbourhood of any point at which the Jacobian matrix $\partial\boldsymbol{f}/\partial{\widehat{\boldsymbol{\kappa}}}$ is non-singular. However, in general, this solution is only implicit and not in a closed-form.
In the following section, we discuss some models for the strain energy function for which it is possible to recover closed-form globally defined solutions.

Once the shear strains have been determined, the helical shear deformation can be obtained by solving the following initial value problems
\begin{equation}\label{inits}
\left\{\begin{array}{ll}
\ph'(\vr)=\dfrac{\kh_\theta(\vr)}{\vr},\\
[3mm]
\ph(\vr_0)=0,
\end{array}\right.
\quad \textrm{and} \quad 
\left\{\begin{array}{ll}
\wh'(\vr)=\kh_z(\vr),\\
[3mm]
\wh(\vr_0)=0,
\end{array}\right.
\end{equation}
where the initial conditions for $\ph$ and $\wh$ have been deduced from the boundary conditions~\eqref{bcin} and the non-dimensionalization~\eqref{nondimz}.



\section{Equilibrium configurations in specific classes of materials}
\label{S:Examples}

In this section, we study the equilibrium problem in some specific classes of incompressible elastic materials in the presence of distributed axial screw dislocations. Specifically, we consider the models discussed in~\S\ref{S:Const_Mod}.

\subsection{Neo-Hookean materials}
\label{S:nHmod}
For the neo-Hookean model~\eqref{nHmod}, the simplest truncation of~\eqref{expansion}, the dimensionless strain energy reads
\begin{equation}
  \Wh^{\textrm{nH}}=\dfrac12(I_1-3)\,.
\end{equation}
The reduced equilibrium equations~\eqref{eqdim} directly yield the shear strains
\begin{equation}\label{ksnh}
  \left\{\begin{array}{ll}
\kh_\theta=\dfrac{\tauh_\theta}{\vr^2}\,,\\
[3mm]
\kh_z=\dfrac{\tauh_z}{\vr}\,.
  \end{array}\right.
\end{equation}
Integrating the Cauchy problems~\eqref{inits} for~\eqref{ksnh} yields the azimuthal and axial displacement fields
\begin{equation}\label{ednh}
 \ph(\vr)=\dfrac{\tauh_\theta}{2}\left(\dfrac{1}{\vr_0^2}-\dfrac{1}{\vr^2}\right) \quad \textrm{and} \quad \wh(\vr)=\tauh_z\ln\left(\dfrac{\vr}{\vr_0}\right)\,. 
\end{equation}

It clearly emerges from the exact solutions above that the equilibrium configuration of a neo-Hookean material is unaffected by the dislocations: the function $\Fh$ occurs in neither the shear strains~\eqref{ksnh} nor the displacement field~\eqref{ednh}, which in fact is identical to that of a dislocation-free neo-Hookean tube.
Indeed, $\Fh$ enters~\eqref{eqdim} only through the dependence of $\Gc_1$ on $\{\widehat{I}_1,\widehat{I}_2\}$ and as a multiplicative factor to $\Gc_2$, both of which degenerate for the neo-Hookean model, so the dislocations drop out of the reduced equilibrium equations before any solving.
The incompressibility Lagrange multiplier is likewise unaffected and, from~\eqref{eq:sol1} and~\eqref{ksnh}, reads
\begin{equation}\label{pnh}
\widehat{p}(\vr)=1 - \dfrac{\tauh_\theta^2}{4}\left(1-\dfrac{1}{\vr^4}\right)\,.
\end{equation}
The stress field, by contrast, captures the dislocation effect: from~\eqref{eq:eg3_stress1}, \eqref{connections}, \eqref{ksnh} and~\eqref{pnh}, exactly two components of the (dimensionless, physical) Cauchy stress tensor are affected
\begin{equation}\label{snh}
\widehat{\sigma}^{\theta z}=\dfrac{\tauh_\theta\tauh_z}{\vr^3}-\beta\Fh(\vr)
\quad \textrm{and} \quad
\widehat{\sigma}^{zz}=\dfrac{\tauh_\theta^2}{4}\left(1-\dfrac{1}{\vr^4}\right)+\dfrac{\tauh_z^2}{\vr^2}+\beta^2\Fh^2(\vr)\,.
\end{equation}
The dislocations' contributions in~\eqref{snh} are additive and do not couple with the applied tractions: at zero load the tube carries the residual stresses $\widehat{\sigma}^{\theta z}=-\beta\Fh$ and $\widehat{\sigma}^{zz}=\beta^2\Fh^2$, the former odd in $\beta$ and hence chiral, the latter even.

\subsection{Generalized neo-Hookean materials}\label{S:gnHmod}
For generalized neo-Hookean materials, the dimensionless strain energy function may be recast as
\begin{equation}
  \Wh^{\textrm{gnH}}=\dfrac12 \mathcal{F}^{\textrm{gnH}}(I_1)\,,
  \quad \textrm{with} \quad
  \mathcal{F}^{\textrm{gnH}}_1 \equiv \dfrac{\d \mathcal{F}^{\textrm{gnH}}}{\d I_1} >0
  \quad \textrm{and} \quad
  \mathcal{F}^{\textrm{gnH}}_1(3)=1\,.
\end{equation}
Thus, the reduced equilibrium equations~\eqref{eqdim} read as follows
\begin{equation}\label{gnhred}
  \left\{\begin{array}{ll}
\kh_\theta\mathcal{F}^{\textrm{gnH}}_1(I_1)=\dfrac{\tauh_\theta}{\vr^2}\,,\\
[5mm]
\kh_z\mathcal{F}^{\textrm{gnH}}_1(I_1)=\dfrac{\tauh_z}{\vr}\,,
  \end{array}\right.
\end{equation}
recalling that $I_1 = 3+\beta^2\Fh^2+\kh_\theta^2+\kh_z^2$. The dislocation field enters~\eqref{gnhred} only through the argument of $\mathcal{F}^{\textrm{gnH}}_1$, and is therefore inert whenever that derivative is constant, i.e. when $\mathcal{F}^{\textrm{gnH}}$ is affine in $I_1$, which is the case of the neo-Hookean model~\eqref{nHmod} addressed in~\S\ref{S:nHmod}. Otherwise, the distributed dislocations affect the equilibrium configuration itself.

The reduced equilibrium equations~\eqref{gnhred} yield the following relation between azimuthal and axial strains
\begin{equation}\label{pseudo}
\tauh_\theta\kh_z=\vr\tauh_z\kh_\theta\,.
\end{equation}
This relation, which is universal in that it does not depend on the constitutive function $\mathcal{F}^{\textrm{gnH}}$ within the generalized neo-Hookean class, tells us that a pure axial (respectively, azimuthal) load, i.e. $\tauh_\theta=0$ (respectively, $\tauh_z=0$), does not induce any azimuthal (respectively, axial) shear, and, in the most general case in which the shear boundary traction has two non-zero components, the ratio between the azimuthal and axial strains is known everywhere in $\mathcal{B}$.
In particular, assuming that $\tauh_\theta\neq0$, the system of two nonlinear equations~\eqref{gnhred} may be reduced to a single equation in terms of $\kh_\theta$ as
\begin{equation}\label{eqred}
 \kh_\theta \mathcal{F}^{\textrm{gnH}}_1(I_1)=\dfrac{\tauh_\theta}{\vr^2}\,,
\end{equation}
where $I_1=3+\beta^2\Fh^2(\vr)+\left(1+(\tauh^2_z/\tauh^2_\theta)\vr^2\right)\kh_\theta^2$, and with $\kh_z=\vr\tauh_z\kh_\theta/\tauh_\theta$.
Closed-form solutions for~\eqref{gnhred}---or~\eqref{eqred}---require a specific constitutive function, and we specialize in the subsequent subsections to three examples of generalized neo-Hookean models.

\subsubsection{Universal relations}

Before proceeding with the solution of the equilibrium equations~\eqref{gnhred} for specific generalized neo-Hookean models, we generalize to the case of distributed axial screw dislocations the universal relations in terms of the Cauchy stress tensor found by \citet{horgan2003helical}---motivated by the fact that the universal relation~\eqref{pseudo} identically coincides with the one obtained in \citep{horgan2003helical} for the undislocated tube.

To this end, we first report the dimensionless physical components of the Cauchy stress tensor~\eqref{eq:eg3_stress1}
\begin{subequations}\label{eq:ur3}
\begin{align}
\label{eq:ur3a}
& \widehat\sigma^{rr} = -\widehat{p} + 2\mathcal{F}^{\textrm{gnH}}_1\,, \\
\label{eq:ur3b}
& \widehat\sigma^{\theta\theta} = -\widehat{p}+2\mathcal{F}^{\textrm{gnH}}_1\left(1 + \kh_\theta^2\right)\,, \\
\label{eq:ur3c}
& \widehat\sigma^{zz} = - \widehat{p}+ 2\mathcal{F}^{\textrm{gnH}}_1\left(1 + \beta^2\Fh^2 + \kh_z^2\right)\,,\\
\label{eq:ur3d}
& \widehat\sigma^{r\theta} = 2\mathcal{F}^{\textrm{gnH}}_1\kh_\theta\,, \\
\label{eq:ur3e}
& \widehat\sigma^{rz} = 2\mathcal{F}^{\textrm{gnH}}_1\kh_z\,, \\
\label{eq:ur3f}
& \widehat\sigma^{\theta z} = 2\mathcal{F}^{\textrm{gnH}}_1\left(\kh_z\kh_\theta-\beta\Fh\right)\,.
\end{align}
\end{subequations}
Since $\mathcal{F}^{\textrm{gnH}}_1>0$, by~\eqref{eq:ur3e} and~\eqref{eq:ur3d}, the shear stresses $\widehat{\sigma}^{r\theta}$ and $\widehat{\sigma}^{rz}$ vanish precisely when the shear strains $\kh_\theta$ and $\kh_z$ do, respectively; they cannot therefore vanish simultaneously unless the tube is undeformed. Without loss of generality we take $\widehat{\sigma}^{r\theta}\neq0$, which amount to assuming $\kh_\theta\neq0$ and equivalently $\tauh_\theta \neq 0$. Under this assumption, by~\eqref{eq:ur3a} and~\eqref{eq:ur3b}, the normal stresses $\widehat{\sigma}^{rr}$ and $\widehat{\sigma}^{\theta\theta}$ cannot coincide, \emph{viz.} $\widehat\sigma^{\theta\theta}-\widehat\sigma^{rr} \neq 0 $.
Next, we solve~\eqref{eq:ur3a} and~\eqref{eq:ur3b} for the Lagrange multiplier $\widehat{p}$ and the $\mathcal{F}^{\textrm{gnH}}_1$, substitute the results in the remaining equations in~\eqref{eq:ur3} and obtain the following identities
\begin{subequations}\label{eq:ur5}
  \begin{align}
  \label{eq:ur5c}
  (\widehat\sigma^{zz}-\widehat\sigma^{rr})\kh_\theta^2&=(\widehat\sigma^{\theta\theta}-\widehat\sigma^{rr})(\kh_z^2+\beta^2\Fh^2)\,,
  \\
  \label{eq:ur5d}
  \widehat\sigma^{r\theta}\kh_\theta&=\widehat{\sigma}^{\theta\theta}-\widehat\sigma^{rr}\,,
  \\
  \label{eq:ur5e}
  \widehat{\sigma}^{rz}\kh_\theta^2&=(\widehat{\sigma}^{\theta\theta}-\widehat\sigma^{rr})\kh_z\,,
  \\
  \label{eq:ur5f}
  \widehat\sigma^{\theta z}\kh_\theta^2&=(\widehat{\sigma}^{\theta\theta}-\widehat\sigma^{rr})(\kh_z\kh_\theta-\beta\Fh)\,.
  \end{align}
\end{subequations}

Now, we express the axial and azimuthal strains in terms of the components of the stress tensor.
From~\eqref{eq:ur5d} and~\eqref{eq:ur5e} one finds
\begin{equation}\label{eq:ur6}
\kh_\theta = \dfrac{\widehat\sigma^{\theta\theta}-\widehat\sigma^{rr}}{\widehat\sigma^{r\theta}}\quad \textrm{and} \quad \kh_z = \dfrac{\widehat\sigma^{rz}(\widehat\sigma^{\theta\theta}-\widehat\sigma^{rr})}{(\widehat\sigma^{r\theta})^2}\,,
\end{equation}
which, substituted in~\eqref{eq:ur5c} and~\eqref{eq:ur5f}, yields the following two nonlinear universal relations
\begin{subequations} \label{eq:ur7}
\begin{align}
\label{eq:ur7a}
 \left(\widehat\sigma^{\theta\theta}-\widehat\sigma^{rr}\right)\left[ \widehat\sigma^{rz}\left(\widehat\sigma^{\theta\theta}-\widehat\sigma^{rr}\right)-\widehat\sigma^{r\theta}\widehat\sigma^{\theta z}\right]&=\beta\Fh\left( \widehat\sigma^{r\theta}\right)^3\,,
\\
\label{eq:ur7b}
\left(\widehat\sigma^{\theta\theta}-\widehat\sigma^{rr}\right)\left[\left( \widehat\sigma^{r\theta}\right)^2\left(\widehat\sigma^{zz}-\widehat\sigma^{rr}\right)-\left( \widehat\sigma^{rz}\right)^2\left(\widehat\sigma^{\theta\theta}-\widehat\sigma^{rr}\right)\right]&=\beta^2\Fh^2\left( \widehat\sigma^{r\theta}\right)^4\,.
\end{align}
\end{subequations}
Unlike~\eqref{pseudo}, the relations~\eqref{eq:ur7} carry the dislocation field explicitly, the eigenstrains entering as the prescribed function $\beta\Fh$ of position. Eliminating the latter between~\eqref{eq:ur7a} and~\eqref{eq:ur7b}, recalling that $\widehat{\sigma}^{r\theta} \neq 0$ and $\widehat\sigma^{\theta\theta}-\widehat\sigma^{rr} \neq 0 $, yields
\begin{equation}\label{eq:ur8}
  \left(\widehat\sigma^{r\theta}\right)^2\left[\left(\widehat\sigma^{r\theta}\right)^2\left(\widehat\sigma^{zz}-\widehat\sigma^{rr}\right)-\left(\widehat\sigma^{rz}\right)^2\left(\widehat\sigma^{\theta\theta}-\widehat\sigma^{rr}\right)\right]=\left(\widehat\sigma^{\theta\theta}-\widehat\sigma^{rr}\right)\left[\widehat\sigma^{rz}\left(\widehat\sigma^{\theta\theta}-\widehat\sigma^{rr}\right)-\widehat\sigma^{r\theta}\widehat\sigma^{\theta z}\right]^2\,,
\end{equation}
which involves neither the constitutive function nor the dislocation density, but this independence should not be overstated. Relation~\eqref{eq:ur8} holds for a generalized neo-Hookean tube with a screw dislocation distribution taken a priori to be axial and radially symmetric. Its universality is thus relative to the class of generalized neo-Hookean tubes carrying such a dislocation field, whatever its density---as any universal relation is relative to the class of deformations within which it is derived.
In the absence of dislocations, i.e. $\Fh=0$, the right-hand sides of~\eqref{eq:ur7} vanish and, dividing by the non-vanishing factor $\widehat\sigma^{\theta\theta}-\widehat\sigma^{rr}$, one recovers the relations of \citet{horgan2003helical}, namely
\begin{equation} \label{eq:ur_HorSac}
\left(\widehat\sigma^{\theta \theta}-\widehat\sigma^{rr}\right) \widehat\sigma^{rz}=\widehat\sigma^{r\theta}\widehat\sigma^{\theta z} \quad \textrm{and} \quad (\widehat\sigma^{zz}-\widehat\sigma^{rr})\widehat\sigma^{r\theta}=\widehat\sigma^{rz}\widehat\sigma^{\theta z}\,,
\end{equation}
so that~\eqref{eq:ur7} extend~\eqref{eq:ur_HorSac} to tubes carrying distributed axial screw dislocations.

\subsubsection{Fung--Demiray materials}

Specializing to the Fung--Demiray model~\eqref{FDmod}, the equilibrium equation~\eqref{eqred} reads
\begin{equation}\label{FDeq}
  \kh_\theta \exp\left[c\left(\beta^2\Fh^2+\left(1+\dfrac{\tauh^2_z}{\tauh^2_\theta}\vr^2\right)\kh^2_\theta\right)\right]=\dfrac{\tauh_\theta}{\vr^2}\,,
\end{equation}
which is solved in closed form for the azimuthal strain:
\begin{equation}\label{lambert}
  \begin{split}
    \kh_\theta(\vr)= \dfrac{\tauh_\theta}{\vr^2} \exp\left[-\dfrac12\mathrm{W}_0\!\left(\dfrac{2c(\tauh^2_z\vr^2+\tauh_\theta^2)}{\vr^4\exp\left(2c\beta^2\Fh^2(\vr)\right)}\right)-c\beta^2\Fh^2(\vr)\right]\,,
  \end{split}
\end{equation}
where $\mathrm{W}_0$ denotes the principal branch of the Lambert W function \citep{corless1996lambert}.
Using~\eqref{pseudo} and~\eqref{lambert}, the Cauchy problems~\eqref{inits} can be solved by quadrature, resulting in
\begin{equation}\label{solfd}
\ph(\vr)=\int_{\vr_0}^\vr\dfrac{ \kh_\theta(\xi)}{\xi}\d \xi \quad \textrm{and} \quad \wh(\vr)=\frac{\tauh_z}{\tauh_\theta}\int_{\vr_0}^\vr\xi \kh_\theta(\xi)\d \xi\,.
\end{equation}

The Fung--Demiray material places no restriction on either the dislocation strength or the load. The argument of $\mathrm{W}_0$ in~\eqref{lambert} being non-negative, $\mathrm{W}_0$ is the only real branch, and~\eqref{lambert} is the unique solution of~\eqref{FDeq}, whose left-hand side is a strictly increasing bijection of $\mathbb{R}$ in $\kh_\theta$. An equilibrium configuration thus exists, and is unique, for every dislocation strength and every load, with no limit load. 
Since $\mathrm{W}_0\geq0$ on $[0,+\infty[$, \eqref{lambert} yields the bound $|\kh_\theta|\leq|\tauh_\theta|\vr^{-2}\exp(-c\beta^2\Fh^2)$, attained asymptotically as the dislocation strength $\beta\to\infty$.
This bound is also asymptotically reached with vanishing tractions, so that to leading order in the tractions, the dislocated Fung--Demiray tube responds as a neo-Hookean tube with a shear modulus raised pointwise by the factor $\exp(c\beta^2\Fh^2)$.
As $c\to0$ the argument of $\mathrm{W}_0$ in~\eqref{lambert} vanishes, so that~\eqref{lambert} tends to the neo-Hookean solution~\eqref{ksnh}, consistent with the fact that the Fung--Demiray model itself tends to the neo-Hookean one in that limit.

\begin{example}

\begin{figure}[t!]
\centering
\begin{subfigure}{0.475\textwidth}
 \centering
 \includegraphics[height=7cm]{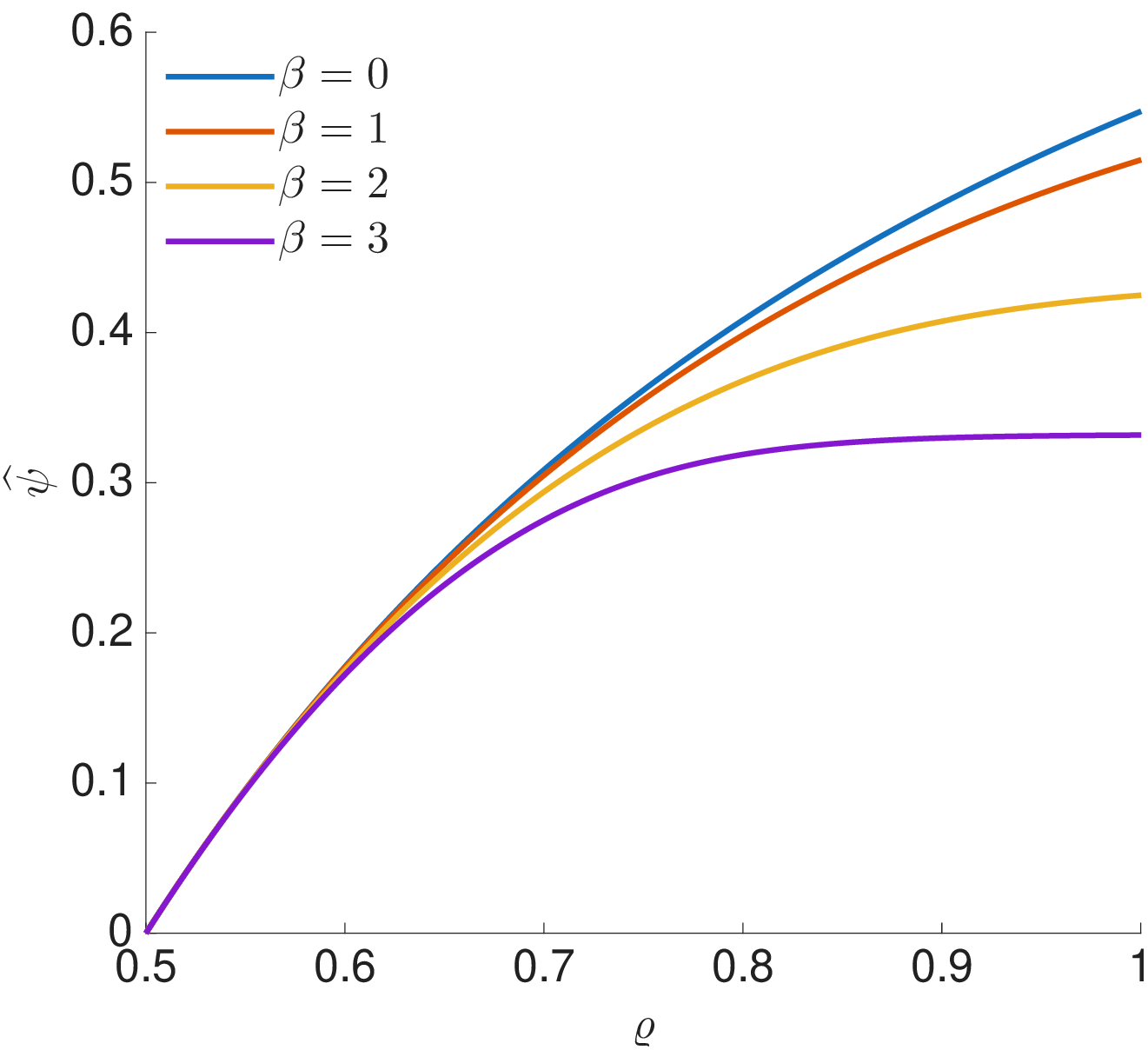}
 \caption{}
 \label{fig:FD-beta-psi}
\end{subfigure}
\hspace{7.5pt}
\begin{subfigure}{0.475\textwidth}
 \centering
 \includegraphics[height=7cm]{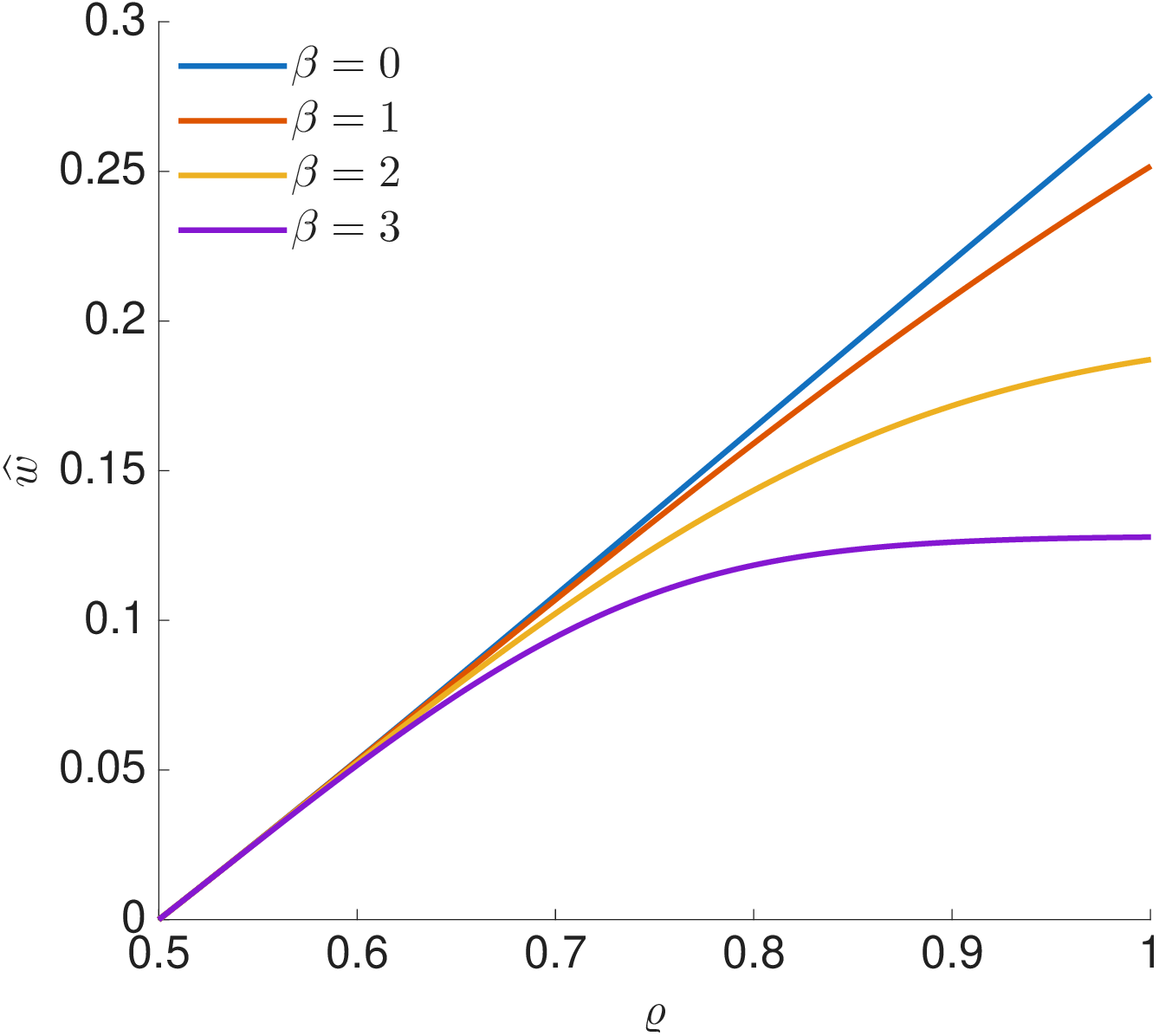}
 \caption{}
 \label{fig:FD-beta-w}
\end{subfigure}
\\[2ex]
\begin{subfigure}{0.475\textwidth}
 \centering
 \includegraphics[height=7cm]{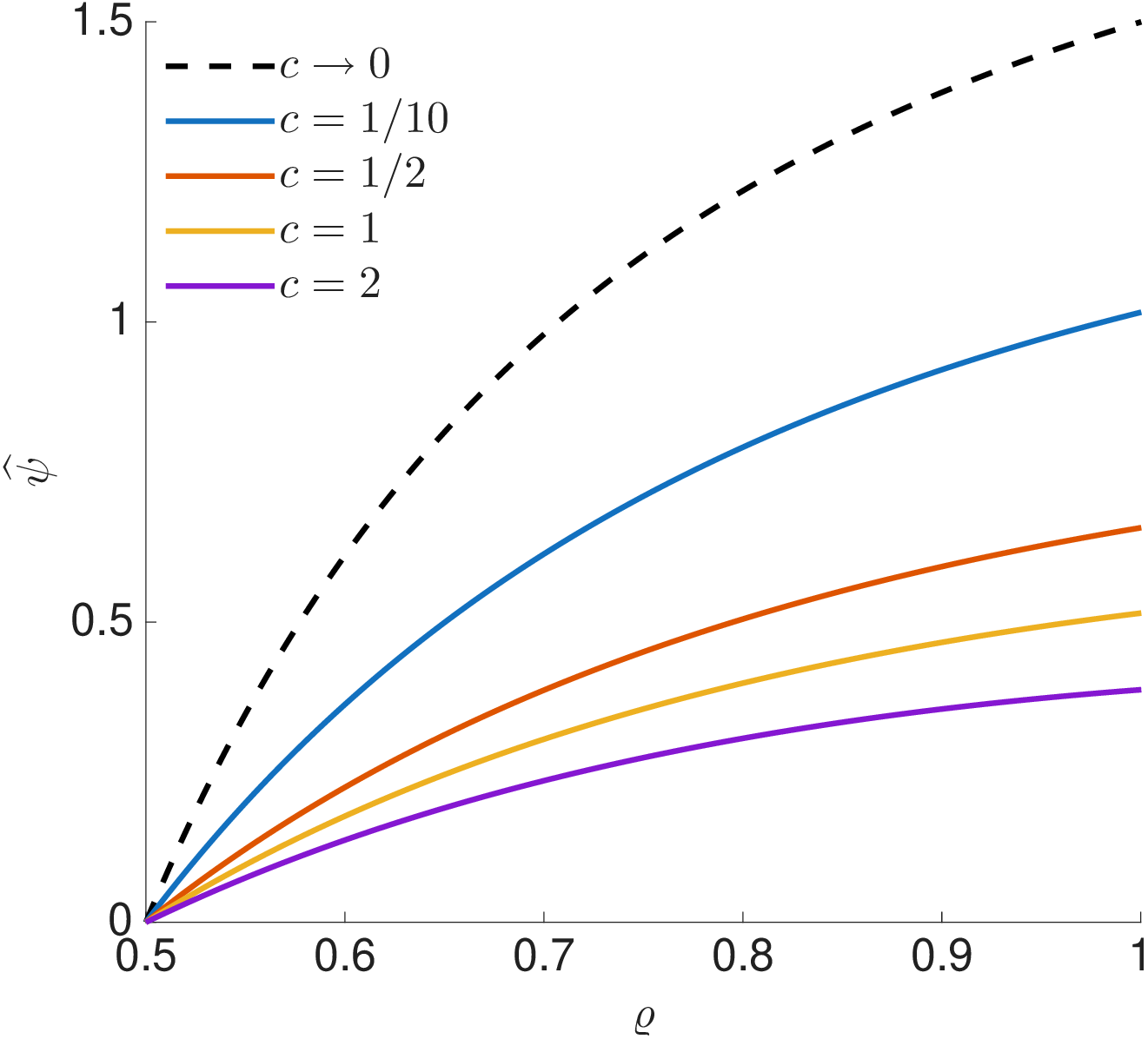}
 \caption{}
 \label{fig:FD-c-psi}
\end{subfigure}
\hspace{7.5pt}
\begin{subfigure}{0.475\textwidth}
 \centering
 \includegraphics[height=7cm]{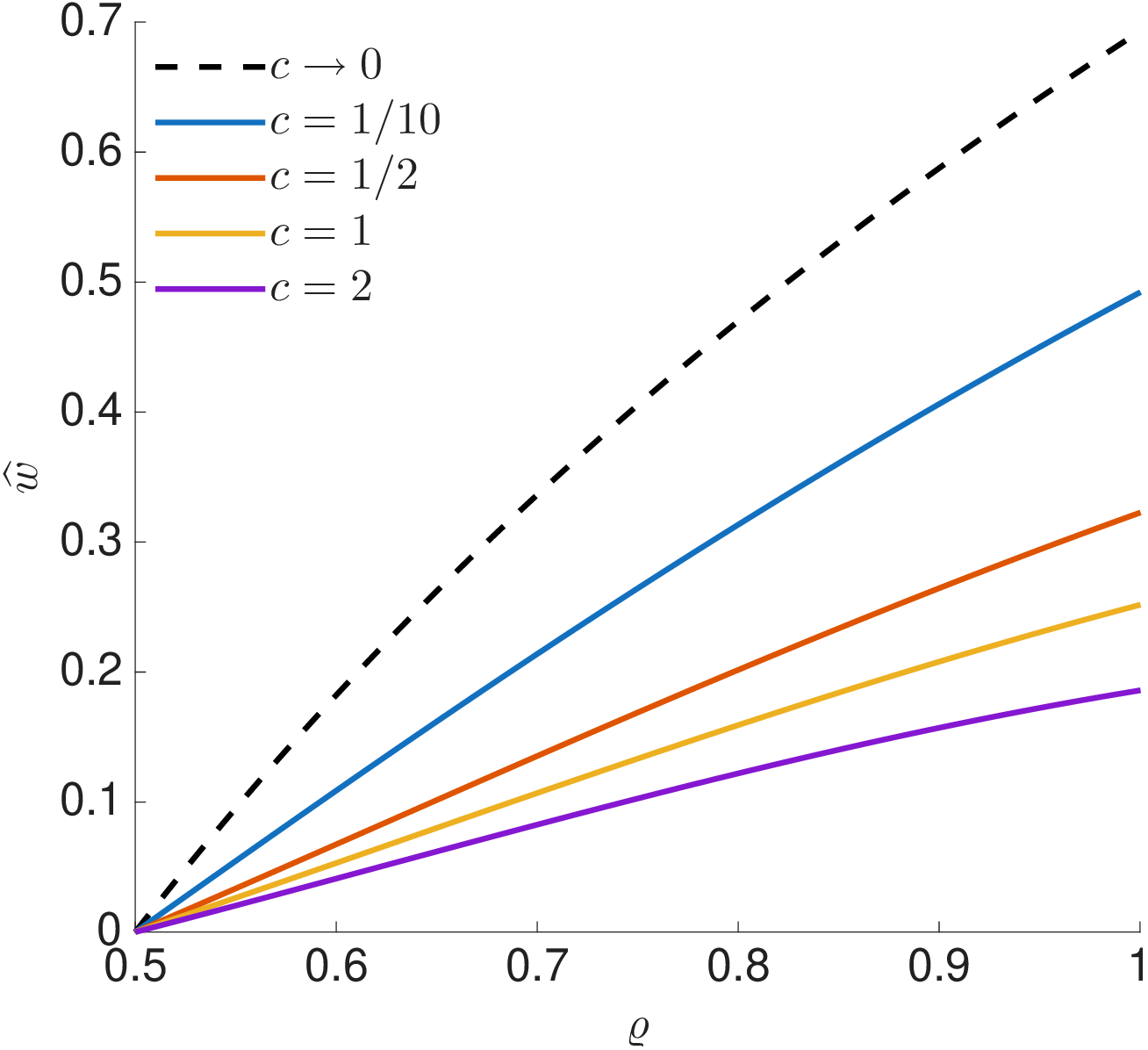}
 \caption{}
 \label{fig:FD-c-w}
\end{subfigure}
\caption{Helical shear of a Fung--Demiray tube with the constant-density profile~\eqref{fhhom}, at $\vr_0=1/2$ and $\tauh_\theta=\tauh_z=1$.
(a) Azimuthal and (b) axial displacements at $c=1$ for $\beta\in\{0,1,2,3\}$, all profiles sharing a common inner-wall slope since $\Fh(\vr_0)=0$.
(c) Azimuthal and (d) axial displacements at $\beta=1$ for $c\in\{1/10,1/2,1,2\}$, with the neo-Hookean limit~\eqref{ednh} as $c\to0$ (dashed).}
\label{fig:FD}
\end{figure}

We further specialize this case to a numerical example by assuming a uniform Burgers density, $\widehat{b}=1$. This yields the following auxiliary dislocation function
\begin{equation}\label{fhhom}
  \Fh(\vr)=\vr-\dfrac{\vr_0^2}{\vr}\,.
\end{equation}
We further set $\vr_0=1/2$ and $\tauh_\theta=\tauh_z=1$.
At fixed stiffening parameter $c=1$, all displacement profiles leave the inner wall with a common slope and separate outward as $\Fh$ grows, see Figs.~\ref{fig:FD-beta-psi} and~\ref{fig:FD-beta-w}. The effect is mild at low dislocation strengths and intensifies with increasing $\beta$, until the factor $\exp(-c\beta^2\Fh^2)$ all but suppresses the shear near the outer wall, where the displacements saturate.
At fixed dislocation strength $\beta=1$, the displacements decrease with $c$ and approach the neo-Hookean ones~\eqref{ednh} as $c\to0$, see Figs.~\ref{fig:FD-c-psi} and~\ref{fig:FD-c-w}. Unlike $\beta$, $c$ also lowers the inner-wall slope, since there the stiffening is driven by the shear alone.
\end{example}

\subsubsection{Gent material}\label{S:Gent}
Assuming a Gent material model~\eqref{gent}, the equilibrium equations~\eqref{gnhred} may be written as
\begin{equation}\label{gentred}
  \dfrac{J_m\kh_\theta}{J_m-\beta^2\Fh^2(\vr)-(\kh_\theta^2 + \kh_z^2)}
  = \dfrac{\tauh_\theta}{\vr^2}\,, \quad \kh_z = \vr \dfrac{\tauh_z\kh_\theta}{\tauh_\theta}\,.
\end{equation}
with the latter as given by~\eqref{pseudo}.
Any solution to~\eqref{gentred} must satisfy the Gent limited-extensibility restriction~\eqref{domainG}, which, by~\eqref{invdim}, reads
\begin{equation}\label{eq:AdmiG}
  \kh_\theta^2(\vr) + \kh_z^2(\vr) < J_m - \beta^2\Fh^2(\vr)\,,\quad\forall\vr\in[\vr_0,1]\,,
\end{equation}
as a pointwise limit on the helical shear strain, with $\kh_\theta^2 + \kh_z^2$ saturating at an extensibility limit eroded by the dislocation field from $J_m$ to $J_m-\beta^2\Fh^2(\vr)$.
Further, following~\eqref{gentred}, at the limit of~\eqref{eq:AdmiG}, one has $\tauh_\theta$ (or, equivalently, $\tauh_z)\to\infty$.
Thus, in spite of the admissibility cap on the strain~\eqref{eq:AdmiG}, the model is well posed for every traction.
Note that deformations beyond~\eqref{eq:AdmiG} would require richer kinematics than the helical ansatz~\eqref{eq:eg4_ansatz}.

Remarkably, this pointwise restriction reduces to a single global condition: an admissible solution to~\eqref{gentred} exists if and only if
\begin{equation}\label{limitGent}
\beta^2\max_{\vr\in[\vr_0,1]}\Fh^2(\vr) < {J_m}\,,
\end{equation}
which is a global condition on the dislocation field alone; unlike the pointwise restriction~\eqref{eq:AdmiG}, it involves neither the shear strain nor the applied tractions.
Indeed, necessity is immediate: by~\eqref{eq:AdmiG}, an admissible solution satisfies $J_m - \beta^2\Fh^2(\vr) > \kh_\theta^2 + \kh_z^2 \geq 0$ at every $\vr\in[\vr_0,1]$, whence~\eqref{limitGent}.
Conversely, assume~\eqref{limitGent}, so that $\beta^2\Fh^2(\vr) < J_m$ for all $\vr\in[\vr_0,1]$. Recalling the working assumption $\tauh_\theta \neq 0$ and $\kh_z=\vr\tauh_z\kh_\theta/\tauh_\theta$, the admissibility condition~\eqref{eq:AdmiG} may be recast as
\begin{equation}\label{domainG_2}
  -\kh_{\theta\textrm{max}}< \kh_\theta < \kh_{\theta\textrm{max}}\,,
  \quad \textrm{with}\quad
  \kh_{\theta\textrm{max}} \equiv \sqrt{\frac{J_m - \beta^2\Fh^2(\vr)}{1 + \vr^2\tauh_z^2/\tauh_\theta^2}}\,.
\end{equation}
Introducing
\begin{equation}\label{Phidef}
  \Phi(\kh_\theta) \equiv \dfrac{J_m\kh_\theta}{[J_m-\beta^2\Fh^2(\vr)]-(1+\vr^2\tauh_z^2/\tauh_\theta^2)\kh_\theta^2}\,,
\end{equation}
equation~\eqref{gentred} reads $\Phi(\kh_\theta)=\tauh_\theta/\vr^2$. On the interval~\eqref{domainG_2}, $\Phi$ is differentiable, with first derivative with respect to $\kh_\theta$
\begin{equation}\label{Phimono}
  \dfrac{\d\Phi}{\d \kh_\theta}=J_m\,\dfrac{J_m-\beta^2\Fh^2(\vr) + (1+\vr^2\tauh_z^2/\tauh_\theta^2) \kh_\theta^2}{\left[J_m-\beta^2\Fh^2(\vr) - (1+\vr^2\tauh_z^2/\tauh_\theta^2) \kh_\theta^2\right]^2}>0\,\quad \forall \kh_\theta \in \left]-\kh_{\theta\textrm{max}}, \kh_{\theta\textrm{max}}\right[\,,
\end{equation}
whence it is strictly increasing. Moreover, since $\Phi(\kh_\theta)\to\pm\infty$ as $\kh_\theta\to\pm\kh_{\theta\textrm{max}}$, the intermediate value theorem and the strict monotonicity of $\Phi$ ensure that it maps $]-\kh_{\theta\textrm{max}},\kh_{\theta\textrm{max}}[$ bijectively onto $\mathbb{R}$, so that~\eqref{gentred} admits exactly one admissible solution $\kh_\theta$ for any traction $\tauh_\theta$, carrying the sign of $\tauh_\theta$ since $\Phi(0)=0$. The Gent model hence places no cap on the applied tractions: under proportional loading, i.e. at fixed ratio $\tauh_z/\tauh_\theta$, the extensibility limit bounds the strain alone, $|\kh_\theta|$ approaching $\kh_{\theta\textrm{max}}$ only as $|\tauh_\theta|\to\infty$; and by~\eqref{gentred}, the same holds for the axial strain, $|\kh_z|$ approaching $\vr|\tauh_z/\tauh_\theta|\,\kh_{\theta\textrm{max}}$ only as $|\tauh_z|\to\infty$.
In the absence of dislocations,~\eqref{limitGent} holds trivially: an equilibrium configuration of the form~\eqref{eq:eg4_ansatz} exists in any tube. In their presence,~\eqref{limitGent} is necessary and sufficient for such a configuration to exist, and caps the dislocation strength accordingly. When~\eqref{limitGent} fails, the Gent model is incompatible with the dislocation content of the body within the helical class~\eqref{eq:eg4_ansatz}.
Provided~\eqref{limitGent} holds, the unique admissible solution of~\eqref{gentred} reads 
\begin{equation}\label{solg}
  \begin{gathered}
  \kh_\theta(\vr)= \tauh_\theta h(\vr;J_m,\beta)\,,
  \quad
  \kh_z = \vr \dfrac{\tauh_z\kh_\theta}{\tauh_\theta}\,,\\
  \textrm{with} \quad h(\vr;J_m,\beta) \equiv \dfrac{-J_m\vr^2+\sqrt{J_m^2\vr^4+4(\tauh_\theta^2+\vr^2\tauh_z^2)[J_m-\beta^2\Fh^2(\vr)]}}{2(\tauh_\theta^2+\vr^2\tauh_z^2)}\,.
  \end{gathered}
\end{equation}
Using~\eqref{pseudo} and~\eqref{solg}, the Cauchy problems~\eqref{inits} may be solved by quadrature, resulting in
\begin{equation}\label{solGent}
  \ph(\vr)=\tauh_\theta\int_{\vr_0}^\vr \dfrac{h(\xi;J_m,\beta)}{\xi}\d \xi \quad \textrm{and} \quad \wh(\vr)=\tauh_z\int_{\vr_0}^\vr \xi h(\xi;J_m,\beta)\d \xi\,.
\end{equation}
Note that $h(\vr;J_m,\beta)$ tends to $1/\vr^2$ as $J_m \to \infty$. Therefore, the solutions~\eqref{solGent} tend to the solutions~\eqref{ednh} for neo-Hookean materials as it is reasonable to expect as the Gent model tends to the neo-Hookean model in the limit as $J_m\rightarrow+\infty$.

\begin{example}

\begin{figure}
\centering
\begin{subfigure}{0.475\textwidth}
 \centering
 \includegraphics[height=7cm]{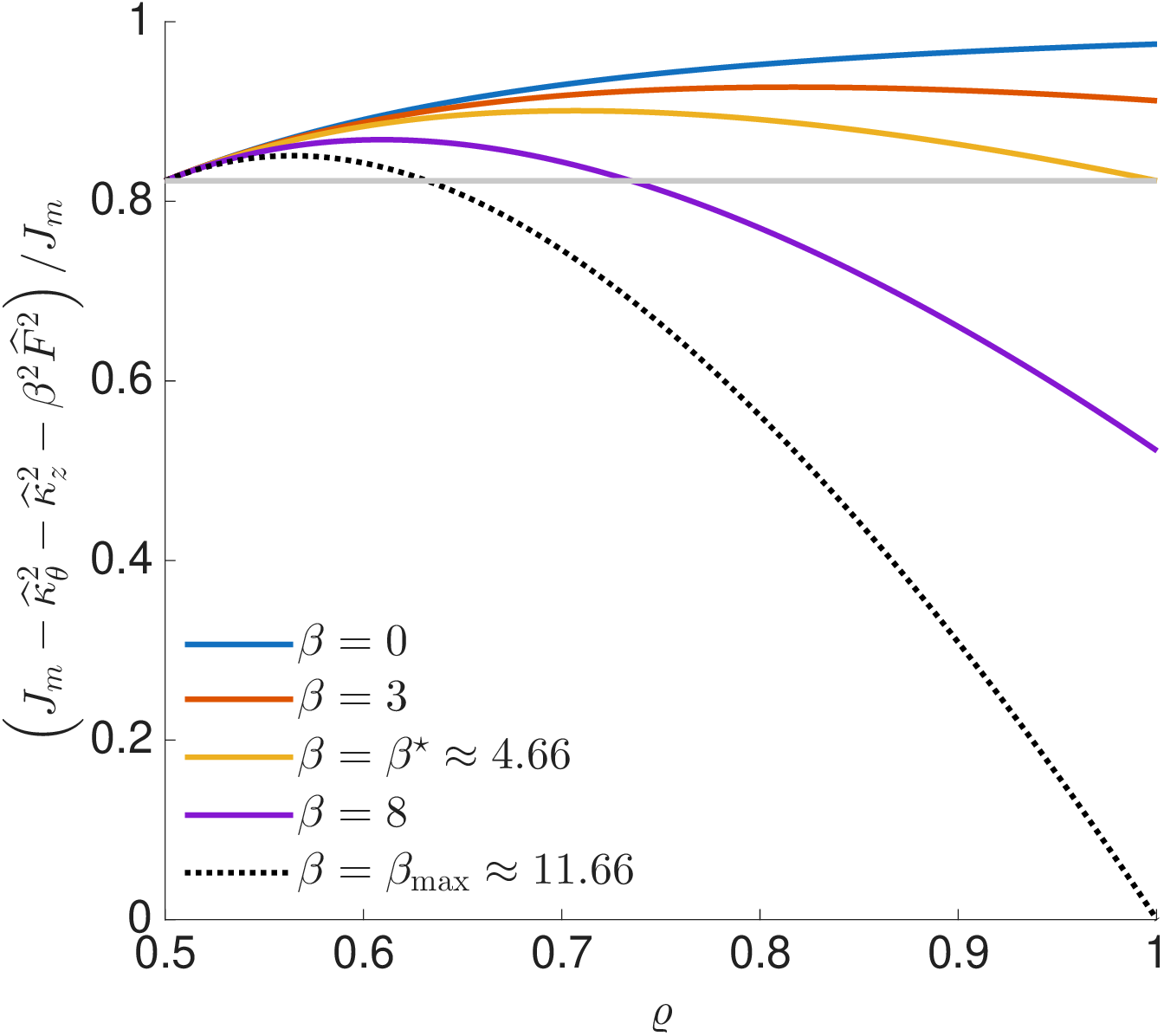}
 \caption{}
 \label{fig:gent-util}
\end{subfigure}
\hspace{7.5pt}
\begin{subfigure}{0.475\textwidth}
 \centering
 \includegraphics[height=7cm]{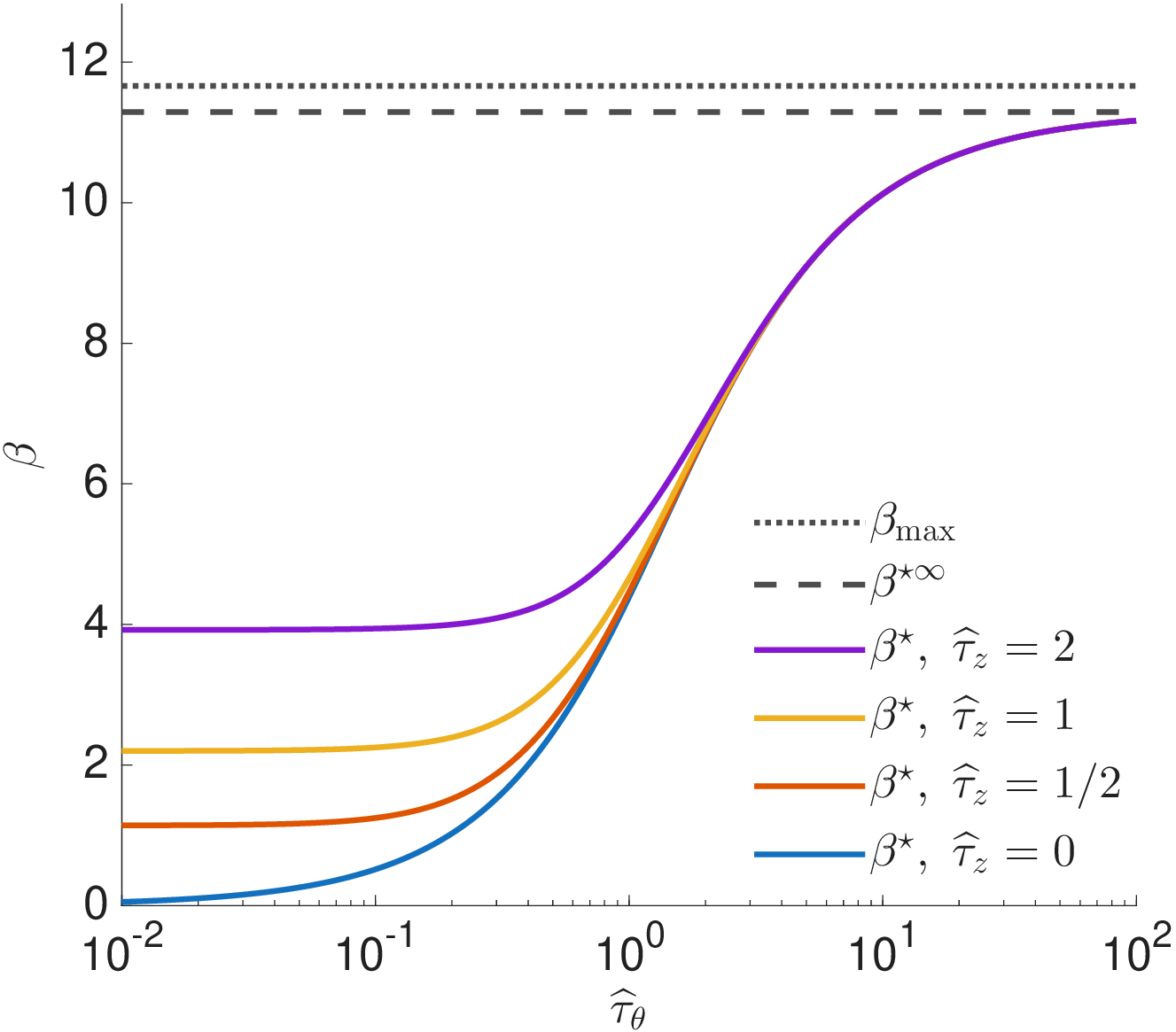}
 \caption{}
 \label{fig:gent-cross}
\end{subfigure}
\caption{Locking crossover in a Gent tube with the constant-density profile~\eqref{fhhom}, at $J_m=76.48$ and $\vr_0=1/2$.
(a) Residual extensibility margin fraction $(J_m-\kh_\theta^2-\kh_z^2-\beta^2\Fh^2)/J_m$ at $\tauh_\theta=\tauh_z=1$, all curves sharing a common inner-wall value since $\Fh(\vr_0)=0$. The narrowest margin lies at the inner wall for $\beta<\beta^\star$, at both walls at $\beta=\beta^\star$, and at the outer wall beyond.
(b) Crossover density $\beta^\star(\tauh_\theta)$ for $\tauh_z\in\{0,1/2,1,2\}$: the inner wall has the narrowest margin below each curve, the outer wall above it. All curves tend with increasing $\tauh_\theta$ towards $\beta^{\star\infty}$.}
\label{fig:gent-cross-pair}
\end{figure}
Let us further specialise this case to a numerical example, focusing on the interplay between the dislocation field and the limited extensibility of the Gent model. To this end, we assume a uniform Burgers vector density, leading to the auxiliary dislocation function~\eqref{fhhom}, i.e. $\Fh(\vr)=\vr-{\vr_0^2}/{\vr}$. The global admissibility condition~\eqref{limitGent} hence reduces to
\begin{equation}\label{limitGent_eg}
  \beta < \dfrac{\sqrt{J_m}}{1-\vr_0^2} \equiv \beta_{\max}\,,
\end{equation}
a cap on the dislocation strength that tightens with a more limited extensibility (smaller $J_m$) and with a thicker annulus (smaller $\vr_0$).
Further, for the uniform dislocation density case, the cap~\eqref{limitGent_eg} represents a necessary and sufficient global condition for the admissibility of the solution~\eqref{solg} independently of the strain state of the tube.
The strain itself, however, remains subject to the pointwise restriction~\eqref{eq:AdmiG}, and we define the residual extensibility margin as
\begin{equation}\label{eq:margin}
  \mathcal{M}(\vr;\beta) \equiv J_m-\kh_\theta^2(\vr)-\kh_z^2(\vr)-\beta^2 \left(\vr-{\vr_0^2}/{\vr}\right)^2
  = \dfrac{J_m \vr^2 \kh_\theta(\vr;\beta)}{\tauh_\theta}
  \,,
\end{equation}
with the second equality following from~\eqref{gentred}.

In light of~\eqref{eq:AdmiG}, dislocations and shear draw on $J_m$ from opposite ends of the annulus: $\beta^2\Fh^2(\vr)$ vanishes at the inner wall and grows outward, while $\kh_\theta^2(\vr)+\kh_z^2(\vr)=(1+\vr^2\tauh_z^2/\tauh_\theta^2)\kh_\theta^2(\vr)$ concentrates at the inner wall and decays outward.\footnote{\label{fn:decay}Differentiating~\eqref{gentred} with respect to $\varrho$, a prime denoting $\d/\d\vr$, and setting $D\equiv J_m\vr^2+2(1+\vr^2\tauh_z^2/\tauh_\theta^2)\tauh_\theta\kh_\theta>0\,$---recalling that $\tauh_\theta\kh_\theta>0$ as discussed following~\eqref{Phimono}---one finds $(\kh_\theta^2)'=-2[2\vr(\tauh_z^2/\tauh_\theta^2)\tauh_\theta\kh_\theta^3+2\beta^2\tauh_\theta\kh_\theta\Fh\Fh'+2J_m\vr\kh_\theta^2]/D$ and $(\kh_\theta^2+\kh_z^2)'=-2[J_m\vr(2+\vr^2\tauh_z^2/\tauh_\theta^2)\kh_\theta^2+{2(1+\vr^2\tauh_z^2/\tauh_\theta^2)\tauh_\theta\kh_\theta\beta^2\Fh\Fh'}]/D$, both strictly negative as $\Fh\Fh'\geq0$.} One may further show that $\mathcal{M}$ admits no interior minimum, so that the margin is narrowest at one of the two walls.\footnote{Differentiating~\eqref{eq:margin} with respect to $\varrho$ and eliminating $\kh_\theta'$, one finds $\mathcal{M}'(\vr)=2J_m\vr[(2+\vr^2\tauh_z^2/\tauh_\theta^2)\kh_\theta^2-\beta^2(\vr^2-\vr_0^4/\vr^2)]/D$, with $D>0$ as in footnote~\ref{fn:decay}. The bracket in $\mathcal{M}'(\vr)$ is the difference between $\kh_\theta^2+(\kh_\theta^2+\kh_z^2)$, positive and strictly decreasing by footnote~\ref{fn:decay}, and $\beta^2(\vr^2-\vr_0^4/\vr^2)$, strictly increasing and vanishing at $\vr_0$. $\mathcal{M}'(\vr)$ is hence positive at $\vr_0$ and changes sign at most once, so that $\mathcal{M}$ increases and then possibly decreases on $[\vr_0,1]$, attaining its minimum at $\vr_0$ or at $\vr=1$.}
Fig.~\ref{fig:gent-util} shows the margin fraction $\mathcal{M}(\vr;\beta)/J_m$ for different dislocation strengths $\beta$. At low dislocation strength $\beta$, the margin is narrowest at the inner wall, where the shear concentrates; as $\beta$ increases, the outer-wall margin closes until the narrowest point migrates there. This suggests the existence of a crossover density $\beta^\star$ such that, for $\beta<\beta^\star$, the inner wall governs the extensibility margin, while for $\beta>\beta^\star$ the outer wall does; it is defined as the dislocation strength at which the residual extensibility margins at the two walls are equal. By the second equality in~\eqref{eq:margin} and~\eqref{solg}, $\beta^\star$ is implicitly defined by\footnote{In~\eqref{solg}, $\beta$ enters only through $\beta^2\Fh^2$, which vanishes at the inner wall, so that $h(\vr_0;J_m)$ carries no dependence on the dislocation strength.}
\begin{equation}\label{eq:gent-crossover}
  h(1;J_m,\beta^\star) = \vr_0^2\,h(\vr_0;J_m)\,.
\end{equation}
For the parameters of Fig.~\ref{fig:gent-util}, $\beta^\star\approx4.66$, against $\beta_{\max}\approx11.66$.
Following~\eqref{solg}, $h(1;J_m,\beta)$ is positive and decreases strictly with $\beta$, tending to $0$ as $\beta\to\beta_{\max}$, while $h(1;J_m,0)>\vr_0^2h(\vr_0;J_m) \geq 0$ since $\mathcal{M}$ increases strictly in the absence of dislocations. The intermediate value theorem thus yields that $\beta^\star$ exists and is unique in $(0,\beta_{\max})$. It depends on the tractions only through $\tauh_\theta^2$ and $\tauh_z^2$, and is found to increase with $\tauh_\theta$ as the inner wall is driven harder, the axial traction raising it at low $\tauh_\theta$ and slightly lowering it at high $\tauh_\theta$, see Fig.~\ref{fig:gent-cross}.
As $|\tauh_\theta|\to\infty$ at fixed $\tauh_z$, one has $\kh_\theta\to\operatorname{sign}(\tauh_\theta)\surd(J_m-\beta^2\Fh^2)$ pointwise, and~\eqref{eq:gent-crossover} yields the asymptotic crossover density
\begin{equation}\label{eq:betastarinf}
  \beta^{\star\infty} = \sqrt{\dfrac{J_m\left(1+\vr_0^2\right)}{1-\vr_0^2}} = \beta_{\max}\sqrt{1-\vr_0^4}<\beta_{\max}\,.
\end{equation}
\end{example}



\subsubsection{Knowles materials}

\begin{figure}
\centering
 \includegraphics[height=7cm]{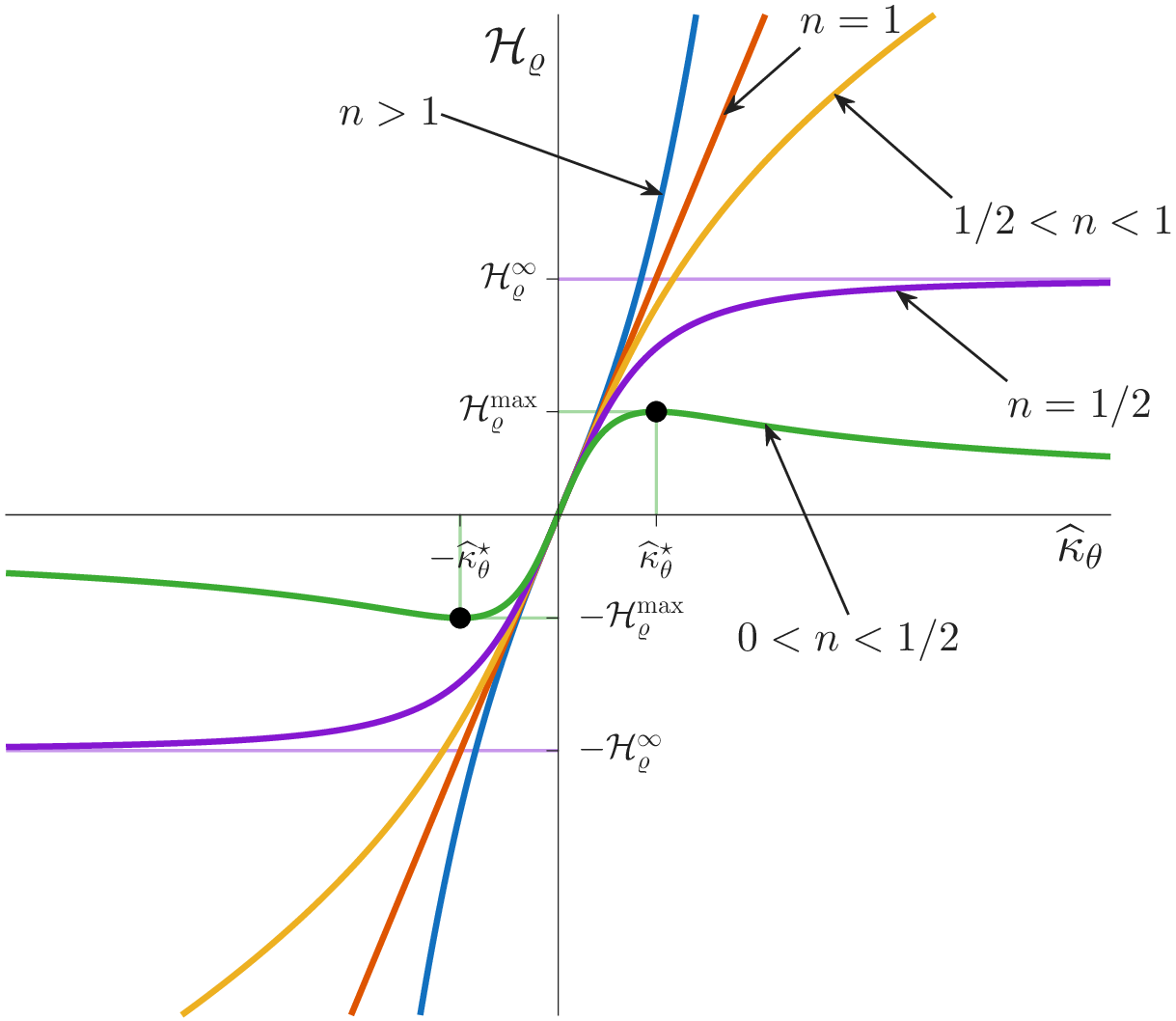}
\caption{Qualitative behaviour of $\mathcal{H}_\vr$ at a fixed radius, for values of the power-law index $n$ spanning the three regimes. For $n>1/2$ the function is strictly increasing and unbounded, degenerating into a straight line at $n=1$, so~\eqref{Kred} admits a unique solution at any traction. At $n=1/2$ it still increases but saturates at $\pm\mathcal{H}_\vr^\infty$, and a solution exists only below that threshold. For $n<1/2$ it attains $\pm\mathcal{H}_\vr^{\max}$ at $\pm\kh^\star_\theta$ and decays to zero thereafter, so there are two solutions below $\mathcal{H}_\vr^{\max}$, exactly one at it, and none above.}
\label{Hrhofig}
\end{figure}

Assuming a Knowles material model~\eqref{knowlesn}, the reduced equilibrium equation~\eqref{eqred} reads
\begin{equation}\label{Kred}
\kh_\theta\left\{\tauh_\theta^2+\dfrac bn\left[\tauh_\theta^2\beta^2\Fh^2(\vr)+\left(\tauh^2_\theta+\vr^2\tauh_z^2\right)\kh_\theta^2\right]\right\}^{n-1}=\operatorname{sign}(\tauh_\theta)\dfrac{|\tauh_\theta|^{2n-1}}{\vr^2}
\,.
\end{equation}
To investigate the existence and uniqueness of solutions to equation~\eqref{Kred} conforming to the helical ansatz~\eqref{eq:eg4_ansatz}, we find it convenient to define, for all $\vr\in[\vr_0,1]$, the odd function 
\begin{equation}
\mathcal{H}_\vr:\zeta\in\mathbb{R}\mapsto \zeta\left\{\tauh_\theta^2+\dfrac bn\left[\tauh_\theta^2\beta^2\Fh^2(\vr)+\left(\tauh^2_\theta+\vr^2\tauh_z^2\right)\zeta^2\right]\right\}^{n-1}\,,  
\end{equation}
so that~\eqref{Kred} reads $\mathcal{H}_\vr(\kh_\theta)=\operatorname{sign}(\tauh_\theta){|\tauh_\theta|^{2n-1}}/{\vr^2}$.
Introducing, for all $\vr\in[\vr_0,1]$, the strictly positive functions
\begin{equation}\label{ABdef}
\mathcal{A}(\vr)\equiv\tauh_\theta^2\left[1+\dfrac bn\beta^2\Fh^2(\vr)\right] \quad \textrm{and} \quad \mathcal{B}(\vr)\equiv\dfrac bn\left(\tauh_\theta^2+\vr^2\tauh_z^2\right)\,,
\end{equation}
so that $\mathcal{H}_\vr(\zeta)=\zeta\left[\mathcal{A}(\vr)+\mathcal{B}(\vr)\zeta^2\right]^{n-1}$, differentiation yields
\begin{equation}\label{Hprime}
\mathcal{H}_\vr'(\zeta)=\left[\mathcal{A}(\vr)+\mathcal{B}(\vr)\zeta^2\right]^{n-2}\left[\mathcal{A}(\vr)+(2n-1)\mathcal{B}(\vr)\zeta^2\right]\,.
\end{equation}
The first bracket in~\eqref{Hprime} is strictly positive, so the sign of $\mathcal{H}_\vr'$ is that of the second, and hence that of $2n-1$. The same quantity fixes the behaviour at infinity, since $\mathcal{H}_\vr(\zeta)\sim\mathcal{B}^{n-1}(\vr)\,\zeta^{2n-1}$ as $\zeta\to+\infty$. The power-law index $n=1/2$ therefore separates three regimes.
\begin{itemize}
  \item[(\emph{i})] If $n>1/2$, the function $\mathcal{H}_\vr$ is strictly increasing and tends to $\pm\infty$ as $\zeta\rightarrow\pm\infty$, hence maps $\mathbb{R}$ bijectively onto itself, and there exists a unique equilibrium configuration of the form~\eqref{eq:eg4_ansatz} for any $\tauh_\theta$. Note that the graph of $\mathcal{H}_\vr$ is convex on $]0,+\infty[$ if $n>1$ and concave if $1/2<n<1$, degenerating into a straight line at $n=1$, where the Knowles model reduces to the neo-Hookean one;
  \item[(\emph{ii})] If $n=1/2$, the function $\mathcal{H}_\vr$ is strictly increasing but bounded by
  \begin{equation}
  \lim_{\zeta\rightarrow\pm\infty} \mathcal{H}_\vr(\zeta)=\pm \dfrac{1}{\sqrt{2b(\tauh_\theta^2+\vr^2\tauh_z^2)}}\equiv\pm\mathcal{H}_\vr^\infty\,.
  \end{equation}
  It follows that $\mathcal{H}_\vr$ maps $\mathbb{R}$ bijectively onto $\left]-\mathcal{H}_\vr^\infty,\mathcal{H}_\vr^\infty\right[$. Since the right-hand side of~\eqref{Kred} reduces to $\operatorname{sign}(\tauh_\theta)/\vr^2$ at $n=1/2$, and $\mathcal{H}_\vr$ being odd, a solution exists if and only if $\mathcal{H}_\vr^\infty>{1}/{\vr^2}$ for all $\vr\in[\vr_0,1]$, that is
  \begin{equation}\label{con121}
  \dfrac{\tauh_\theta^2}{\vr^4}+\dfrac{\tauh_z^2}{\vr^2}<\dfrac{1}{2b} \quad \textrm{for all} \quad \vr\in[\vr_0,1]\,.
  \end{equation}
  The left-hand side decreases with $\vr$, so the constraint binds at the inner wall and reduces to the requirement that the boundary shear lies in the elliptical disk
  \begin{equation}\label{con12}
  \dfrac{\tauh_\theta^2}{\vr_0^4}+\dfrac{\tauh_z^2}{\vr_0^2}<\dfrac{1}{2b}\,.
  \end{equation}
  \item[(\emph{iii})] If $0<n<1/2$, the function $\mathcal{H}_\vr$ is strictly increasing on $]-\kh^\star_\theta,\kh^\star_\theta[$, with
  \begin{equation}
  \kh^\star_\theta=|\tauh_\theta|\sqrt{\dfrac{n+b\beta^2\Fh^2(\vr)}{(1-2n)b(\tauh_\theta^2+\vr^2\tauh_z^2)}}\,,
  \end{equation}
  strictly decreasing on $]-\infty,-\kh^\star_\theta[$ and on $]\kh^\star_\theta,+\infty[$, attains at $\pm\kh^\star_\theta$ its extremal values
  \begin{equation}
  \pm\mathcal{H}_\vr^{\max}=\pm\left[\dfrac{2(1-n)}{n}\right]^{n-1}\left[\dfrac{n+b\beta^2\Fh^2(\vr)}{1-2n}\right]^{n-1/2}\dfrac{\left|\tauh_\theta\right|^{2n-1}}{\sqrt{b(\tauh_\theta^2+\vr^2\tauh_z^2)}}\,,
  \end{equation}
  and $\mathcal{H}_\vr(\zeta)$ tends to zero as $\zeta\rightarrow\pm\infty$. The image of $\mathcal{H}_\vr$ hence being $\left[ -\mathcal{H}_\vr^{\max}, \mathcal{H}_\vr^{\max} \right]$, equation~\eqref{Kred} admits two solutions if $\mathcal{H}_\vr^{\max}>|\tauh_\theta|^{2n-1}/\vr^2$, exactly one if the two sides coincide, and none otherwise. Introducing
  \begin{equation}\label{Sndef}
  S_n(\vr;\beta)\equiv \sqrt 2 C_n\left[n+b\beta^2\Fh^2(\vr)\right]^{n-1/2}\,, \quad \textrm{with} \quad C_n\equiv\left[\dfrac{n}{2(1-n)}\right]^{1-n}(1-2n)^{1/2-n}\,,
  \end{equation}
  the requirement $\mathcal{H}_\vr^{\max}>|\tauh_\theta|^{2n-1}/\vr^2$ reads, at each radius, as the condition that the outer-boundary tractions belong to the open elliptical disk
  \begin{equation}\label{Endef}
  E_n(\vr;\beta)\equiv\left\{(\tauh_\theta,\tauh_z)\ :\ \dfrac{\tauh_\theta^2}{\vr^4}+\dfrac{\tauh_z^2}{\vr^2}<\dfrac{S_n^2(\vr;\beta)}{2 b}\right\}\,,
  \end{equation}
  whose semi-axes $\vr^2S_n(\vr)/\sqrt{2b}\equiv\tauh_{\theta\textrm{max}}(\vr)$ and $\vr S_n(\vr)/\sqrt{2b}\equiv\tauh_{z\textrm{max}}(\vr)$ are the limit loads that the radius $\vr$ sustains in pure azimuthal and pure axial shear, respectively. Imposing this at every radius, there exist \emph{two} equilibrium configurations of the form~\eqref{eq:eg4_ansatz} if and only if the tractions lie in the admissible region $\mathcal{E}_n(\beta)\equiv\bigcap_{\vr\in[\vr_0,1]}E_n(\vr;\beta)$, that is
  \begin{equation}\label{conK}
  \max_{\vr\in[\vr_0,1]} \left( \dfrac{\tauh_\theta^2}{\vr^4 S_n^2(\vr;\beta)}+\dfrac{\tauh_z^2}{\vr^2 S_n^2(\vr;\beta)}\right)< \dfrac{1}{2b}\,.
  \end{equation}
  This pointwise erosion need not reach $\mathcal{E}_n(\beta)$. Being an intersection, the admissible region is fixed by whichever radii are active on its boundary, and so long as the core disk $E_n(\vr_0)$---itself independent of $\beta$---stays contained in every other constituent disk, one has $\mathcal{E}_n(\beta)=E_n(\vr_0)$: the dislocation field leaves the admissible tractions untouched even though it lowers the ceiling at every interior radius. Erosion sets in only once some interior disk first cuts inside the core one.
  As an intersection of compact convex sets, $\mathcal{E}_n(\beta)$ is compact and convex; however, once more than one radius is active, it is no longer an elliptical disk, as distinct arcs of its boundary being supplied by distinct radii.
\end{itemize}
The analysis above is summarised in Fig.~\ref{Hrhofig}, which shows $\mathcal{H}_\varrho$ at representative values of $n$ spanning the three regimes.

Note that the dislocation field shapes the equilibrium at every $n$, entering $\mathcal{H}_\vr$ through $\beta^2\Fh^2(\vr)$, but governs its existence only below the critical exponent. For $n>1/2$ no restriction arises at all; at $n=1/2$ the condition~\eqref{con12} carries no $\beta$; and only for $n\in\,]0,1/2[$ does~\eqref{conK} depend on the dislocation content. To gauge that dependence, we compare the pointwise condition~\eqref{Endef} with its dislocation-free reduction
\begin{equation}\label{Endef_free}
E_n(\vr;0)\equiv\left\{(\tauh_\theta,\tauh_z)\ :\ \dfrac{\tauh_\theta^2}{\vr^4}+\dfrac{\tauh_z^2}{\vr^2}<\dfrac{C_n^2n^{2n-1}}{b}\right\}\,.
\end{equation}
These disks are nested, the innermost being that of the bonded core, so that $\mathcal{E}_n(0)=E_n(\vr_0)$.
Since $n-1/2<0$, it follows from~\eqref{Sndef} that $S_n(\vr;\beta)\leq\sqrt2\,C_nn^{n-1/2}$ throughout the annulus, with equality only at the bonded core, where $\Fh$ vanishes. Dislocations thus contract every constituent disk $E_n(\vr;\beta)$ with $\vr>\vr_0$, strictly so as $\beta$ grows, while leaving the core disk $E_n(\vr_0)$ untouched. Being their intersection, the global admissible region $\mathcal{E}_n(\beta)$ equals $\mathcal{E}_n(0)$ so long as $E_n(\vr_0)$ is contained in every other constituent disk; once one of them ceases to contain it, the dislocation field erodes $\mathcal{E}_n(\beta)$.
The two cases converge continuously at the critical exponent. As $n\rightarrow1/2^-$ the exponent in~\eqref{Sndef} vanishes and $C_n\rightarrow1/\sqrt2$, so that $S_n\rightarrow1$ independently of both $\beta$ and $\vr$, and~\eqref{conK} reduces to~\eqref{con12}.
The $\beta$-independence at $n=1/2$ is therefore the limiting form of the admissibility region erosion mechanism above, the contraction of the constituent disks weakening as $n\rightarrow1/2^-$ and switching off with the exponent that carries it.

\begin{example}
Let us further specialise this case to an example featuring the effect of the dislocation field on the admissible traction region, assuming $n<1/2$ and a uniform Burgers vector density---resulting in the auxiliary dislocation function~\eqref{fhhom}.
Recall that, in the absence of dislocations, $S_n$ is constant and both semi-axes of~\eqref{Endef} increase with $\vr$, so that the disks are nested and $\mathcal{E}_n(0)=E_n(\vr_0)$.
A dislocation field acts solely through $S_n$, which by the exponent $n-1/2<0$ decreases outward, in competition with the growth of $\vr$ and $\vr^2$. The admissible traction region is hence eroded precisely when this competition breaks the nesting, and it does so one semi-axis at a time: nesting holds in the azimuthal and axial directions if and only if $\tauh_{\theta\textrm{max}}(\vr_0) \leq \tauh_{\theta\textrm{max}}(\vr)$ and $\tauh_{z\textrm{max}}(\vr_0) \leq \tauh_{z\textrm{max}}(\vr)$ respectively,\footnote{Recall that $\tauh_{\theta\textrm{max}}(\vr)=\vr^2S_n(\vr;\beta)/\sqrt{2b}$ and $\tauh_{z\textrm{max}}(\vr)=\vr S_n(\vr;\beta)/\sqrt{2b}$ are the semi-axes of the open elliptical disk~\eqref{Endef}.} which, following~\eqref{Sndef}, read
\begin{subequations}\label{eq:Knowles-G}
\begin{align}
  \beta^2 &\leq \dfrac nb\,\dfrac{\left(\vr/\vr_0\right)^{4/(1-2n)}-1}{\Fh^2(\vr)} \equiv G^\theta_n(\vr)\,, \label{eq:Knowles-Gth}\\
  \beta^2 &\leq \dfrac nb\,\dfrac{\left(\vr/\vr_0\right)^{2/(1-2n)}-1}{\Fh^2(\vr)} \equiv G^z_n(\vr)\,, \label{eq:Knowles-Gz}
\end{align}
\end{subequations}
for all $\vr\in\,]\vr_0,1]$. Both $G_n^\theta$ and $G_n^z$ diverge at the bonded core, $\Fh(\vr_0)$ vanishing, and attain their minima at interior radii, yielding the two thresholds $\beta_n^\theta\equiv\min_{]\vr_0,1]}\surd G_n^\theta$ and $\beta_n^z\equiv\min_{]\vr_0,1]}\surd G_n^z$---see Fig.~\ref{fig:knowles-thresholds_a}. Since $\vr/\vr_0>1$, one has $G_n^\theta>G_n^z$ on $]\vr_0,1]$, whence $\beta_n^z<\beta_n^\theta$, the two thresholds separating three cases.

\begin{figure}
\centering
\begin{subfigure}{0.475\textwidth}
 \centering
 \includegraphics[height=7cm]{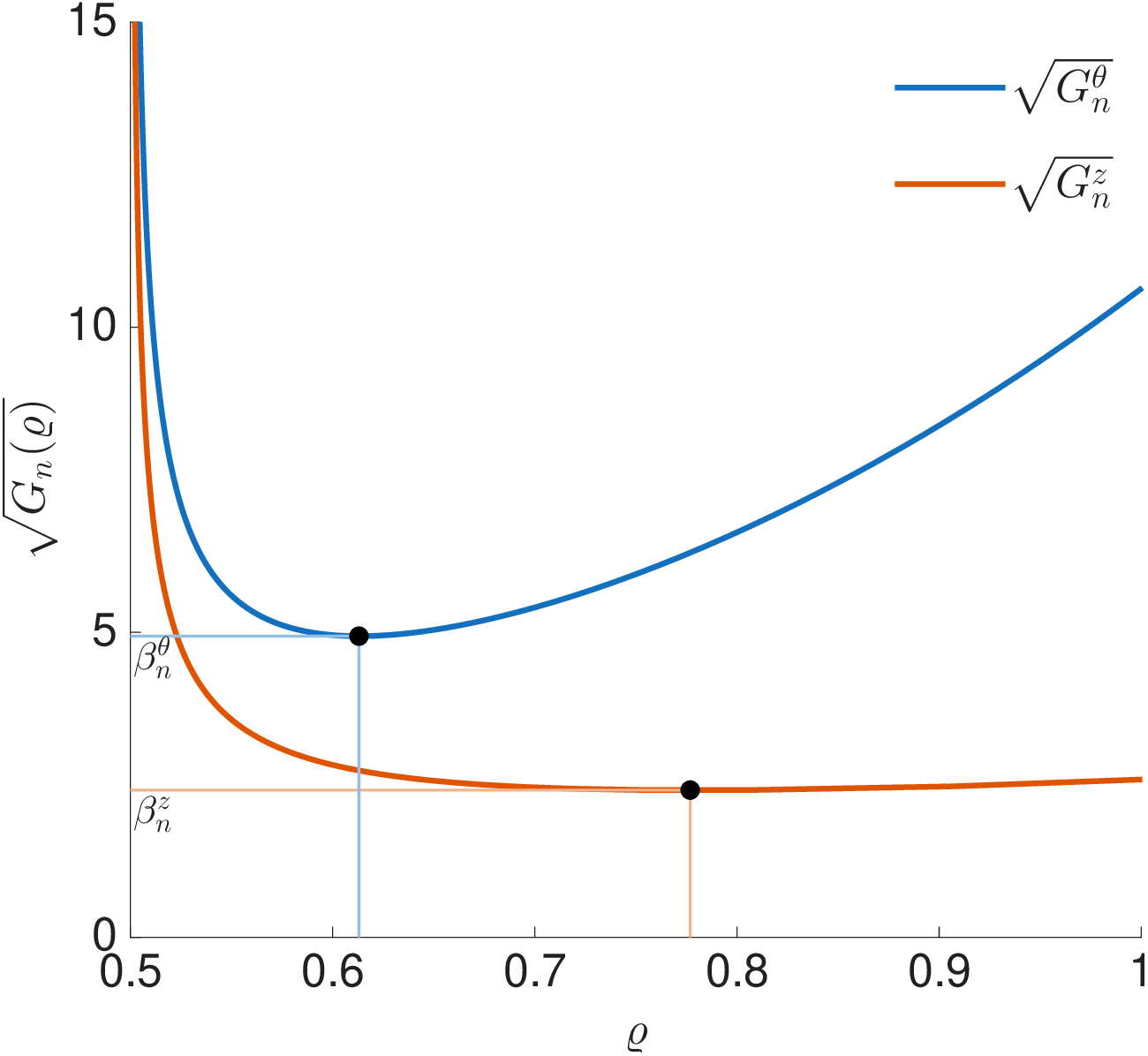}
 \caption{}
 \label{fig:knowles-thresholds_a}
\end{subfigure}
\hspace{7.5pt}
\begin{subfigure}{0.475\textwidth}
 \centering
 \includegraphics[height=7cm]{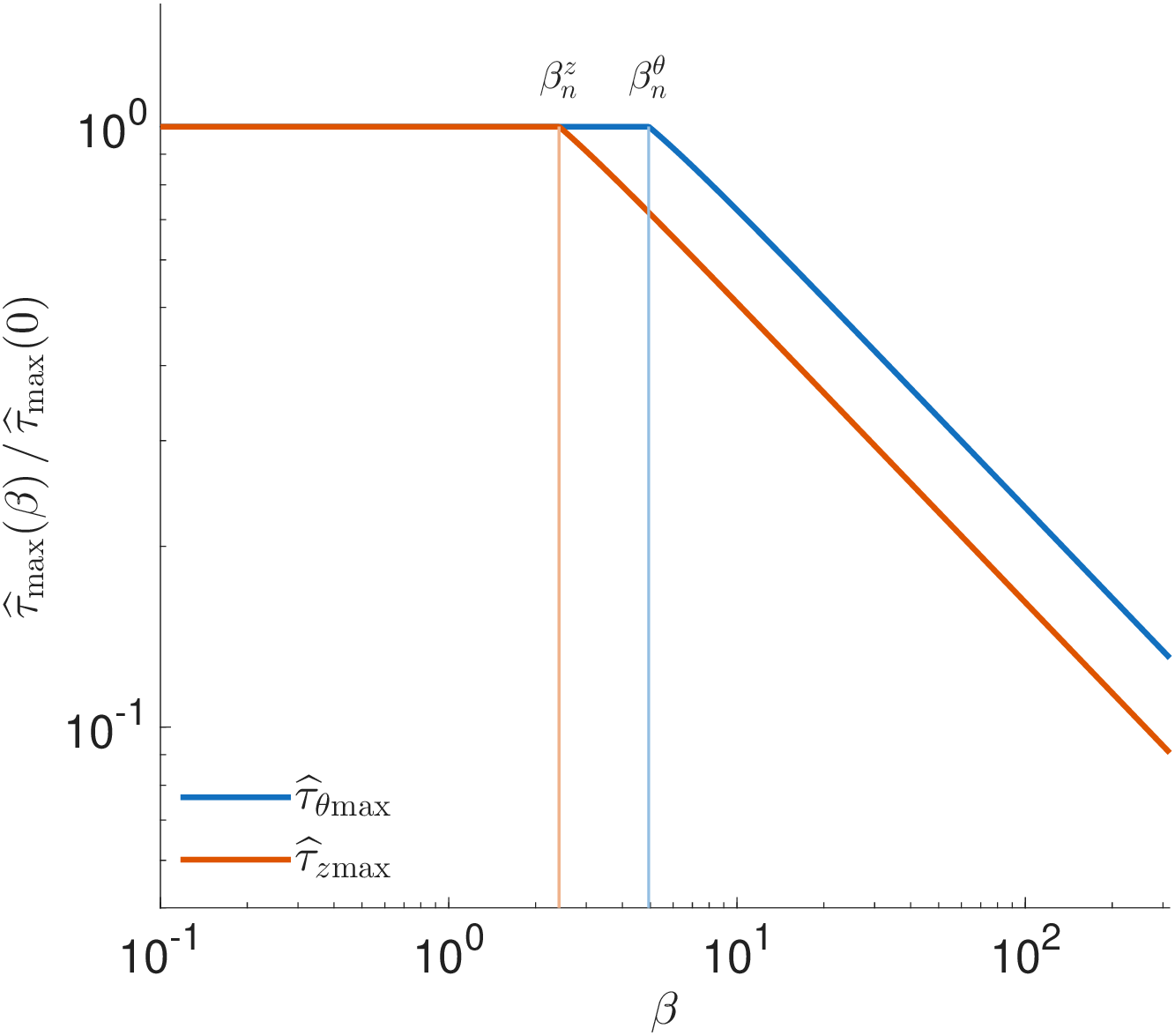}
 \caption{}
 \label{fig:knowles-thresholds_b}
\end{subfigure}
\caption{Nesting thresholds and limit loads for a Knowles tube with $n=1/4$, $b=1$ and $\vr_0=1/2$, with the constant-density profile~\eqref{fhhom}. (a) The nesting bounds~\eqref{eq:Knowles-G}, plotted as $\surd G_n$ so that nesting holds in the corresponding direction precisely where $\beta$ lies below the curve: both diverge at the bonded core and attain their minima at interior radii, yielding $\beta^z_n\approx2.41$ at $\vr\approx0.78$ and $\beta^\theta_n\approx4.94$ at $\vr\approx0.61$. (b) The limit loads, normalised by their dislocation-free values, versus $\beta$: each stays at unity until its own threshold is crossed and decays thereafter, the straight tails on the doubly logarithmic axes exhibiting the common rate $\beta^{2n-1}$, here $\beta^{-1/2}$.}
\label{fig:knowles-thresholds}
\end{figure}

\begin{itemize}
  \item[$\bullet$] \emph{Nested disks}, $\beta\leq\beta_n^z$. Both nesting requirements hold, so that $\mathcal{E}_n(\beta)=E_n(\vr_0)$ and both limit loads retain their dislocation-free values $\vr_0^2S_n(\vr_0)/\sqrt{2b}$ and $\vr_0S_n(\vr_0)/\sqrt{2b}$. Although the ceiling in~\eqref{Endef} is lowered pointwise at every $\vr>\vr_0$, the global admissible traction region is left unchanged;
  \item[$\bullet$] \emph{Asymmetric erosion}, $\beta_n^z<\beta\leq\beta_n^\theta$. The axial nesting requirement fails on a subinterval of $]\vr_0,1]$ while the azimuthal one persists: the dislocation field lowers the limit load in pure axial shear while leaving the pure azimuthal one untouched. Accordingly, $\mathcal{E}_n(\beta)\subsetneq E_n(\vr_0)$, the two regions still sharing the azimuthal extremities of their boundaries;
  \item[$\bullet$] \emph{Full erosion}, $\beta>\beta_n^\theta$. Both requirements fail, both limit loads drop below their dislocation-free values, the two regions no longer share any boundary point, and the admissible traction region shrinks in every direction as $\beta$ grows.
\end{itemize}
The three cases are illustrated in Fig.~\ref{fig:knowles-region_a}, and read quantitatively off the limit loads in Fig.~\ref{fig:knowles-thresholds_b}, each of which stays at its dislocation-free value until its own threshold is crossed and decays thereafter at the common asymptotic rate $\beta^{2n-1}$. Being an intersection of compact convex sets, $\mathcal{E}_n(\beta)$ is itself compact and convex, but outside the first case it is no longer an elliptical disk, its boundary being assembled from contributions of distinct radii. Both thresholds increase strictly with $n$, diverging as $n\rightarrow1/2^-$ and vanishing as $n\rightarrow0^+$---see Fig.~\ref{fig:knowles-region_b}: erosion switches off continuously at the critical exponent, consistent with the $\beta$-independence of~\eqref{con12}, whereas for a strongly softening material an arbitrarily weak dislocation field suffices to erode the admissible traction region.

\begin{figure}
\centering
\begin{subfigure}{0.475\textwidth}
 \centering
 \includegraphics[height=7cm]{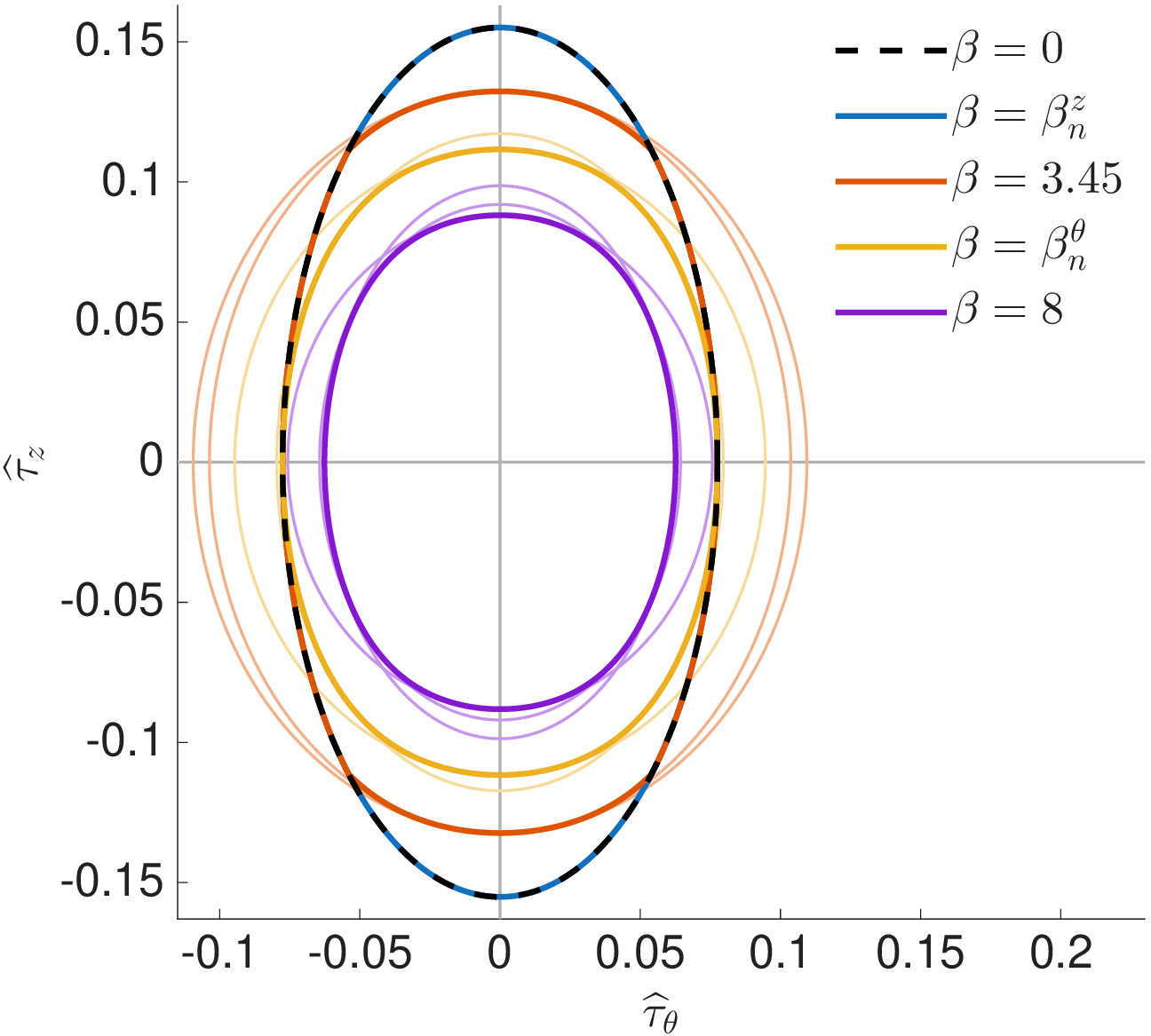}
 \caption{}
 \label{fig:knowles-region_a}
\end{subfigure}
\hspace{7.5pt}
\begin{subfigure}{0.475\textwidth}
 \centering
 \includegraphics[height=7cm]{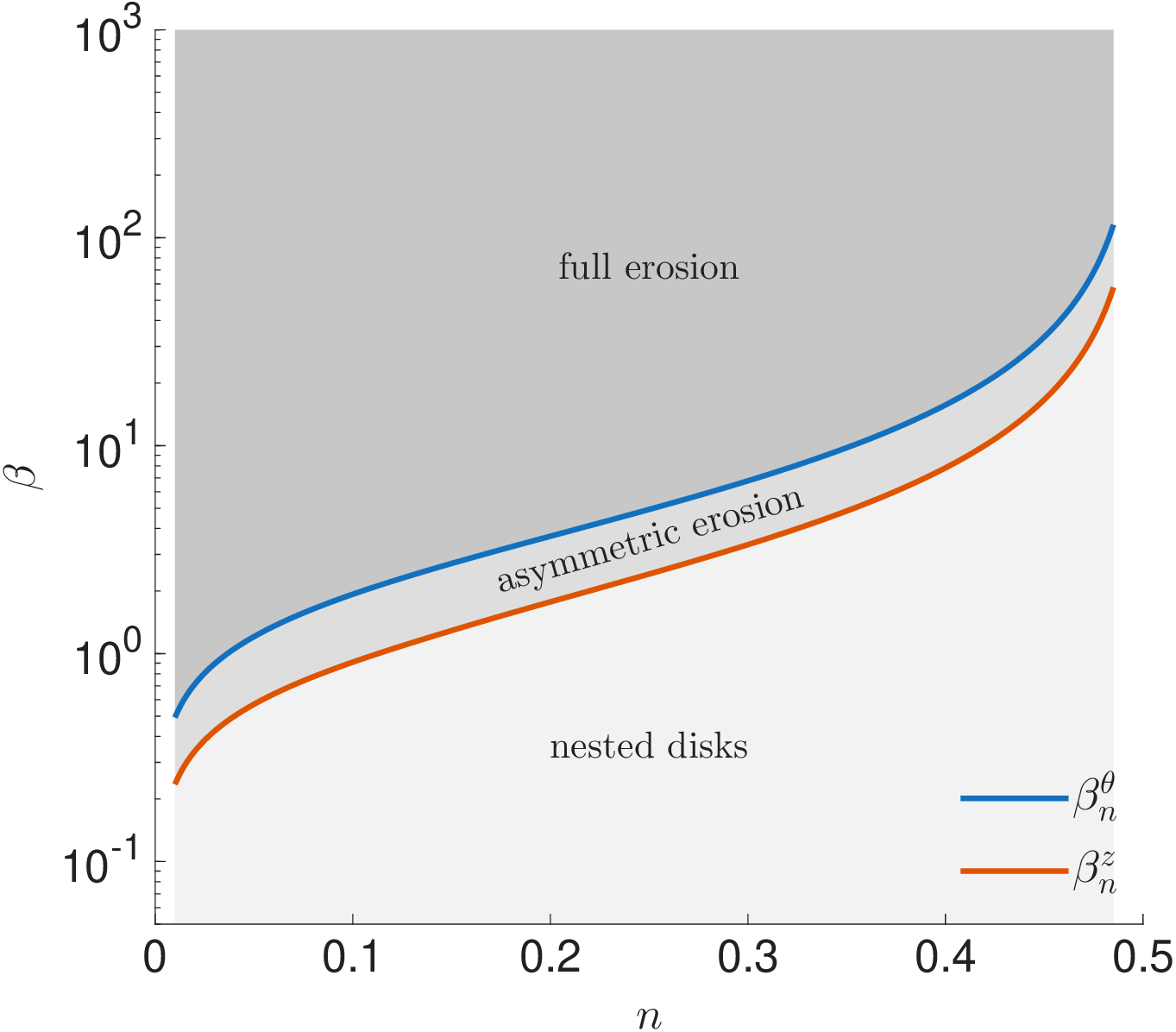}
 \caption{}
 \label{fig:knowles-region_b}
\end{subfigure}
\caption{Erosion of the admissible traction region of a Knowles tube, at $n=1/4$, $b=1$ and $\vr_0=1/2$. (a) The region $\mathcal{E}_n(\beta)$ for one dislocation strength in each case, against the dislocation-free region $E_n(\vr_0)$ (black dashed); the thin tinted curves are the constituent disks active on the corresponding boundary. At $\beta=2$ the sole active constituent is $E_n(\vr_0)$ and the two regions coincide; at $\beta=3.5$ the bonded core still supplies the azimuthal extremity while interior radii supply the rest; at $\beta=8$ the core is inactive and the two regions share no boundary point. Only the first quadrant is shown, the region being symmetric in both tractions. (b) The two thresholds versus the power-law index $n$, partitioning the $(n,\beta)$ plane into the three cases (shaded, from light to dark): both increase strictly with $n$, diverging as $n\rightarrow1/2^-$ and vanishing as $n\rightarrow0^+$.}
\label{fig:knowles-region}
\end{figure}
\end{example}

\subsection{Mooney–Rivlin materials}

For the Mooney–Rivlin model~\eqref{mrmod}, $\widehat{Q}_1=\alpha/2$ and $\widehat{Q}_2=(1-\alpha)/2$ are constant, so the coefficients of~\eqref{eqdim} do not depend on the shear strains and the reduced equilibrium equations yield the following linear system
\begin{equation}\label{eq:MR_system}
\left\{\begin{array}{l}
\left[1+(1-\alpha)\beta^2\Fh^2(\vr)\right]\kh_\theta+(1-\alpha)\beta\Fh(\vr)\kh_z=\dfrac{\tauh_\theta}{\vr^2}\,,\\[5mm]
(1-\alpha)\beta\Fh(\vr)\kh_\theta+\kh_z=\dfrac{\tauh_z}{\vr}\,.
\end{array}\right.
\end{equation}
Its determinant is $1+\alpha(1-\alpha)\beta^2\Fh^2(\vr) \ge 1$. Hence, for any dislocation density and any applied helical traction, \eqref{eq:MR_system} admits the unique solution
\begin{subequations}\label{eq:MR_strains}
\begin{align}
\label{eq:MR_ktheta}
\kh_\theta(\vr)&=\dfrac{\tauh_\theta-(1-\alpha)\beta\vr\Fh(\vr)\tauh_z}{\vr^2\left[1+\alpha(1-\alpha)\beta^2\Fh^2(\vr)\right]}\,,\\[2.5mm]
\label{eq:MR_kz}
\kh_z(\vr)&=\dfrac{\left[1+(1-\alpha)\beta^2\Fh^2(\vr)\right]\vr\tauh_z-(1-\alpha)\beta\Fh(\vr)\tauh_\theta}{\vr^2\left[1+\alpha(1-\alpha)\beta^2\Fh^2(\vr)\right]}\,.
\end{align}
\end{subequations}
Inserting~\eqref{eq:MR_strains} into~\eqref{inits} yields the Cauchy problems
\begin{subequations}\label{cauchyMR}
\begin{align}
&\left\{\begin{array}{r@{}l}
\ph'(\vr)&{}=\dfrac{\tauh_\theta-(1-\alpha)\beta\vr\Fh(\vr)\tauh_z}{\vr^3\left[1+\alpha(1-\alpha)\beta^2\Fh^2(\vr)\right]}\,,\\[5mm]
\ph(\vr_0)&{}=0\,,
\end{array}\right.
\\[2.5mm]
&\left\{\begin{array}{r@{}l}
\wh'(\vr)&{}=\dfrac{\left[1+(1-\alpha)\beta^2\Fh^2(\vr)\right]\vr\tauh_z-(1-\alpha)\beta\Fh(\vr)\tauh_\theta}{\vr^2\left[1+\alpha(1-\alpha)\beta^2\Fh^2(\vr)\right]}\,,\\[5mm]
\wh(\vr_0)&{}=0\,,
\end{array}\right.
\end{align}
\end{subequations}
whose solutions give the equilibrium configuration of the form~\eqref{eq:eg4_ansatz} by quadrature,
\begin{subequations}\label{eq:sol_MR}
\begin{align}
\label{eq:sol_MR_psi}
\ph(\vr)&=\int_{\vr_0}^{\vr}\dfrac{\tauh_\theta-(1-\alpha)\beta\xi\Fh(\xi)\tauh_z}{\xi^3\left[1+\alpha(1-\alpha)\beta^2\Fh^2(\xi)\right]}\d\xi\,,\\[2.5mm]
\label{eq:sol_MR_w}
\wh(\vr)&=\int_{\vr_0}^{\vr}\dfrac{\left[1+(1-\alpha)\beta^2\Fh^2(\xi)\right]\xi\tauh_z-(1-\alpha)\beta\Fh(\xi)\tauh_\theta}{\xi^2\left[1+\alpha(1-\alpha)\beta^2\Fh^2(\xi)\right]}\d\xi\,.
\end{align}
\end{subequations}

One may see that, according to the solutions above, dislocations introduce a coupling of the axial and azimuthal shear modes in Mooney–Rivlin materials: for $\alpha\neq1$ and $\beta\neq0$, an axial traction $\tauh_z\neq0$ induces an azimuthal shear $\kh_\theta\neq0$ even when $\tauh_\theta=0$, and conversely, namely
\begin{equation}
\label{eq:MR_induced}
\kh_\theta\big|_{\tauh_\theta=0}=-\dfrac{(1-\alpha)\beta\Fh(\vr)\tauh_z}{\vr\left[1+\alpha(1-\alpha)\beta^2\Fh^2(\vr)\right]}\,,
\quad
\kh_z\big|_{\tauh_z=0}=-\dfrac{(1-\alpha)\beta\Fh(\vr)\tauh_\theta}{\vr^2\left[1+\alpha(1-\alpha)\beta^2\Fh^2(\vr)\right]}\,.
\end{equation}
Both induced shears are odd in $\beta$ and the coupling is hence chiral: reversing the orientation of the dislocations reverses the induced shear. This coupling has its origins in the off-diagonal entries $(1-\alpha)\beta\Fh$ of~\eqref{eq:MR_system}, and two cases of decoupling follow. First, in the absence of dislocations, $\beta=0$, one recovers the purely elastic Mooney–Rivlin solution, in which the two shear strains are independent \citep{horgan2003helical}.
This decoupling is not specific to the Mooney–Rivlin model: in the absence of dislocations, the off-diagonal terms of~\eqref{eqdim} vanish and $\widehat{I}_1=\widehat{I}_2$, so that, for every material, neither traction induces shear in the other mode. Conversely, since the Mooney–Rivlin model coincides with the most general incompressible isotropic model of third-order elasticity \citep{rivlin1951,destrade2010onset}, the induced shears~\eqref{eq:MR_induced} capture, to leading order, the dislocation-induced coupling of every incompressible isotropic solid.
Second, in the neo-Hookean limit $\alpha=1$, the modes decouple regardless of the dislocation field (cf.~\eqref{ednh}).

In the remainder of this section, we specialize to the case of uniformly distributed dislocations, i.e. $\Fh(\vr)=(\vr^2-\vr_0^2)/\vr$ by~\eqref{fhhom}. The Cauchy problems~\eqref{cauchyMR} hence reduce to
\begin{subequations}\label{cauchyMR_uniform}
\begin{align}
&\left\{\begin{array}{r@{}l}
\ph'(\vr)&{}=\dfrac{\tauh_\theta-(1-\alpha)\beta\tauh_z(\vr^2-\vr_0^2)}{\vr\left[\vr^2+\alpha(1-\alpha)\beta^2(\vr^2-\vr_0^2)^2\right]}\,,\\[5mm]
\ph(\vr_0)&{}=0\,,
\end{array}\right.
\\[2.5mm]
&\left\{\begin{array}{r@{}l}
\wh'(\vr)&{}=\dfrac{\tauh_z\left[\vr^2+(1-\alpha)\beta^2(\vr^2-\vr_0^2)^2\right]-(1-\alpha)\beta\tauh_\theta(\vr^2-\vr_0^2)}{\vr\left[\vr^2+\alpha(1-\alpha)\beta^2(\vr^2-\vr_0^2)^2\right]}\,,\\[5mm]
\wh(\vr_0)&{}=0\,.
\end{array}\right.
\end{align}
\end{subequations}
If $\alpha=1$ or $\beta=0$, the system~\eqref{cauchyMR_uniform} reduces to~\eqref{inits} with the strains~\eqref{ksnh}, whose solution is given by~\eqref{ednh}.
In what follows, we investigate the case where $\alpha\in\,]0,1[$ and $\beta\neq0$.
Setting $x \equiv \vr^2$, ${\widetilde{\psi}(x) \equiv \ph(\vr)}$, and ${\widetilde{w}(x) \equiv \wh(\vr)}$, the Cauchy problems~\eqref{cauchyMR_uniform} recast as
\begin{subequations}\label{eq:MR_pf}
\begin{align}
&\left\{\begin{array}{r@{}l}
\widetilde{\psi}'(x)&{}=\dfrac{a_1}{x}-\dfrac{a_1}{2}\dfrac{Q'(x)}{Q(x)}-\dfrac12\left(a_1-\dfrac{\tauh_\theta}{\vr_0^2}\right)\dfrac{1}{Q(x)}\,,\\[5mm]
\widetilde{\psi}(\vr_0^2) &{}=0\,,
\end{array}\right.
\\[2.5mm]
&\left\{\begin{array}{r@{}l}
\widetilde{w}'(x)&{}=\left(\dfrac{\tauh_z+a_3}{2}+a_2\right)\dfrac1x-a_2\dfrac{Q'(x)}{Q(x)}-\dfrac{a_3}{2}\dfrac{1}{Q(x)}\,,\\[5mm]
\widetilde{w}(\vr_0^2)&{}=0\,,
\end{array}\right.
\end{align}
\end{subequations}
where
\begin{equation}\label{eq:MR_Q}
Q(x) \equiv x+\alpha(1-\alpha)\beta^2\left(x-\vr_0^2\right)^2\,,
\end{equation}
and
\begin{equation}\label{eq:MR_coeffs}
a_1 \equiv \dfrac{\tauh_\theta+(1-\alpha)\beta\tauh_z\vr_0^2}{2\alpha(1-\alpha)\beta^2\vr_0^4}\,,\qquad
a_2 \equiv \dfrac{\tauh_\theta}{4\alpha\beta\vr_0^2}\,,\qquad
a_3 \equiv \dfrac{\tauh_\theta+2(1-\alpha)\beta\tauh_z\vr_0^2}{2\alpha\beta\vr_0^2}\,.
\end{equation}
It follows by integration of~\eqref{eq:MR_pf} that, back in terms of $\vr$,
\begin{subequations}\label{eq:MR_uniform}
\begin{align}
\ph(\vr) &= a_1\left[ \Lambda(\vr^2) - \ln\dfrac{\vr_0 \sqrt{Q(\vr^2)}}{\vr^2}\right] -\dfrac{\tauh_\theta}{\vr_0^2}\Lambda(\vr^2) \,,\\
\wh(\vr) &= \tauh_z\ln\dfrac{\vr}{\vr_0} - a_2\ln\dfrac{Q(\vr^2)}{\vr^2} + a_3\left[\ln\dfrac{\vr}{\vr_0}+\Lambda(\vr^2)\right] \,,
\end{align}
\end{subequations}
with
\begin{equation}\label{eq:MR_Lambda_int}
\Lambda(\vr^2)=-\dfrac12\int_{\vr_0^2}^{\vr^2}\dfrac{\d\xi}{Q(\xi)}=-\int_1^{Q'(\vr^2)}\dfrac{\d v}{v^2-\Delta}\,,
\end{equation}
where the second equality follows by the change of variable $v=Q'(\xi)$.\footnote{Using the identity ${{Q'}^2-4\alpha(1-\alpha)\beta^2Q=\Delta}$---where $\Delta=1-4\alpha(1-\alpha)\beta^2\vr_0^2$ is the discriminant of $Q$---and $Q''=2\alpha(1-\alpha)\beta^2$, it follows that
\begin{equation}
\dfrac{\d\xi}{Q(\xi)}=\dfrac{4\alpha(1-\alpha)\beta^2}{{Q'}^2 - \Delta}\d\xi=\dfrac{4\alpha(1-\alpha)\beta^2}{\nu^2 - \Delta}\frac{\d\nu}{2\alpha(1-\alpha)\beta^2}=\dfrac{2 \d\nu}{\nu^2 - \Delta}\,,
\end{equation}
}
We may hence compute $\Lambda$ as follows depending on the sign of $\Delta$\footnote{If $\Delta>0$, we perform the change of variable $\tanh(\zeta)=\sqrt{\Delta}/\nu$; and if $\Delta<0$, we perform the change of variable $\tan(\zeta)=\sqrt{|\Delta|}/\nu$.}
\begin{equation}\label{eq:MR_Lambda}
\Lambda(\vr^2)=\left\{\begin{array}{ll}
	\dfrac{1}{\sqrt{\Delta}}\left[\operatorname{arctanh}\dfrac{\sqrt{\Delta}}{Q'(\vr^2)}-\operatorname{arctanh}\sqrt{\Delta}\right]\,, & \Delta>0\,,\\[3mm]
	\dfrac{1}{Q'(\vr^2)}-1\,, & \Delta=0\,,\\[3mm]
	\dfrac{1}{\sqrt{|\Delta|}}\left[\arctan\dfrac{\sqrt{|\Delta|}}{Q'(\vr^2)}-\arctan\sqrt{|\Delta|}\right]\,, & \Delta<0\,,
\end{array}\right.
\end{equation}
where, following~\eqref{eq:MR_Q}, $Q'(\vr^2)=1+2\alpha(1-\alpha)\beta^2\left(\vr^2-\vr_0^2\right)$.
Note that, although the coefficients~\eqref{eq:MR_coeffs} diverge as $\alpha\to1^-$ or $\beta\to0$, one may however find that~\eqref{eq:MR_uniform} tends to the neo-Hookean solution~\eqref{ednh} in both limit cases.

\subsection{Gent--Gent materials}\label{S:GG}

For a Gent--Gent material~\eqref{GGmodel}, recalling~\eqref{ggcal}, the dimensionless response functions $\Gc_1$ and $\Gc_2$ appearing in the equilibrium equations~\eqref{eqdim} read
\begin{subequations}\label{gcgg}
\begin{align}
  \Gc_1(\kh_\theta,\kh_z;\Fh) &= \dfrac{\alpha J_m}{2[J_m-\beta^2\Fh^2-\kh_\theta^2-\kh_z^2]}\,,\\
  \Gc_2(\kh_\theta,\kh_z;\Fh) &= \dfrac{3(1-\alpha) }{2[3+\beta^2\Fh^2+(\beta\Fh\kh_\theta+\kh_z)^2+\kh_\theta^2]}\,.
\end{align}
\end{subequations}
Inserting~\eqref{gcgg} into the equilibrium equations~\eqref{eqdim} yields a highly nonlinear system of equations; unlike the generalized neo-Hookean case~\S\ref{S:gnHmod}, the off-diagonal terms $\beta\Gc_2\Fh$ no longer vanish and the system does not reduce to a single scalar equation. Similarly to the Gent case~\S\ref{S:Gent}, the response function $\Gc_1$ diverges as $\widehat{I}_1\to J_m+3$, and any solution must satisfy the limited-extensibility restriction
\begin{equation}\label{admgg}
  \kh_\theta^2(\vr)+\kh_z^2(\vr) < J_m-\beta^2\Fh^2(\vr)\,,\quad\forall\vr\in[\vr_0,1]\,.
\end{equation}
The response function $\Gc_2$ imposes no further restriction: since $\widehat{I}_2\geq3$, it remains bounded, ${0\leq\Gc_2\leq(1-\alpha)/2}$, and the extensibility limit is precisely that of the Gent model, i.e.~\eqref{eq:AdmiG}.
As for the Gent model, \eqref{admgg} implies the global condition
\begin{equation}\label{admgg_max}
  \beta^2\max_{\vr\in[\vr_0,1]}\Fh^2(\vr)<J_m\,,
\end{equation}
i.e.~\eqref{limitGent}, which is therefore necessary for the existence of an admissible solution and caps the dislocation strength independently of the applied tractions. For the Gent model, this condition is also sufficient, the proof resting on the reduction of the equilibrium system~\eqref{gentred} to the single scalar equation $\Phi(\kh_\theta)=\tauh_\theta/\vr^2$ with $\Phi$ strictly increasing---cf.~\eqref{Phidef}. For the Gent--Gent case, this reduction is no longer available, and we restrict ourselves to necessity.

\begin{example}

\begin{figure}[t!]
\centering
\begin{subfigure}{0.475\textwidth}
 \centering
 \includegraphics[height=7cm]{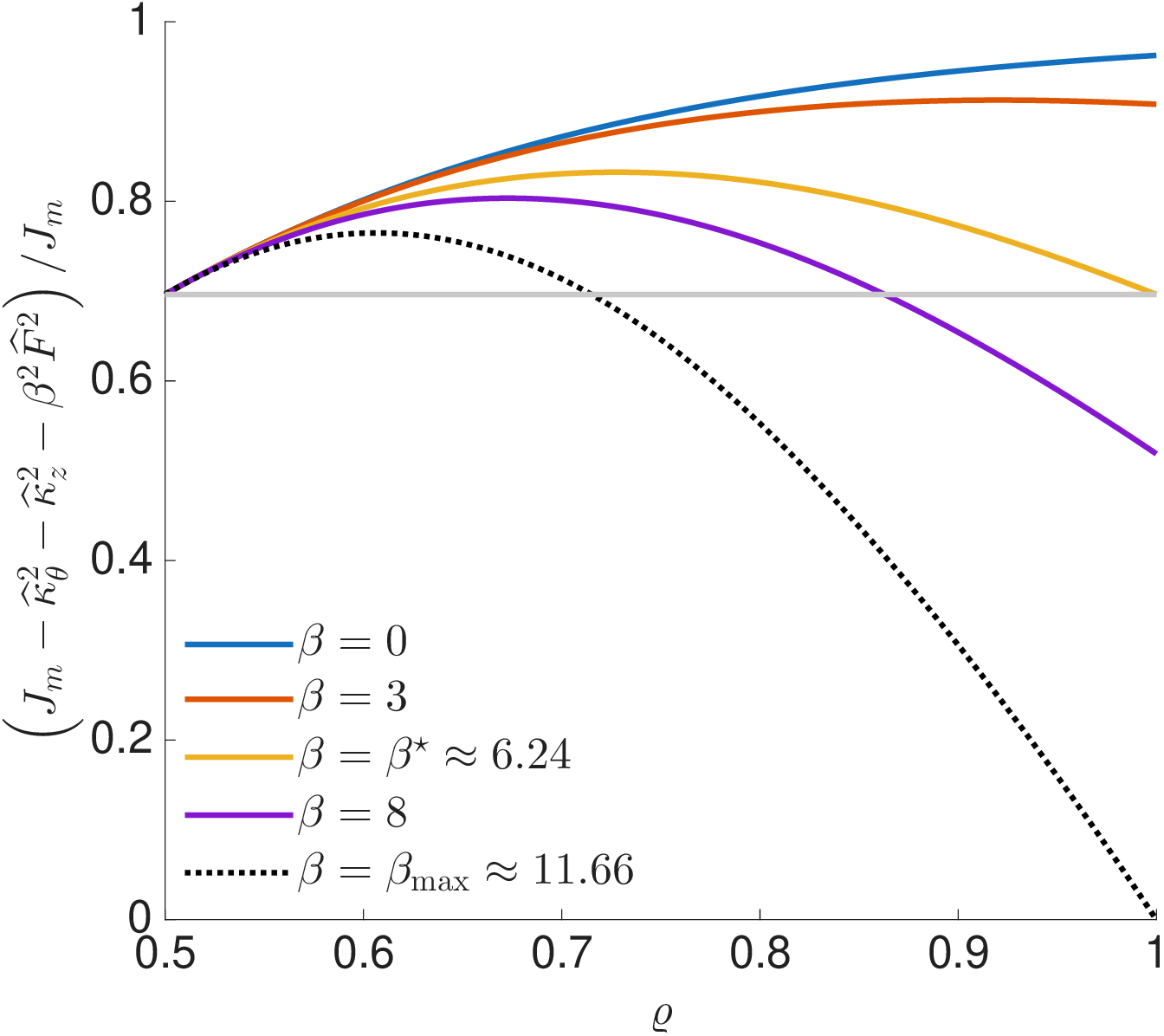}
 \caption{}
 \label{fig:admiss}
\end{subfigure}\\[6pt]
\begin{subfigure}{0.475\textwidth}
 \centering
 \includegraphics[height=7cm]{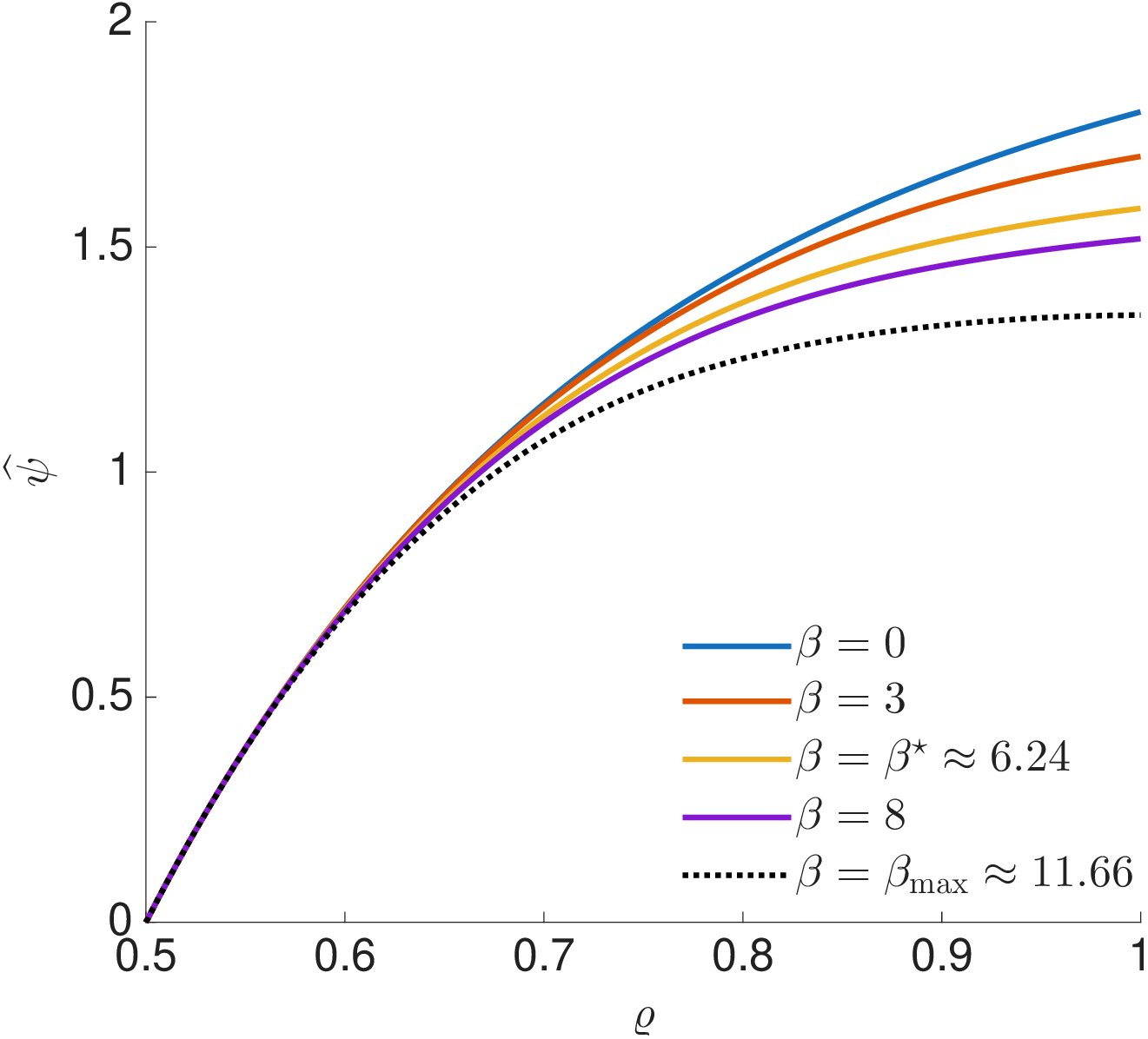}
 \caption{}
 \label{fig:psigg}
\end{subfigure}
\hspace{7.5pt}
\begin{subfigure}{0.475\textwidth}
 \centering
 \includegraphics[height=7cm]{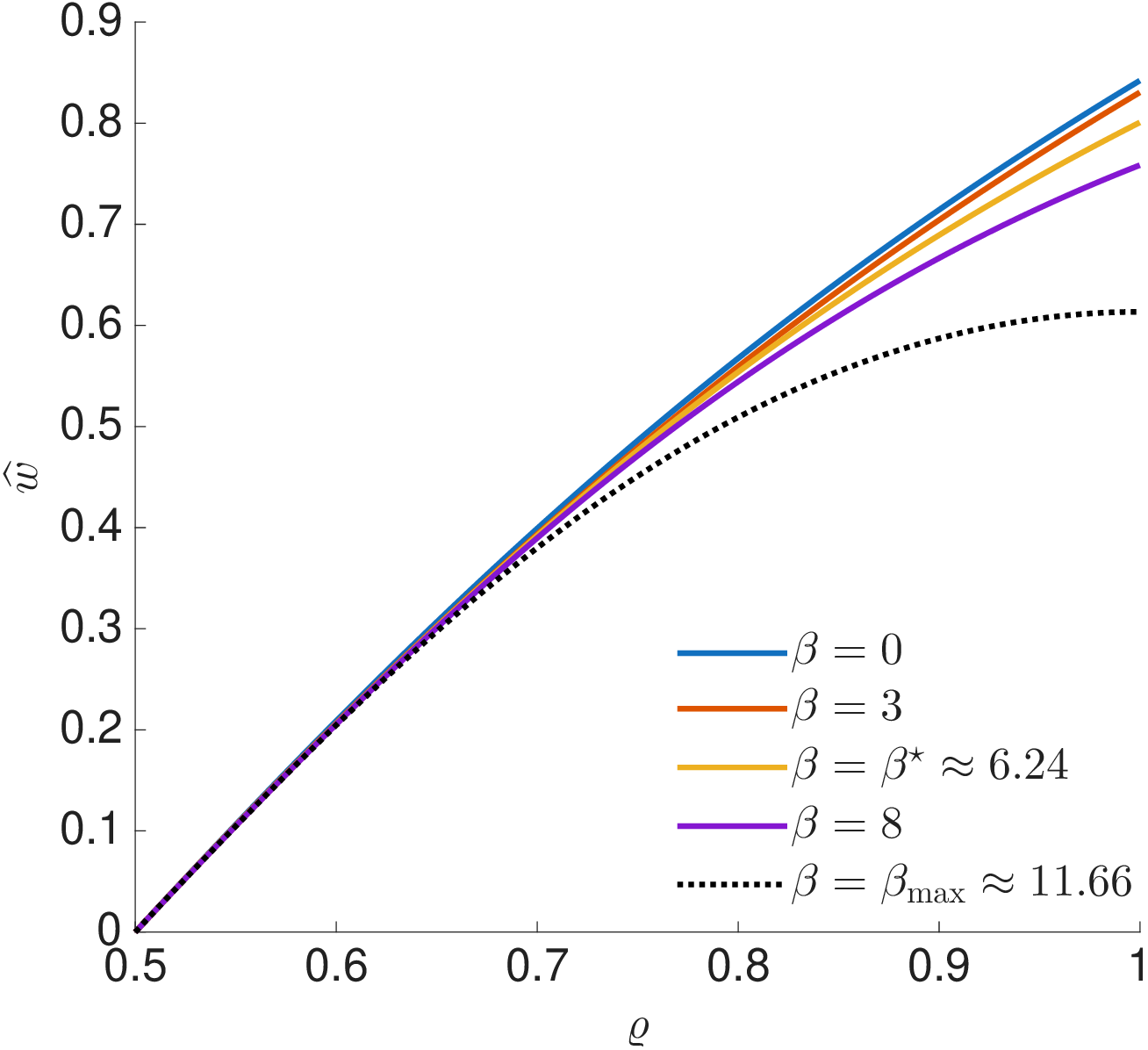}
 \caption{}
 \label{fig:upsilongg}
\end{subfigure}
\caption{Helical shear of a Gent--Gent tube with the fit~\eqref{rubbergg}, the constant-density profile~\eqref{fhhom}, $\vr_0=1/2$ and $\tauh_\theta=\tauh_z=1$, for $\beta\in\{0,3,\beta^\star,8,\beta_{\max}\}$, with $\beta^\star\approx6.24$ and $\beta_{\max}\approx11.66$. (a)~Residual extensibility margin fraction $\mathcal{M}(\vr;\beta)/J_m$, all curves sharing a common inner-wall value since $\Fh(\vr_0)=0$. The narrowest margin lies at the inner wall for $\beta<\beta^\star$, at both walls at $\beta=\beta^\star$, and at the outer wall beyond, closing there as $\beta\to\beta_{\max}$. (b)~Radial profile of~$\ph$ at different dislocation strengths. (c)~Radial profile of~$\wh$ at different dislocation strengths.}
\label{GGfig}
\end{figure}

In the absence of a closed-form solution of~\eqref{eqdim}, even for the uniform dislocation density~\eqref{fhhom}, we resort to numerical solutions. Unlike the Mooney--Rivlin model, accurate up to moderate amounts of shear, the Gent--Gent model captures the response of rubber-like materials up to the strain-hardening regime: fitting the uniaxial-extension data of~\cite{dobrynin2011universality}, \cite{destrade_penelope} found
\begin{equation}\label{rubbergg}
C_1=\alpha\mu=4.7472\,\textrm{N mm}^{-2}\,,\qquad C_2=(1-\alpha)\mu=2.9650\,\textrm{N mm}^{-2}\,,\qquad J_m=76.48\,,
\end{equation}
so that $\alpha=C_1/(C_1+C_2)\approx0.616$. We adopt the same extensibility limit, annulus $\vr_0=1/2$, uniform dislocation density~\eqref{fhhom} and tractions $\tauh_\theta=\tauh_z=1$ as in the Gent example of~\S\ref{S:Gent}, so that the two tubes differ only through the $I_2$ term, and the necessary condition~\eqref{admgg_max} again reads $\beta<\beta_{\max}=4\sqrt{J_m}/3\approx11.66$.

Fig.~\ref{GGfig} shows the numerical solutions for different values of $\beta$, including the crossover density $\beta^\star$ and the limiting case $\beta=\beta_{\max}$. The residual extensibility margin $\mathcal{M}(\vr;\beta)\equiv J_m-\kh_\theta^2-\kh_z^2-\beta^2\Fh^2$, as in~\eqref{eq:margin}, remains positive throughout the admissible solutions (Fig.~\ref{fig:admiss}), and its profiles reproduce the structure found for the Gent tube.
They share a common inner-wall value, since at $\vr_0$, where $\Fh$ vanishes, the margin $\mathcal{M}$ is blind to the dislocation field. Towards the outer wall, the dislocations erode the margin, which closes at $\vr=1$ as $\beta\to\beta_{\max}$: the eroded limit $J_m-\beta^2\Fh^2(1)$ vanishes there, and~\eqref{admgg} forces $\kh_\theta,\kh_z\to0$ while $\Gc_1$ diverges: the outer layer locks under the dislocation field alone, carrying any applied tractions with vanishing shear.
Accordingly, the outer-wall displacements $\ph$ and $\wh$ decrease with $\beta$, so that the dislocations stiffen the tube in both shear modes; ultimately the displacements profiles flatten near $\vr=1$ as $\beta$ approaches $\beta_{\max}$ (Figs.~\ref{fig:psigg} and~\ref{fig:upsilongg}).
As for the Gent tube, the margin is found to be narrowest at one of the two walls, and a crossover density $\beta^\star$, at which the margins at the two walls coincide, separates inner-wall from outer-wall locking. Here $\beta^\star\approx6.24$, against $\beta^\star\approx4.66$ for the Gent tube under the same conditions. Indeed, the Gent--Gent tube is softer in shear, the stiffening Gent term carrying only the fraction $\alpha$ of the shear modulus, so that the inner wall, where the shear is largest, draws more heavily on the extensibility limit, $\mathcal{M}(\vr_0)/J_m\approx0.70$ for the Gent--Gent tube against $0.82$ for the Gent tube, and the outer wall takes over only at a higher dislocation strength.

\end{example}

\section{Weakly nonlinear analysis of the equilibrium problem}
\label{S:WNL}

In the absence of dislocations, the cross terms in~\eqref{eqdim} vanish, so that for every material, neither traction induces shear in the other mode, and any coupling between the two modes may only appear in the presence of a dislocation field.\footnote{\label{fn:coupling}By \emph{coupling}, we mean that a traction in one mode induces shear in the other, e.g. $\kh_\theta\neq0$ under $\tauh_\theta=0$ and $\tauh_z\neq0$. This is to be distinguished from an \emph{interaction}, whereby each mode alters the stiffness of the other without inducing it, so that $\kh_\theta$ depends on $\tauh_z$ but vanishes whenever $\tauh_\theta=0$, and vice versa.} In~\S\ref{S:Examples}, we explore this coupling for some specific elastic energies in the finite strain regime. In the present section, however, we concern ourselves with the case of small dislocation strengths and strains. In this case, the response of any material is governed by a few elastic moduli, and the origin of the coupling can be identified independently of the model. Here, we perform a perturbation analysis of the reduced equilibrium equations~\eqref{eqdim} under the hypothesis that the dislocation strength and dimensionless shear stresses at the outer boundary of $\mathcal{B}$ are small in the sense that
\begin{equation}\label{eps4th}
  \veps = \max\{\beta,|\tauh_\theta|,|\tauh_z|\}\ll 1\,,
\end{equation}
retaining terms up to $O(\veps^3)$ in the shear strains, consistently with the weakly nonlinear theory of the fourth order, in which the strain energy enters only through $\mu$, $\A$ and $\D$ via~\eqref{eq:mu} and~\eqref{ad}.\footnote{A fourth-order theory in the energy is a third-order theory in the stress: the stress, and hence the equilibrium equations~\eqref{eqdim}, are retained up to cubic terms in the strain.
}
We search for solutions to~\eqref{eqdim} in the form
\begin{equation}\label{powG}
\kh_\alpha=\sum_{m=0}^{+\infty}\kh_\alpha^{(m)} \quad \textrm{with} \quad \kh_\alpha^{(m)}=O(\veps^m) \quad \textrm{for all } m\in\mathbb{N}_0 \textrm{ and } \alpha\in\{\theta,z\}\,.
\end{equation}

Inserting~\eqref{powG} into~\eqref{eqdim}, recalling from~\eqref{fhin} that $\Fh$ is bounded, and collecting terms of order $O(1)$ produces the following system
\begin{equation}
\left\{\begin{array}{ll}
\left[\Gc_1\left(\kh_\theta^{(0)},\kh_z^{(0)};0\right)+\Gc_2\left(\kh_\theta^{(0)},\kh_z^{(0)};0\right)\right]\kh_\theta^{(0)}=0,\\
[3mm]
\left[\Gc_1\left(\kh_\theta^{(0)},\kh_z^{(0)};0\right)+\Gc_2\left(\kh_\theta^{(0)},\kh_z^{(0)};0\right)\right]\kh_z^{(0)}=0\,,
\end{array}\right.
\end{equation}
which, under the assumption of strong ellipticity of the strain energy function,\footnote{Strong ellipticity of $\Wc$ yields $\Gc_1(\kh_\theta,\kh_z;0)+\Gc_2(\kh_\theta,\kh_z;0)>0$ for all $(\kh_\theta,\kh_z)$, as can be seen from the first condition provided in~\citep[p. 437, (6.3.124)]{Ogden84}.} admits the unique trivial solution $\kh_\theta^{(0)}=\kh_z^{(0)}=0$.

At $O(\veps)$, the off-diagonal terms $\beta\Gc_2\Fh\,\kh_\alpha$ are of higher order, and recalling $\kh_\theta^{(0)}=\kh_z^{(0)}=0$, \eqref{eqdim} yields
\begin{equation}
\left\{\begin{array}{ll}
\left[\Gc_1(0,0;0)+\Gc_2(0,0;0)\right]\kh_\theta^{(1)}=\dfrac{\tauh_\theta}{2\vr^2}\,,\\
[5mm]
\left[\Gc_1(0,0;0)+\Gc_2(0,0;0)\right]\kh_z^{(1)}=\dfrac{\tauh_z}{2\vr}\,.
\end{array}\right.
\end{equation}
Since $\Gc_1(0,0;0)+\Gc_2(0,0;0)=1/2$ by~\eqref{eq:mu}, it follows that
\begin{equation}\label{wnl1}
\kh_\theta^{(1)}(\vr)=\dfrac{\tauh_\theta}{\vr^2}\,,\qquad\kh_z^{(1)}(\vr)=\dfrac{\tauh_z}{\vr}\,,
\end{equation}
in which the two modes remain uncoupled and the dislocation field invisible: at first order, every material responds as the neo-Hookean tube~\eqref{ednh}.

At $O(\veps^2)$, the off-diagonal terms now survive the expansion, so that~\eqref{eqdim} yields
\begin{equation}
\left\{\begin{array}{ll}
\dfrac12\,\kh_\theta^{(2)}+\beta\,\Gc_2(0,0;0)\,\Fh\,\kh_z^{(1)}=0\,,\\
[5mm]
\dfrac12\,\kh_z^{(2)}+\beta\,\Gc_2(0,0;0)\,\Fh\,\kh_\theta^{(1)}=0\,.
\end{array}\right.
\end{equation}
With $\Gc_2(0,0;0)=\Wc_2(3,3)/\mu=-\left(1+\A/(4\mu)\right)/2$ by~\eqref{eq:mu} and~\eqref{alan}, it follows that
\begin{equation}\label{wnl2}
\left\{\begin{array}{ll}
\kh_\theta^{(2)}(\vr)=\beta\tauh_z\left(1+\dfrac{\A}{4\mu}\right)\dfrac{\Fh(\vr)}{\vr}\,,\\
[5mm]
\kh_z^{(2)}(\vr)=\beta\tauh_\theta\left(1+\dfrac{\A}{4\mu}\right)\dfrac{\Fh(\vr)}{\vr^2}\,.
\end{array}\right.
\end{equation}
The dislocation field couples the two modes at this order, each traction inducing a shear in the other mode. This coupling is chiral, being odd in $\beta$, and is carried by $\Wc_2(3,3)$: it is absent in the dislocation-free tube and in materials with $\Wc_2(3,3)=0$, such as the generalized neo-Hookean ones of~\S\ref{S:gnHmod}. Since $1+\A/(4\mu)=\alpha-1$ with $\alpha$ as in~\eqref{mrmod},~\eqref{wnl2} coincides with the leading-order expansion of the Mooney--Rivlin induced shears~\eqref{eq:MR_induced}: to second order, every material couples the two modes as the Mooney--Rivlin tube with the same $\mu$ and $\A$.

Further proceeding with the collection of $O(\veps^3)$ terms, \eqref{eqdim} yields
\begin{equation}\label{eq:eps3}
\left\{\begin{array}{ll}
\dfrac12\,\kh_\theta^{(3)}+\left[\mathcal{K}\left(\beta^2\Fh^2+(\kh_\theta^{(1)})^2+(\kh_z^{(1)})^2\right)+\beta^2\,\Gc_2(0,0;0)\,\Fh^2\right]\kh_\theta^{(1)}+\beta\,\Gc_2(0,0;0)\,\Fh\,\kh_z^{(2)}=0\,,\\
[5mm]
\dfrac12\,\kh_z^{(3)}+\mathcal{K}\left(\beta^2\Fh^2+(\kh_\theta^{(1)})^2+(\kh_z^{(1)})^2\right)\kh_z^{(1)}+\beta\,\Gc_2(0,0;0)\,\Fh\,\kh_\theta^{(2)}=0\,,
\end{array}\right.
\end{equation}
where
\begin{equation}\label{wnlK}
\mathcal{K}\equiv\dfrac{\left(\Wc_{11}+2\Wc_{12}+\Wc_{22}\right)(3,3)}{\mu}=\dfrac12\left(1+\dfrac{\A}{2\mu}+\dfrac{\D}{\mu}\right)\,,
\end{equation}
by~\eqref{eq:mu} and~\eqref{ad}.
Using~\eqref{wnl1} and~\eqref{wnl2}, it follows from~\eqref{eq:eps3} that
\begin{equation}\label{wnl3}
\left\{\begin{array}{ll}
\kh_\theta^{(3)}(\vr)=\dfrac{\tauh_\theta}{\vr^2}\left[-\left(1+\dfrac{\A}{2\mu}+\dfrac{\D}{\mu}\right)\left(\dfrac{\tauh_\theta^2}{\vr^4}+\dfrac{\tauh_z^2}{\vr^2}\right)+\beta^2\left(1+\dfrac{\A}{4\mu}+\dfrac{\A^2}{16\mu^2}-\dfrac{\D}{\mu}\right)\Fh^2(\vr)\right],\\
[5mm]
\kh_z^{(3)}(\vr)=\dfrac{\tauh_z}{\vr}\left[-\left(1+\dfrac{\A}{2\mu}+\dfrac{\D}{\mu}\right)\left(\dfrac{\tauh_\theta^2}{\vr^4}+\dfrac{\tauh_z^2}{\vr^2}\right)+\beta^2\left(\dfrac{\A^2}{16\mu^2}-\dfrac{\D}{\mu}\right)\Fh^2(\vr)\right].
\end{array}\right.
\end{equation}
At this order, the two shear modes interact, yet neither traction induces shear in the other mode: $\tauh_\theta=0$ still yields $\kh_\theta^{(3)}=0$, and $\tauh_z=0$ still yields $\kh_z^{(3)}=0$. The interaction resides entirely in the dislocation-free contribution, the first terms in~\eqref{wnl3}, which correct the stiffness of each mode in proportion to the combined shear traction quantity $\tauh_\theta^2/\vr^4+\tauh_z^2/\vr^2$. The dislocation contribution, the terms in $\beta^2$, corrects the stiffness of each mode in proportion to its own traction only.

Integrating~\eqref{inits} with $\kh_\alpha=\kh_\alpha^{(1)}+\kh_\alpha^{(2)}+\kh_\alpha^{(3)}+O(\veps^4)$, $\alpha\in\{\theta,z\}$ yields the helical solution~\eqref{eq:eg4_ansatz} within the weakly nonlinear theory of the fourth order
\begin{subequations}\label{4thdef}
\begin{align}
\ph(\vr)&=\dfrac{\tauh_\theta}{2}\left(\dfrac{1}{\vr_0^2}-\dfrac{1}{\vr^2}\right)+\beta\tauh_z\left(1+\dfrac{\A}{4\mu}\right)\int_{\vr_0}^{\vr}\dfrac{\Fh(\bar\vr)}{\bar\vr^2}\,\d\bar\vr\nonumber\\
&\quad-\tauh_\theta\left(1+\dfrac{\A}{2\mu}+\dfrac{\D}{\mu}\right)\left[\dfrac{\tauh_\theta^2}{6}\left(\dfrac{1}{\vr_0^6}-\dfrac{1}{\vr^6}\right)+\dfrac{\tauh_z^2}{4}\left(\dfrac{1}{\vr_0^4}-\dfrac{1}{\vr^4}\right)\right]\nonumber\\
&\quad+\beta^2\tauh_\theta\left(1+\dfrac{\A}{4\mu}+\dfrac{\A^2}{16\mu^2}-\dfrac{\D}{\mu}\right)\int_{\vr_0}^{\vr}\dfrac{\Fh^2(\bar\vr)}{\bar\vr^3}\,\d\bar\vr+O(\veps^4)\,,\\
\wh(\vr)&=\tauh_z\ln\dfrac{\vr}{\vr_0}+\beta\tauh_\theta\left(1+\dfrac{\A}{4\mu}\right)\int_{\vr_0}^{\vr}\dfrac{\Fh(\bar\vr)}{\bar\vr^2}\,\d\bar\vr\nonumber\\
&\quad-\tauh_z\left(1+\dfrac{\A}{2\mu}+\dfrac{\D}{\mu}\right)\left[\dfrac{\tauh_\theta^2}{4}\left(\dfrac{1}{\vr_0^4}-\dfrac{1}{\vr^4}\right)+\dfrac{\tauh_z^2}{2}\left(\dfrac{1}{\vr_0^2}-\dfrac{1}{\vr^2}\right)\right]\nonumber\\
&\quad+\beta^2\tauh_z\left(\dfrac{\A^2}{16\mu^2}-\dfrac{\D}{\mu}\right)\int_{\vr_0}^{\vr}\dfrac{\Fh^2(\bar\vr)}{\bar\vr}\,\d\bar\vr+O(\veps^4)\,.
\end{align}
\end{subequations}

\begin{example}

Let us assess the weakly nonlinear theory on the Gent--Gent tube of~\S\ref{S:GG}, with the annulus $\vr_0=1/2$ and the uniform dislocation density~\eqref{fhhom}. For this density, the integrals in~\eqref{4thdef} evaluate to
\begin{equation}\label{appgg}
\begin{gathered}
\int_{\vr_0}^{\vr}\dfrac{\Fh(\bar\vr)}{\bar\vr^2}\,\d\bar\vr=\ln\dfrac{\vr}{\vr_0}-\dfrac{\vr^2-\vr_0^2}{2\vr^2}\,,\quad
\int_{\vr_0}^{\vr}\dfrac{\Fh^2(\bar\vr)}{\bar\vr^3}\,\d\bar\vr=\ln\dfrac{\vr}{\vr_0}+\dfrac{(\vr^2-\vr_0^2)(\vr_0^2-3\vr^2)}{4\vr^4}\,,
\\
\int_{\vr_0}^{\vr}\dfrac{\Fh^2(\bar\vr)}{\bar\vr}\,\d\bar\vr=\dfrac{\vr^4-\vr_0^4}{2\vr^2}-2\vr_0^2\ln\dfrac{\vr}{\vr_0}\,.
\end{gathered}
\end{equation}
For the Gent--Gent model~\eqref{GGmodel} with the experimental fit~\eqref{rubbergg}, the relations~\eqref{ggcal} yield the Landau moduli
\begin{equation}
\mu=7.7122\,\textrm{N mm}^{-2}\,,\qquad\A=-42.7088\,\textrm{N mm}^{-2}\,,\qquad\D=12.7159\,\textrm{N mm}^{-2}\,.
\end{equation}

\begin{table}
\centering
\caption{Sup-norm difference between the weakly nonlinear solution~\eqref{4thdef} and the numerical solution of~\eqref{eqdim}, for a Gent--Gent tube with the fit~\eqref{rubbergg}, the constant-density profile~\eqref{fhhom}, $\vr_0=1/2$ and $\beta=\tauh_\theta=\tauh_z=\veps$.}
\label{tab:wnl}
\begin{tabular}{ccc}
\hline
$\veps$ & $\|\ph_{\mathrm{num}}-\ph_{\mathrm{approx}}\|_\infty$ & $\|\wh_{\mathrm{num}}-\wh_{\mathrm{approx}}\|_\infty$\\
\hline
$1\times10^{-1}$ & $1.447\times10^{-5}$ & $1.071\times10^{-5}$\\
$5\times10^{-2}$ & $9.513\times10^{-7}$ & $6.860\times10^{-7}$\\
$1\times10^{-2}$ & $1.512\times10^{-9}$ & $1.104\times10^{-9}$\\
$5\times10^{-3}$ & $9.439\times10^{-11}$ & $6.904\times10^{-11}$\\
$1\times10^{-3}$ & $1.509\times10^{-13}$ & $1.105\times10^{-13}$\\
\hline
\end{tabular}
\end{table}

\begin{figure}
\centering
\includegraphics[height=7cm]{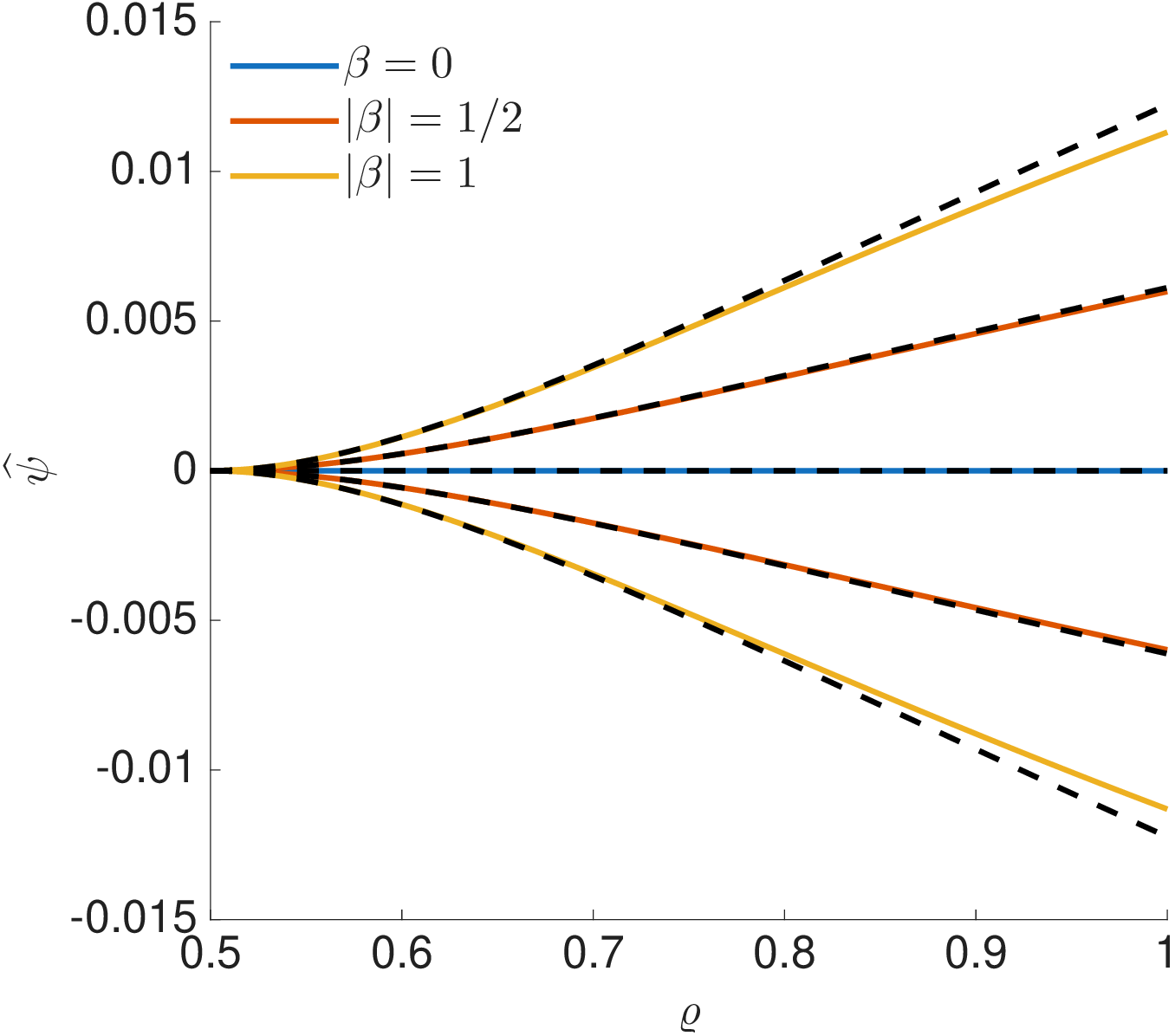}
\caption{Twist induced by a pure axial traction, $\tauh_\theta=0$ and $\tauh_z=1/10$, in a Gent--Gent tube with the fit~\eqref{rubbergg}, the constant-density profile~\eqref{fhhom} and $\vr_0=1/2$, for $\beta=0$, $\pm1/2$ and $\pm1$: numerical solution of~\eqref{eqdim} (solid) and weakly nonlinear prediction~\eqref{4thdef} (black dashed). The twist vanishes without dislocations and reverses with them.}
\label{fig:wnl-chiral}
\end{figure}

Table~\ref{tab:wnl} compares the solution~\eqref{4thdef} with the numerical solutions of~\eqref{eqdim}, for $\beta=\tauh_\theta=\tauh_z=\veps$, in the sup-norm $\|f\|_\infty=\max_{\vr\in[\vr_0,1]}|f(\vr)|$. The differences decrease as $\veps^4$, confirming that~\eqref{4thdef} is accurate to $O(\veps^3)$, with an $O(\veps^4)$ remainder.
Fig.~\ref{fig:wnl-chiral} illustrates the chiral coupling~\eqref{wnl2}: under a pure axial traction, the tube twists although no torque is applied, in a direction set by the sign of $\beta$. At $\tauh_\theta=0$, the twist in~\eqref{4thdef} reduces to its $O(\veps^2)$ term, which matches the numerical solution at $|\beta|=1/2$ and departs from it near the outer wall at $|\beta|=1$.

\end{example}

\section{Conclusion}\label{S:Conclusion}

In this work we have studied, to the best of our knowledge for the first time, the problem of helical shear in a nonlinear anelastic framework. We considered an incompressible, isotropic, hyper-anelastic thick-walled circular tube containing a distributed field of axial screw dislocations, modelled through the Bilby–Kröner–Lee multiplicative decomposition of the deformation gradient and the associated material metric. The reduced equilibrium equations form, at each radius, an algebraic system for the azimuthal and axial shear strains, which we solved in closed form for neo-Hookean, generalized neo-Hookean (Fung–Demiray, Gent, and Knowles), and Mooney–Rivlin materials, and numerically for Gent–Gent materials.

The analysis reveals that the presence of distributed dislocations profoundly affects the equilibrium configuration, though in ways that depend crucially on the constitutive class. For neo-Hookean materials the dislocation field is inert: the kinematics of the deformation and the shear strains are identical to those of a defect-free tube, and dislocations manifest themselves only through residual stress components. For generalized neo-Hookean materials, the dislocation density enters the equilibrium equations through the first invariant and produces a strain-stiffening effect, the helical shear decreasing as the dislocation strength increases. For Gent materials, the dislocation field erodes the extensibility limit, yielding a global condition on the dislocation strength that is necessary and sufficient for the existence of an admissible solution, independently of the applied tractions; a crossover density separates regimes in which the extensibility margin is narrowest at the inner wall from regimes in which it is narrowest at the outer wall, where the tube locks as the cap is approached. For Knowles materials, the power-law index governs existence: for $n>1/2$ a unique equilibrium configuration exists for any traction, for $n=1/2$ the tractions are bounded by limit loads independent of the dislocation field, and for $n<1/2$ the dislocation field erodes the admissible traction region beyond threshold dislocation strengths. For Mooney–Rivlin materials, dislocations induce a chiral coupling between the axial and azimuthal shear modes, so that an axial traction alone produces azimuthal shear and vice versa; this coupling vanishes in the neo-Hookean limit and in the absence of defects, and is therefore a genuine anelastic effect. For Gent–Gent materials, the same cap on the dislocation strength remains necessary, and the numerical solutions reproduce the erosion, outer-wall locking and crossover of the Gent tube, the softer response in shear shifting the crossover to higher dislocation strengths.
We have also extended the universal relations known for generalized neo-Hookean defect-free tubes to dislocated generalized neo-Hookean tubes. These relations retain the dislocation field explicitly and reduce to the classical ones in its absence.

A weakly nonlinear analysis, to third order in the applied tractions and dislocation strength, then identifies the origin of the mode coupling independently of the constitutive model.
In the absence of dislocations, although the two shear modes do interact, they do not couple for any material and at any order, neither traction inducing shear in the other mode.
Axial screw dislocations, by contrast, leave the shear response unaffected at first order, i.e. within the linear theory of elasticity, while carrying a residual stress, and couple the modes chirally at second order, i.e. within third-order elasticity, so that every material behaves there as the Mooney–Rivlin tube with the same second- and third-order elastic moduli.
At third order, no further coupling arises, but the modes interact through stiffness corrections involving the fourth-order elastic modulus: a dislocation-free correction, proportional to the combined traction of both modes, which vanishes for neo-Hookean and Mooney–Rivlin materials, and a dislocation correction, acting on each mode independently of the other, which vanishes for neo-Hookean materials but not for Mooney–Rivlin materials.
Compared with the numerical Gent–Gent solutions, the expansion is accurate up to an $O(\veps^4)$ error and captures the twist induced in the tube by an imposed traction on the outer wall. Notably, this twist is observed under pure axial traction, which isolates the mode coupling unambiguously.

Taken together, these results show that helical shear, despite its kinematic simplicity, is a remarkably rich probe of anelastic behaviour. It allows the effects of eigenstrains to be isolated, their interplay with constitutive nonlinearity to be quantified, and the limits of validity of the models to be identified. The present work can be regarded as a first step towards a systematic extension of the universal program of nonlinear elasticity to defective solids: natural directions for future research include the analysis of other families of general solutions in anelasticity, the study of compressible and anisotropic materials, the incorporation of more general defect distributions and dislocation densities, and the comparison of the theoretical predictions with experiments on soft materials with engineered eigenstrains.

A more challenging class of problems is represented by deformations, such as anti-plane shear, in which the balance equations in the inelastic regime reduce to a single partial differential equation rather than to a system of ordinary differential equations. These problems require mathematical tools that are considerably more sophisticated than those used here — for instance, the theory of fully nonlinear elliptic equations, maximum principles, and regularity results for quasilinear systems — and they raise genuinely new questions concerning existence, uniqueness, and the possible formation of singularities in the presence of eigenstrains. Extending the anelastic universal program to this setting would constitute a natural and ambitious next step and would bring the analysis of defective solids closer to the mathematical standards achieved in the classical theory of nonlinear elasticity.

\section*{Declarations}

\paragraph{Funding.} S.S. was invited to Perugia under a Visiting Fellowship of the GNFM of INDAM. G.S. and L.V. are supported by the GNFM of INDAM.

\paragraph{Competing interests.} The authors have no relevant financial or non-financial conflicting interests to disclose.

\paragraph{Data availability.} No new data were generated in this study. The experimental data used in~\S\ref{S:GG} are those of~\citet{dobrynin2011universality}, as fitted by~\citet{destrade_penelope}.
  
\paragraph{Code availability.} The MATLAB scripts used to produce the numerical results, Figs.~\ref{fig:FD}–\ref{fig:wnl-chiral}, and Table~\ref{tab:wnl} are available from the corresponding author upon reasonable request.

\paragraph{Author contributions.} All authors contributed to the conception of the study, the analysis, the numerical computations, and the writing of the manuscript, and reviewed and approved the final version.

\bibliography{HelShear_ref}

\end{document}